\documentclass[11pt]{article}
\usepackage{jheppub}
\usepackage[table]{xcolor}
\usepackage{amsmath}
\usepackage{amssymb}
\usepackage{physics}
\usepackage{slashed}
\usepackage{graphicx}
\graphicspath{{./images}}
\usepackage{subcaption}
\usepackage{array}
\usepackage{tikz}
\usepackage{url}
\usepackage{multirow}
\usepackage{multicol}
\usepackage{tabularx}
\usepackage{float}
\restylefloat{table}
\usepackage{longtable}
\usepackage{tikz-feynman}
\tikzfeynmanset{compat=1.1.0}
\usepackage{tikz}
\usepackage{tikz-cd}
\usetikzlibrary{positioning, arrows.meta}
\usepackage{multicol}
\usepackage[normalem]{ulem}
\usepackage{tabularx, booktabs, makecell}

\renewcommand{\arraystretch}{1.25}

\newcolumntype{Y}{>{\centering\arraybackslash}X}

\def\nn{\nonumber\\ }
\def\rd{{\rm d}}

\title{LFV in flavourful SMEFT: Dimension-Six Running versus Dimension-Eight Mixing}

\author[a]{Md Isha Ali,}
\author[b]{Siddhartha Karmakar,}
\author[c]{N Rajeev,}
\author[d]{and Sudhir K. Vempati}

\affiliation[a]{School of Physical Sciences, Indian Association for the Cultivation of Science, Kolkata - 700032, India}
\affiliation[b]{Department of Physics, Indian Institute of Technology Kanpur, Kanpur - 208016, India}
\affiliation[c]{Bogoliubov Laboratory of Theoretical Physics, Joint Institute for Nuclear Research, Dubna - 141980, Russia}
\affiliation[d]{Centre for High Energy Physics, Indian Institute of Science, Bengaluru - 560012, India}

\emailAdd{isha.ali080@gmail.com}
\emailAdd{siddharthak@iitk.ac.in}
\emailAdd{rajeevneutrino@gmail.com}
\emailAdd{vempati@iisc.ac.in}

\abstract{We study how charged-lepton flavour violation (CLFV) generated in the $\tau$ sector propagates into the $\mu\to e$ sector within the Standard Model Effective Field Theory (SMEFT). Assuming that new physics at a scale $\Lambda\gg v$ produces only dimension-6 operators that mediate $\tau\to e$ and $\tau\to\mu$ transitions, we identify and compare on an equal footing the three mechanisms that induce $\mu\to e$ observables. All three originate from double insertions of the two $\tau\to\ell$ operators and are proportional to $C_6^{\tau e}C_6^{\tau\mu}/\Lambda^4$ up to couplings, loop factors and logarithms. (T1) Renormalization-group mixing misaligns the charged-lepton mass matrix, and the rotation to the mass basis generates dimension-6 $\mu\to e$ operators. (T2) A finite, non-logarithmic remainder, which we find to be negligible. (T3) The overall divergence renormalizes dimension-8 operators, contributing after electroweak symmetry breaking. Since T1 and T3 carry no relative power of $\Lambda$, neither can be neglected at any scale, and only their direct comparison reveals which mechanism, and hence which $\mu\to e$ observable, most strongly probes a given operator pair. Surveying the Warsaw-basis operators, we classify representative $(\tau\to e,\tau\to\mu)$ pairs into T1-dominated, T3-dominated, and T1--T3-comparable regimes, and explain the hierarchy through the underlying 
Yukawa-, gauge- and Higgs-sector mixing chains and mass-basis rotations. Translating current $\mu\to e\gamma$, $\mu\to eee$ and $\mu\to e$ conversion limits into bounds on products of $\tau\to\ell$ Wilson coefficients, we find that they exceed the direct $\tau$-decay constraints by factors of ten to several hundred for a wide class of operators. Excluding contribution from either of the mechanisms T1 or T3 would miss these limits by orders of magnitude for some pairs, and shift them by factors of two to three in others through the neglected contribution. Interpreting $\mu\to e$ data as a probe of $\tau$ flavour violation therefore requires both the dimension-6 RG running and the dimension-8 mixing contributions.}

\begin{document}
\maketitle

\newpage
\section{Introduction} \label{sec:intro}

The Standard Model (SM) of particle physics, based on the gauge symmetry $SU(3)_C \times SU(2)_L \times U(1)_Y$, together with the Higgs mechanism and the Cabibbo--Kobayashi--Maskawa (CKM) mechanism for quark mixing~\cite{Cabibbo:1963yz,Kobayashi:1973fv}, provides an extraordinarily successful description of strong, weak and electromagnetic interactions. 
Its predictions have been tested with remarkable precision, reaching the per-mille level in several observables~\cite{ALEPH:2005ab}, and the discovery of the Higgs boson~\cite{ATLAS:2012gk,CMS:2012qbp} completed the SM picture. Nevertheless, the SM is widely understood to be an effective theory rather than a fundamental description: it does not explain the origin of neutrino masses~\cite{Super-Kamiokande:1998kpq,SNO:2002tuh}, the nature of dark matter, the baryon asymmetry of the Universe, or the stability of the electroweak scale. These shortcomings motivate the search for new physics (NP) at a scale $\Lambda$ above the electroweak scale.

Flavour physics provides one of the most sensitive indirect probes of such NP. In the quark sector, measurements of neutral meson mixing ($K$, $D$, $B_d$, and $B_s$) and rare flavour-changing neutral current (FCNC) decays such as $b\to s\ell^+\ell^-$ have already placed strong constraints on generic flavour-violating interactions, probing scales ranging from multi-TeV to the PeV regime~\cite{Isidori:2010kg}. The charged-lepton sector is particularly interesting: in the renormalisable SM, individual lepton numbers $L_e$, $L_\mu$, and $L_\tau$ arise as accidental symmetries, rendering charged-lepton flavour violation (CLFV) negligibly small even after neutrino masses are introduced. Consequently, any observation of processes such as $\mu\to e\gamma$, $\mu\to eee$, $\tau\to\ell\gamma$, or coherent $\mu\to e$ conversion would provide an unambiguous indication of physics beyond the SM.

The experimental sensitivity to CLFV has improved dramatically over the past decades. In the muon sector, the MEG experiment has established the stringent bound $\mathrm{BR}(\mu^+\to e^+\gamma)<4.2\times10^{-13}$~\cite{MEG:2016leq}, with MEG~II aiming at a further order-of-magnitude improvement~\cite{MEGII:2018kmf}. The decay $\mu^+\to e^+e^-e^+$ is constrained at the level of $\mathcal{O}(10^{-12})$ by SINDRUM~\cite{SINDRUM:1987nra}, while Mu3e is expected to reach sensitivities of $10^{-16}$--$10^{-17}$~\cite{Blondel:2013ia,Mu3e:2014nml}. Similarly, coherent $\mu\to e$ conversion has been constrained by SINDRUM~II~\cite{SINDRUMII:2006dvw} and will be explored at $10^{-17}$ sensitivity by Mu2e and COMET~\cite{Mu2e:2014fns,Adamov:2018vin}. Complementary searches in the $\tau$ sector from BaBar and Belle constrain $\tau\to\ell\gamma$ and $\tau\to\ell\ell'\ell''$ decays at the $\mathcal{O}(10^{-8})$ level~\cite{Aubert:2009ag,BaBar:2009hkt,Belle:2007qzi,Hayasaka:2010np}, with Belle~II expected to significantly improve these limits~\cite{Belle-II:2018jsg}. Direct probes at the LHC through LFV $Z$ and Higgs decays provide additional constraints on lepton-flavour-violating interactions~\cite{CMS:2017con,ATLAS:2019pmk, ATLAS:2021bdj}.

A systematic interpretation of these searches is naturally provided by the Standard Model Effective Field Theory (SMEFT)~\cite{Buchmuller:1985jz,Grzadkowski:2010es}, where NP at a scale $\Lambda\gg v$ is integrated out and its effects are described by higher-dimensional operators invariant under the SM gauge symmetry. The SMEFT framework enables a unified treatment of different CLFV processes and allows operator mixing induced by renormalization-group evolution (RGE) to be consistently included. One-loop analyses of CLFV processes with the complete set of dimension-6 operators were given in Refs.~\cite{Crivellin:2013hpa,Pruna:2014asa}, while complete analyses incorporating RG-improved matching between SMEFT and the low-energy effective theory (LEFT) for $\mu\to e$ transitions were presented in Refs.~\cite{Crivellin:2017rmk,Davidson:2016edt}. The role of higher-order EFT effects, including dimension-8 operators generated by double insertions of dimension-6 operators, was investigated in Refs.~\cite{Ardu:2021koz,Ardu:2022pzk}. More recently, global analyses of LFV observables~\cite{Davidson:2020ord,Davidson:2022eff,Plakias:2023esq,Ali:2023kua} and SMEFT studies of semileptonic processes~\cite{Greljo:2023bdy,Karmakar:2024gla,Ali:2025xkw} have emphasized the importance of operator mixing in obtaining consistent bounds on Wilson coefficients.

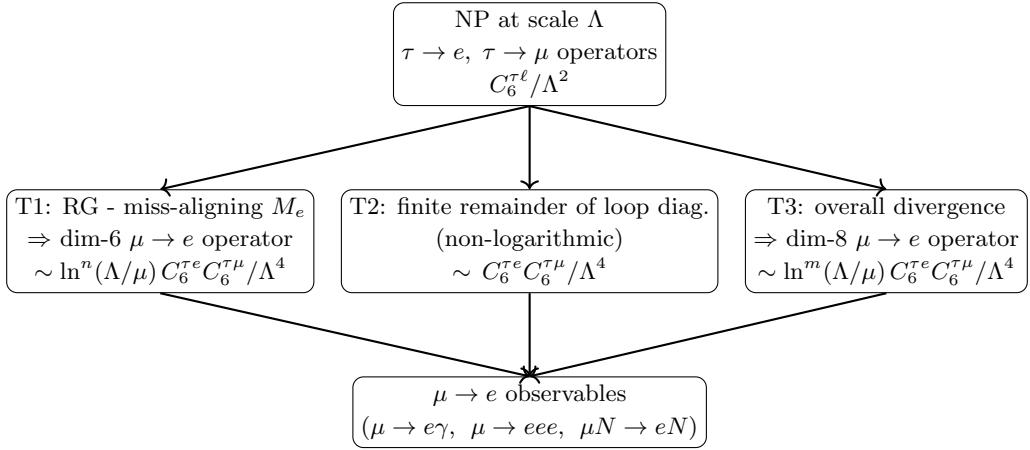
\begin{figure}[t]
\centering
\begin{tikzpicture}[
    node distance=1.1cm and 0.35cm,
    every node/.style={font=\footnotesize},
    box/.style={draw, rectangle, rounded corners, align=center,
                minimum width=2.6cm, minimum height=0.9cm, inner sep=3pt},
    arrow/.style={->, thick}
]

\node[box] (NP) {NP at scale $\Lambda$\\
                  $\tau \to e,\;\tau \to \mu$ operators\\
                  $C^{\tau\ell}_6/\Lambda^2$};

\node[box, below=of NP]        (T2) {T2: finite remainder of loop diag.\\
                                      (non-logarithmic)\\ 
                                      $\sim \, C_6^{\tau e} C_6^{\tau\mu}/\Lambda^4$};

\node[box, left=of T2]         (T1) {T1: RG - miss-aligning $M_e$\\
                                      $\Rightarrow$ dim-6 $\mu\to e$ operator\\
                                      $\sim  \ln^{n}(\Lambda/\mu)\, C_6^{\tau e} C_6^{\tau\mu}/\Lambda^4$};

\node[box, right=of T2]        (T3) {T3: overall divergence\\
                                      $\Rightarrow$ dim-8 $\mu\to e$ operator\\
                                      $\sim \ln^{m}(\Lambda/\mu)\, C_6^{\tau e} C_6^{\tau\mu}/\Lambda^4$};

\node[box, below=of T2]        (obs) {$\mu \to e$ observables\\
                                       ($\mu\to e\gamma$,\;
                                        $\mu\to eee$,\;
                                        $\mu N \to e N$)};

\draw[arrow] (NP.south)  -- (T1.north);
\draw[arrow] (NP.south)  -- (T2.north);
\draw[arrow] (NP.south)  -- (T3.north);
\draw[arrow] (T1.south)  -- (obs.north);
\draw[arrow] (T2.south)  -- (obs.north);
\draw[arrow] (T3.south)  -- (obs.north);

\end{tikzpicture}
\caption{Schematic illustration of the three classes of contributions to $\mu \to e$ transitions induced by high-scale $\tau\to\ell$ operators. Each estimate holds up to the couplings and loop factors of the corresponding mixing chain, with $n=1$--$3$ and $m=1,2$. $M_e$ is the charged lepton mass matrix.}
\label{fig:T1T2T3_schematic}
\end{figure}

In this work we study how high-scale $\tau\to\ell$ interactions generate $\mu\to e$ transitions in the SMEFT, working in the Warsaw basis for dimension-6 flavour off-diagonal dipole, four-fermion, and modified gauge-coupling operators. Since RGE running mixes operators of different flavour structure~\cite{Jenkins:2013zja,Jenkins:2013wua,Alonso:2013hga}, NP generating only $\tau\to\ell$ operators at the matching scale induces $\mu\to e$ operators at lower scales, making $\mu\to e$ observables indirect probes of the $\tau e$ and $\tau\mu$ sectors. Beyond this RGE-induced dimension-6 effect, double insertions of two dimension-6 $\tau\to\ell$ operators contribute to $\mu\to e$ transitions already at order $C_6^{\tau e}C_6^{\tau\mu}/\Lambda^4$: their divergences renormalise dimension-6 operators as well as genuine dimension-8 operators, leaving a finite remainder besides. The dimension-8 piece was analysed in Ref.~\cite{Ardu:2022pzk}, while the global-analysis frameworks of Refs.~\cite{Davidson:2020ord,Davidson:2022eff} bound the dimension-6 $\mu\to e$ Wilson coefficients directly from $\mu\to e\gamma$, $\mu\to eee$ and $\mu\to e$ conversion, without addressing how these coefficients could themselves arise radiatively from $\tau\to\ell$ operators or their accompanying dimension-8 piece. To our knowledge, no prior work treats the RGE-induced dimension-6 and dimension-8 effects together, even though, as we show below, they are parametrically of the same order and only their sum gives a consistent leading-order amplitude. We provide this unified treatment, computing the three contributions illustrated in Fig.~\ref{fig:T1T2T3_schematic}. (T1) The RG evolution of one dimension-6 operator misaligns the charged-lepton mass matrix,  and the rotation to the mass basis acts on the other operator, generating dimension-6 $\mu\to e$ operators enhanced by up to three powers of $\ln(\Lambda/\mu)$. (T2) The finite, non-logarithmic remainder from the one-loop diagram with the double insertion. (T3) The overall divergence renormalizing dimension-8 operators that contribute after electroweak symmetry breaking. All the three contributions scale as $C_6^{\tau e}C_6^{\tau\mu}/\Lambda^4$ and differ in coulplings, loop factors and powers of logarithms. Dropping either logarithmic piece is therefore inconsistent at this order. The resulting effective dimension-6 $\mu\to e$ coefficients, with the T1--T3 interference included, can feed directly into the global-fit machinery of Refs.~\cite{Davidson:2020ord,Davidson:2022eff}, extending its reach to a combined interpretation of $\mu\to e$ searches in terms of $\tau$ flavour violation. 

We use \texttt{wilson}~\cite{Aebischer:2018bkb} for the SMEFT and LEFT evolution and matching, including the mass-basis rotation that yields T1, and \texttt{flavio}~\cite{Straub:2018kue} for the low-energy $\mu\to e$ observables, extending this pipeline to the double-insertion effects by injecting the dimension-8 contribution as an effective dimension-6 coefficient at the electroweak scale. The required anomalous dimensions of the dimension-8 operators for our T3 calculation are obtained from Ref.~\cite{Ardu:2022pzk}.
For light external quarks the dimension-8 coefficient is matched onto an effective dimension-6 coefficient suppressed by $v_T^2/2\Lambda^2$. For top-flavoured operators, whose direct matching vanishes, the leading contribution is a double logarithm through mixing into Higgs-lepton or dipole operators. T2 is computed analytically and shown to be numerically negligible for the observables considered.
Since both T1 and T3 scale with the same product of coefficients, the dominant mechanism is a property of the operator pair rather than of the parameter region.
In our analysis, we find that T1 dominates when a single rotation step transfers the flavour violation and T3 dominates when the T1 chain needs additional loop steps. The two are comparable when both are one-loop, single-log effects. The resulting indirect bounds on products of $\tau\to\ell$ Wilson coefficients improve on the direct $\tau$-decay bounds by one to three orders of magnitude, and dropping either T1 or T3 shifts them by factors of two to three in some cases and by orders of magnitude in others.

The paper is organised as follows. Section~\ref{sec:framework} introduces the SMEFT and LEFT operator bases and defines the T1, T2, T3 decomposition of the $\mu\to e$ amplitude. Section~\ref{directbound} presents  the current and projected future bounds on the SMEFT Wilson coefficients for LFV operators.  Section~\ref{sec:indirect} presents the indirect bounds for T1-dominated, T3-dominated, and T1--T3 comparable scenarios. Conclusions are given in the final section Sec.~\ref{conclusion}, and complete tables of operator combinations are collected in the appendices.

\section{Lepton flavor violation in SMEFT}\label{sec:framework}

In this section we introduce the effective field theory framework used to describe charged-lepton flavour violation induced by heavy new physics. We work within the SMEFT, where the effects of degrees of freedom at a high scale $\Lambda \gg v$ are encoded in higher-dimensional operators constructed from SM fields and respecting the $SU(3)_C \times SU(2)_L \times U(1)_Y$ gauge symmetry. The SMEFT Lagrangian can be written as
\begin{align}
\mathcal{L}_\text{eff} = \mathcal{L}_\text{SM} + \sum_i \frac{C_i^{(6)}}{\Lambda^2}\,\mathcal{O}_i^{(6)} + \sum_j \frac{C_j^{(8)}}{\Lambda^4}\,\mathcal{O}_j^{(8)} + \cdots,
\end{align}
where $\mathcal{O}_i^{(d)}$ are operators of mass dimension $d$, and $C_i^{(d)}$ denote the corresponding Wilson coefficients encoding the strength and flavour structure of the underlying new physics. Operators of odd dimension violate lepton number and are not relevant for the processes considered here. Throughout we adopt the Warsaw basis~\cite{Grzadkowski:2010es} for the
dimension-6 operators in the WCxf conventions~\cite{Aebischer:2017ugx} employed by \texttt{wilson} and \texttt{flavio}, and the Murphy basis~\cite{Murphy:2020rsh, Li:2020gnx} for the dimension-8 operators. We work in the down-aligned
weak basis, in which $Y_d$ and $Y_e$ are diagonal and the CKM matrix resides
in the up-type Yukawa, $Y_u = V_{\rm CKM}^\dagger\,\hat{Y}_u$.

In this work we focus on the dimension-6 operators that provide the leading contributions to CLFV processes. Lepton-flavour violation arises from operators with off-diagonal lepton
flavour indices, broadly classified into dipole, Higgs-current, and four-fermion operators; those relevant to our study are collected in Table~\ref{tab:operators}. 
A single such operator induces $\tau \to e$ or $\tau \to \mu$ transitions, while $\mu \to e$ processes arise only from pairs of them, at order $C_6^{\tau e}C_6^{\tau\mu}/\Lambda^4$. 

\begin{table}[h] 
\centering
\renewcommand{\arraystretch}{1.4}
\begin{tabular}{|c|c|} 
\hline 
Operator type & Most relevant operators  \\
\hline \hline
Dipole (D6) 
& $\mathcal{O}_{eB}$, $\mathcal{O}_{eW}$ \\
\hline
Penguin (P6) 
& $\mathcal{O}_{\phi \ell}^{(1)}$, $\mathcal{O}_{\phi \ell}^{(3)}$, $\mathcal{O}_{\phi e}$ \\
\hline
Yukawa (Y6) 
& $\mathcal{O}_{e \phi}$ \\
\hline
Four-lepton (4l6) 
& $\mathcal{O}_{ee}$, $\mathcal{O}_{\ell e}$, $\mathcal{O}_{\ell \ell}$ \\
\hline
Two-lepton–two-quark (4f6) 
& $\mathcal{O}_{\ell q}^{(1,3)}$, $\mathcal{O}_{\ell u}$, $\mathcal{O}_{\ell d}$, 
$\mathcal{O}_{eq}$, $\mathcal{O}_{eu}$, $\mathcal{O}_{ed}$, 
$\mathcal{O}_{\ell edq}$, $\mathcal{O}_{\ell equ}$ \\
\hline \hline
Dipole (D8) 
& $\mathcal{O}_{\ell eW\phi^3}^{(1)}$, $\mathcal{O}_{\ell eW \phi^3}^{(2)}$, $\mathcal{O}_{\ell eB\phi^3}$ \\
\hline
Penguin (P8) 
& $\mathcal{O}_{\ell^2 \phi^4 D}^{(1)}$, $\mathcal{O}_{\ell^2 \phi^4 D}^{(2)}$, $\mathcal{O}_{e^2 \phi^4 D}$ \\
\hline
Four-lepton (4l8) 
& $\mathcal{O}_{\ell^2 e^2 \phi^2}^{(1,2,3,4)}$, $\mathcal{O}_{\ell^4 \phi^2}^{(1,2)}$, $\mathcal{O}_{e^4 \phi^2}$ \\
\hline
Two-lepton–two-quark (4f8) 
& $\mathcal{O}_{\ell^2 q^2 \phi^2}^{(1,2,3,4)}$, $\mathcal{O}_{\ell^2 u^2 \phi^2}^{(1,2)}$, $\mathcal{O}_{\ell^2 d^2 \phi^2}^{(1,2)}$, \\
& $\mathcal{O}_{e^2 q^2 \phi^2}^{(1,2)}$, $\mathcal{O}_{e^2 u^2 \phi^2}$, $\mathcal{O}_{e^2 d^2 \phi^2}$, \\
& $\mathcal{O}_{\ell edq \phi^2}^{(1,2,3,4)}$, $\mathcal{O}_{\ell equ \phi^2}^{(1,2,3,4,5)}$ \\
\hline
\end{tabular}
\caption{Relevant dimension--6~\cite{Grzadkowski:2010es} and dimension--8~\cite{Murphy:2020rsh} SMEFT operators contributing to charged-lepton flavour violation. For dimension--8 operators, only operator classes are shown, while explicit forms are omitted for brevity.}
\label{tab:operators}
\end{table}

We consider a scenario in which two dimension-6 lepton flavour violating operators, $O^{\tau \to e}_6$ and $O^{\tau \to \mu}_6$, are generated at the NP scale $\Lambda$. Even in the absence of a direct $\mu \to e$ operator at this scale, such interactions induce $\mu \to e$ transitions at low energies through three mechanisms. The resulting NP contribution to the $\mu \to e$ amplitude can be expressed schematically as

\begin{align}
\mathcal{A}^{\rm NP} = 
T1\!\left(\frac{C_6^{\mu\to e}(\mu)}{\Lambda^2}\right)
+T2
+T3\!\left(\frac{v^2}{2}\,\frac{C_8^{\mu\to e}(\mu)}{\Lambda^4}\right),
\qquad
C_6^{\mu\to e},\;\frac{v^2}{2}\frac{C_8^{\mu\to e}}{\Lambda^2}
\;\propto\;\frac{v^2}{\Lambda^2}\,C_6^{\tau e}C_6^{\tau\mu}\,,
\label{amp1}
\end{align}

where $\mu$ denotes the low-energy scale and the second relation exhibits the common scaling of the two logarithmic contributions. The three terms in eq.~(\ref{amp1}) arise from qualitatively different sources:
\begin{itemize}
    \item[(T1)]  The renormalization-group contribution.
    The RG evolution renormalizes dimension-6 operators and $Y_e$ at linear order, and the product of coefficients arises upon integrating the coupled RG system and rotating to the charged-lepton mass basis. On operator misaligns the mass matrix, and the rotation acts on the other. 
    Since a dimension-6 counter-term requires a coefficient of mass dimension $-2$, this contribution  is proportional to the electroweak mass scale, entering either as $m_\phi^2$ through the running of $Y_e$ or as $v^2$ through the mass-matrix misalignment induced by $C_{e\phi}$. It is the leading-log part of the double insertion, enhanced by powers of $L = \ln(\Lambda/\mu)$.

    \item[(T2)]  
    The finite, non-logarithmic remainder of the one-loop double insertion once its divergence has been subtracted and the local finite pieces have been absorbed into the T3 matching.
    T2 depends on the external invariants rather than being a constant, and reduces to a local coefficient only when expanded at a fixed kinematic point, as we do
    in Sec.~\ref{sec:T2} at the kinematics of the $\mu\to e$ observables.

    \item[(T3)] 
    The one-loop double insertion possesses an irreducible overall divergence, whose divergent structure carries dimension-8 field content and which is not generated by the iteration of single insertions.
    It is absorbed by local dimension-8 counterterms with coefficients $C^{\mu\to e}_8 \propto C^{\tau\to e}_6 C^{\tau\to\mu}_6$, requiring no $m_\phi^2$ factor. After electroweak symmetry breaking these operators contribute to $\mu\to e$ amplitudes with a factor $v^2/2\Lambda^2$ relative to a dimension-6 coefficient, so that T3, like T1, scales as $v^2C_6^{\tau e}C_6^{\tau\mu}/\Lambda^4$ and carries no relative power of $\Lambda$~\footnote{The separation between T1 and T3 is unambiguous once the operator bases are fixed: we work in the fully reduced Warsaw and Murphy bases, with all redundant dimension-8 operators removed by field redefinitions, and the anomalous dimensions of Ref.~\cite{Ardu:2022pzk} are given in this reduced basis. T1 resides entirely in the running and rotation of the dimension-6 coefficients, T3 in the coefficients of non-redundant dimension-8 operators, so no contribution is counted
    twice; only their sum is basis independent.}.
\end{itemize}

We assume throughout that the new physics at $\Lambda$ generates only dimension-6 operators, so that all dimension-8 coefficients relevant here are radiatively generated. They arise as counterterms for the ultraviolet divergences of double insertions of two dimension-6 operators, rather than being present at the matching scale. In a UV completion that also matches directly onto a $\mu\to e$ dimension-8 coefficient, the contributions computed here constitute an irreducible radiative floor, and the indirect bounds derived below are correspondingly conservative.

Throughout, we work in dimensional regularization with $\overline{\rm MS}$ subtraction: the dimension-6 anomalous dimensions of
Refs.~\cite{Jenkins:2013zja,Jenkins:2013wua,Alonso:2013hga}, the dimension-8 anomalous dimensions of Ref.~\cite{Ardu:2022pzk}, and the finite T2 remainder are all defined in this scheme. The Wilson coefficients are specified at the matching scale $\Lambda$, and the three contributions are evolved to low energies as follows.

Our analysis proceeds as follows. For each pair of operators $(O^{\tau \to e}_6,\, O^{\tau \to \mu}_6)$, we compute the three contributions T1, T2, and T3 defined above, and evaluate their impact on $\mu \to e$ observables. By confronting these predictions with current experimental limits, we derive indirect constraints on the product of Wilson coefficients $C^{\tau \to e}_6(\Lambda)\, C^{\tau \to \mu}_6(\Lambda)$. We find that, in several cases, these indirect bounds are more stringent than the direct limits on the individual Wilson coefficients.
In the following, we illustrate the origin and structure of the three contributions with a representative example. The full set of results is presented in Sec.~\ref{sec:indirect}.   

\subsection{Contribution from operator mixing (T1)}\label{sec:T1}

A pair of dimension-6 operators $\mathcal{O}^{\tau\to\mu}$ and $\mathcal{O}^{\tau\to e}$ can generate a dimension-6 $\mu\to e$ operator through RG evolution combined with the rotation to the charged-lepton mass basis. 
In the flavour basis in which $Y_e(\Lambda)$ is diagonal, the $\tau\ell$ Wilson coefficients carry no $\mu e$ flavour, and gauge and top-Yukawa mixing preserve lepton flavour. Therefore, the transfer proceed through the charged-lepton Yukawa sector, the only lepton-flavour-violating spurion available to the RG.  The running of $Y_e$ in the presence of dimension-6 operators is~\cite{Alonso:2013hga}
\begin{align}
\mu \frac{\rd}{\rd \mu}  [Y_e]_{rs} &=  \frac{m_\phi^2 }{16\pi^2 {\Lambda^2}}  \biggl[3 C_{\substack{e\phi \\ sr}}^* - C_{\phi \Box} [Y_e]_{rs} + \frac12 C_{\phi D} [Y_e]_{rs} +
[Y_e]_{rt} \left( C^{(1)}_{\substack{\phi l \\ ts}} +3 C^{(3)}_{\substack{\phi l \\ ts}}\right) -C_{\substack{\phi e \\ rt}}  [Y_e]_{ts} \nn
&-2 C_{\substack{le \\ psrt}}[Y_e]_{tp} + N_c C^{*}_{\substack{ledq \\ srpt}}[Y_d]_{pt}
-N_c C^{(1)*}_{\substack{lequ \\ srpt}} [Y_u^\dagger]_{pt} \biggr]\,,
\label{yukawarg}
\end{align}
so a $\mu e$ operator can be seeded only by pairs in which at least one operator feeds $Y_e$ or $C_{e\phi}$, directly or through mixing; this includes the four-lepton (4l6), four-fermion (4f6), penguin (P6) and Yukawa (Y6) classes of Table~\ref{tab:operators}.
As a concrete example we take $\mathcal{O}_{eu}^{1333}$ and $\mathcal{O}_{eu}^{2333}$, a 4f6-4f6  pair.
At $\Lambda = 4$ TeV the direct constraints on the corresponding WCs are
\begin{align}
    C_{eu}^{1333}/\Lambda^2 < 4.3\times 10^{-7} ~ {\rm GeV}^{-2},\quad C_{eu}^{2333}/\Lambda^2 < 5\times 10^{-7} ~ {\rm GeV}^{-2} 
\end{align}
This pair generates the $\mu e$ coefficients $C^{1233}_{eu}(\mu)$ and $C^{12}_{\phi e}(\mu)$, which we now derive at the leading log. We treat $y_t$, $y_\tau$, $\lambda$, and $m_\phi^2$ as constants over the running interval, define $L \equiv \ln(\Lambda/\mu) > 0$, $y_t^2 \equiv [Y_uY_u^\dagger]_{33}$, $y_\tau \equiv [Y_e]_{33}$, and impose the boundary conditions that $C^{1333}_{eu}(\Lambda)$ and $C^{2333}_{eu}(\Lambda)$ are the only nonzero Wilson coefficients, with $Y_e(\Lambda)$ diagonal. Hermiticity of $\mathcal{O}_{eu}$ implies
$C^{3233}_{eu} = (C^{2333}_{eu})^{*}$ and
$C^{3133}_{eu} = (C^{1333}_{eu})^{*}$.
Three coefficients are generated at one loop from $C^{i333}_{eu}(\Lambda)$, $i=1,2$.  The $y_t^2$-enhanced term $-2N_c\,C_{eu,prst}[Y_uY_u^\dagger]_{ts}$ of the $C_{\phi e}$ anomalous dimension~\cite{Jenkins:2013wua} gives
\begin{align}
	C^{i3}_{\phi e}(\mu) \simeq \frac{2N_c\, y_t^2}{16\pi^2}\, L\;
	C^{i333}_{eu}(\Lambda)\,;
	\label{eq:CHe_i3_LL}
\end{align}
Similarly from $\dot C^{(1)}_{lequ,3i33} \simeq 2y_\tau y_t\,(C^{i333}_{eu})^{*}$ we get
\begin{align}
	C^{(1)}_{lequ,\,3i33}(\mu) =
	-\,\frac{2\, y_\tau\, y_t}{16\pi^2}\, L\;
	\big(C^{i333}_{eu}(\Lambda)\big)^{*};
	\label{eq:lequLL}
\end{align}
The generated $C^{(1)}_{lequ}$ in turn sources $C_{e \phi}$ through its top-Yukawa
and Higgs-quartic anomalous dimension,
$\dot C^{3i}_{e \phi} \simeq (4N_c y_t^3 - 6\lambda y_t)\,
C^{(1)}_{lequ,3i33}$
[Eq.~(A.18) of Ref.~\cite{Jenkins:2013wua} together with the
$\lambda$-dependent terms of Ref.~\cite{Jenkins:2013zja}], so that at
$\mathcal{O}(L^2)$
\begin{align}
	C^{3i}_{e \phi}(\mu) =
	\frac{y_\tau\, y_t^2\,\big(4N_c\, y_t^2 - 6\lambda\big)}
	{(16\pi^2)^2}\; L^2\;
	\big(C^{i333}_{eu}(\Lambda)\big)^{*}.
	\label{eq:CeHLL}
\end{align}
The first two coefficients from Eqs.\,\eqref{eq:CHe_i3_LL} and \eqref{eq:lequLL} feed back into Eq.~\eqref{yukawarg} through the terms $-C_{\phi e}[Y_e]$ and $-N_cC^{(1)*}_{lequ}[Y_u^\dagger]$, and their contributions cancel identically at leading log,
{\small
\begin{align}
	\mu\frac{\mathrm{d}}{\mathrm{d}\mu}\,[Y_e]_{i3}
	= \frac{m_\phi^2}{16\pi^2\Lambda^2}
	\Big[ -\,y_\tau\, C^{i3}_{\phi e}
	- N_c\, y_t \big(C^{(1)}_{lequ,\,3i33}\big)^{*} \Big]
	= \frac{m_\phi^2\, y_\tau\, y_t^2\, L}{(16\pi^2)^2\Lambda^2}
	\big(\!-2N_c + 2N_c\big)\, C^{i333}_{eu}(\Lambda)
	= 0\,,
	\label{eq:Yecancel}
\end{align}
}
so no off-diagonal Yukawa coupling is generated at this order. The flavour transfer nevertheless survives, because in SMEFT the charged-lepton mass matrix is not $Y_e$ alone but
[cf.\ Eq.~(3.2) of Ref.~\cite{Jenkins:2013wua}]
\begin{align}
	[M_e]_{rs} = \frac{v}{\sqrt2}
	\Big( [Y_e]_{rs} - \frac{v^2}{2\Lambda^2}\, C^{*}_{e \phi,\,sr} \Big),
	\label{eq:massmatrix}
\end{align}
and Eq.~\eqref{eq:CeHLL} misaligns it even though $Y_e$ remains diagonal: $[M_e]_{i3} \simeq -\frac{v^3}{2\sqrt2\,\Lambda^2} (C^{3i}_{e \phi})^{*} \neq 0$, while $[M_e]_{3i} = 0$.  Physical amplitudes are computed in the basis in which $M_e$ is diagonal, reached by a rotation of the right-handed (singlet) fields alone, $e = U_e\, e'$
with $U_e \simeq 1 + A$,
$A_{i3} = \theta_i = -A^{*}_{3i}$, where
\begin{align}
	\theta_i \equiv \frac{[M_e]_{i3}}{[M_e]_{33}}
	= -\,\frac{v^2}{2\, y_\tau\,\Lambda^2}\,\big(C^{3i}_{e \phi}\big)^{*}
	= -\,\frac{v^2\, y_t^2\,\big(2N_c\, y_t^2 - 3\lambda\big)}
	{(16\pi^2)^2\,\Lambda^2}\; L^2\;
	C^{i333}_{eu}(\Lambda)\,.
	\label{eq:theta}
\end{align}
The re-diagonalization generates the $\mu e$ operators at tree level, as follows.  In terms of the mass-basis fields,
\begin{align}
	e_i = e'_i + \theta_i\, e'_3\,, \qquad
	e_3 = e'_3 - \theta_1^{*}\, e'_1 - \theta_2^{*}\, e'_2\,,
	\label{eq:fieldrot}
\end{align}
so substituting Eq.\,\eqref{eq:fieldrot} into the $\tau\ell$ operators rotates their $\tau$-singlet legs into light-lepton legs:
\begin{align}
\mathcal{O}_{eu}^{1333}	= (\bar e_1\gamma^\mu e_3)(\bar u_3\gamma_\mu u_3) \;\supset\; -\,\theta_2^{*}\, (\bar e'_1\gamma^\mu e'_2)(\bar u_3\gamma_\mu u_3)\,,
\end{align}
with $\theta_2^{*} \propto (C^{2333}_{eu})^{*}$, while the conjugate operator $\mathcal{O}_{eu}^{3233}$ contributes $-\theta_1\, C^{3233}_{eu}$ through its $\bar e_3$ leg.  

The double-insertion structure of T1 is thus explicit. One operator tilts the mass matrix through the chain $C_{eu}\to C^{(1)}_{lequ}\to C_{e\phi}$, and the other supplies the remaining flavour indices when its $\tau$ leg is rotated away. Collecting both terms, and applying the same rotation to the $\mathcal{O}(L)$ coefficients $C^{i3}_{\phi e}$ of
Eq.\,\eqref{eq:CHe_i3_LL},
\begin{align}
	C^{1233}_{eu}(\mu) &=
	-\,\theta_1\, C^{3233}_{eu} - \theta_2^{*}\, C^{1333}_{eu}
	=
	\frac{2\, v^2\, y_t^2\,\big(2N_c\, y_t^2 - 3\lambda\big)}
	{(16\pi^2)^2\,\Lambda^2}\; L^2\;
	C^{1333}_{eu}(\Lambda)\,\big(C^{2333}_{eu}(\Lambda)\big)^{*},
	\label{eq:T1eu}
	\\[2pt]
	C^{12}_{\phi e}(\mu) &=
	-\,\theta_1\, C^{32}_{\phi e} - \theta_2^{*}\, C^{13}_{\phi e}
	=
	\frac{4 N_c\, v^2\, y_t^4\,\big(2N_c\, y_t^2 - 3\lambda\big)}
	{(16\pi^2)^3\,\Lambda^2}\; L^3\;
	C^{1333}_{eu}(\Lambda)\,\big(C^{2333}_{eu}(\Lambda)\big)^{*}.
	\label{eq:T1phie}
\end{align}
We have validated Eqs.~\eqref{eq:T1eu} and \eqref{eq:T1phie} against the full numerical solution of \texttt{wilson}, which performs precisely the re-diagonalization of Eq.~\eqref{eq:fieldrot} when matching at the electroweak scale.

\subsection{Contribution from the dimension 8 operator (T3)}\label{sec:T3}

As discussed in Sec.~\ref{sec:framework}, the divergences of single insertions of the $\tau\to\ell$ operators are linear in the Wilson coefficients and are re-summed by the dimension-6 RG evolution that drives T1. The one-loop graphs with one insertion of each operator carry, in addition, an irreducible overall divergence proportional to the product of the two coefficients. This divergence is not generated by any iteration of single insertions, and its divergent structure carries dimension-8 field content: four fermions and two Higgs fields. It is therefore absorbed by local dimension-8 counter-terms with coefficients proportional to $C_6^{\tau e}C_6^{\tau\mu}/\Lambda^4$, and the associated logarithm constitutes T3.

For the operator pair considered there, the relevant anomalous dimension was computed in Ref.~\cite{Ardu:2022pzk} and after converting to our normalization, it gives
\begin{align}
  16\pi^{2}\,\mu\frac{d}{d\mu}\,C^{\,e\mu ii}_{e^{2}u^{2}\phi^{2}}
  \;=\;
  -\,2\,y_{t}^{2}\;C^{\,e\tau 3i}_{eu}\,C^{\,\tau\mu i3}_{eu}
  \;+\;8\,y_{t}^{2}\;C^{\,e\tau i3}_{eu}\,C^{\,\tau\mu 3i}_{eu}\,,
  \label{eq:dim8RGE}
\end{align}
where $i$ labels the up-type flavour of the external quark bilinear and the third-generation index is contracted inside the top loop. With the boundary condition that the dimension-8 coefficients vanish at the scale $\Lambda$, the solution of Eq.\,\eqref{eq:dim8RGE} at
leading-log accuracy is
\begin{align}
  \frac{C^{\,e\mu ii}_{e^{2}u^{2}\phi^{2}}(\mu)}{\Lambda^{4}}
  =
  \frac{L}{16\pi^{2}}
  \left[\,2\,y_{t}^{2}\;C^{\,e\tau 3i}_{eu}\,C^{\,\tau\mu i3}_{eu}
  - 8\,y_{t}^{2}\;C^{\,e\tau i3}_{eu}\,C^{\,\tau\mu 3i}_{eu}\,\right]
  \frac{1}{\Lambda^{4}}\,.
  \label{eq:dim8sol}
\end{align}
For the input pair of Sec.~\ref{sec:T1}, in which every up-type index is in the third generation, the two flavour structures of Eq.~\eqref{eq:dim8sol} coincide and
\begin{align}
  \frac{C^{\,e\mu 33}_{e^{2}u^{2}\phi^{2}}(\mu)}{\Lambda^{4}}
  =
  -\,\frac{6\,y_{t}^{2}}{16\pi^{2}}\,L\;
  \frac{C^{1333}_{eu}(\Lambda)\,\big(C^{2333}_{eu}(\Lambda)\big)^{*}}{\Lambda^{4}}\,.
  \label{eq:dim8top}
\end{align}
In order to calculate the T3 contribution in \texttt{wilson} and \texttt{flavio}, we take the following two approaches depending on the up-type quark flavor in the dimension-8 operator. 

\paragraph{Light-flavour dimension-8 operators:}
After electroweak symmetry breaking the Higgs fields are replaced by their vacuum expectation value, and dimension-8 operators of the form $O_{6}\times(\phi^{\dagger}\phi)$ contribute to the same WET structures as the corresponding dimension-6 operators \cite{Hamoudou:2022tdn}.
For light up-type flavours the matching relation is
\begin{align}
  \frac{4G_{F}}{\sqrt{2}}\;C^{V,RR}_{eu,\,prst}
  =
  \frac{1}{\Lambda^{2}}
  \left[\,C^{\,prst}_{eu}
  +\frac{v_{T}^{2}}{2\Lambda^{2}}\,C^{\,prst}_{e^{2}u^{2}\phi^{2}}\,\right],
  \qquad s,t\in\{1,2\}\,,
  \label{eq:matching}
\end{align}
with $v_{T}\simeq 246$~GeV, and with the superscript $V,RR$ reflecting that both currents of $O_{eu}$ are right-handed. Eq.~\eqref{eq:matching} shows that, as far as the matching onto WET is concerned, such a dimension-8 operator acts as the corresponding dimension-6 operator multiplied by $v_{T}^{2}/2\Lambda^{2}$. Since \texttt{wilson} and \texttt{flavio} accept dimension-6 input only, we exploit this equivalence and evaluate
T3 by substituting an effective dimension-6 coefficient,
\begin{align}
  \frac{C^{\,\mathrm{eff}}_{6}(\mu_{\rm inj})}{\Lambda^{2}}
  =
  \frac{v_{T}^{2}}{2}\;\frac{C_{8}(\mu_{\rm inj})}{\Lambda^{4}}\,,
  \qquad
  \mu_{\rm inj}\simeq 100~\mathrm{GeV}\,,
  \label{eq:effC6}
\end{align}
injected into \texttt{wilson} just above the electroweak matching scale. The logarithm of Eq.\,\eqref{eq:dim8sol} already resums the running of the dimension-8 coefficient from $\Lambda$ down to $\mu_{\rm inj}$, so injecting the effective coefficient at a higher scale and evolving it would double count this logarithm. Below $\mu_{\rm inj}$ the coefficient is processed by the standard pipeline, namely tree-level matching onto WET at $m_{Z}$ followed by QED and QCD evolution to the observable scale. 

\paragraph{Top-flavoured dimension-8 operators:}
When both up-type indices of the dimension-8 operator are in the third generation, as in Eq.~\eqref{eq:dim8top}, the substitution of Eq.~\eqref{eq:effC6} would produce a $\mu e$ contact interaction with top quarks. Such a coefficient has no correspondence in the low-energy theory with the same generation indices. The top quark is integrated out at the electroweak scale, its tree-level WET matching vanishes. The single-log term of Eq.~\eqref{eq:dim8top} therefore does not contribute to $\mu\to e$ observables.  The leading non-vanishing contribution arises at the next order in $L$, from the top-loop mixing of the top-flavoured dimension-8 operator into the dimension-8 Higgs-current operator~\cite{Li:2020gnx, Wu:2025qto, Murphy:2020rsh} 
$O^{\,e\mu}_{e^{2}\phi^{4}D}
 = i(\bar e_{e}\gamma^{\alpha}e_{\mu})
   (\phi^{\dagger}i\overleftrightarrow{D}_{\!\alpha}\phi)(\phi^{\dagger}\phi)$.
The required dimension-8 anomalous dimension has been recently computed in the literature~\cite{Bakshi:2024wzz,DasBakshi:2026ief}. However, for our purpose, we approximate it by the corresponding dimension-6 entry, $\gamma_{\phi e\leftarrow eu}=-2N_{c}\,y_{t}^{2}$ of Eq.~\eqref{eq:CHe_i3_LL}, since the one-loop diagram is the same top loop with the additional Higgs pair acting as a spectator. The same strategy was adopted in Ref.~\cite{Ardu:2022pzk} for the mixing of top-flavoured dimension-8 tensors into the dimension-8 dipole.
The mixing then gives
\begin{align}
16\pi^{2}\,\mu\frac{d}{d\mu}\,C^{\,e\mu}_{e^{2}\phi^{4}D} = -\,2N_{c}\,y_{t}^{2}\;C^{\,e\mu 33}_{e^{2}u^{2}\phi^{2}}(\mu)\,,
\label{eq:dim8mix}
\end{align}
and the nested solution of Eqs.~\eqref{eq:dim8RGE} and \eqref{eq:dim8mix} at leading-log accuracy, with both dimension-8 coefficients vanishing at $\Lambda$, is
\begin{align}
\frac{C^{\,e\mu}_{e^{2}\phi^{4}D}(\mu)}{\Lambda^{4}} = -\,\frac{6N_{c}\,y_{t}^{4}}{(16\pi^{2})^{2}}\;L^{2}\; \frac{C^{1333}_{eu}(\Lambda)\,\big(C^{2333}_{eu}(\Lambda)\big)^{*}}{\Lambda^{4}}\,,
\label{eq:T3toploop}
\end{align}
The dimension-8 Higgs-current operator matches after electroweak symmetry breaking on the penguin operator $O_{\phi e}$ with the factor $v_{T}^{2}/2\Lambda^{2}$ as in the previous scenario.
T3 is again evaluated through the substitution of Eq.~\eqref{eq:effC6},
\begin{align}
  \frac{C^{\,\mathrm{eff},12}_{\phi e}(\mu_{\rm inj})}{\Lambda^{2}}
  &=
  \frac{v_{T}^{2}}{2}\,
  \frac{C^{\,e\mu}_{e^{2}\phi^{4}D}(\mu_{\rm inj})}{\Lambda^{4}}\nonumber\\
  &=
  -\,\frac{3N_{c}\,v_{T}^{2}\,y_{t}^{4}}{(16\pi^{2})^{2}\,\Lambda^{4}}\;L^{2}\;
  C^{1333}_{eu}(\Lambda)\,\big(C^{2333}_{eu}(\Lambda)\big)^{*}\,.
  \label{eq:effphie}
\end{align}
As in the previous case the effective dimension-6 effective WC is injected into \texttt{wilson} at $\mu_{\rm inj}$ to evaluate the T3 contribution.

Each contribution is evaluated at its leading non-vanishing logarithm. 
For T1 this ranges from a single logarithm, when the mass-matrix misalignment is generated at one loop, to $L^3$; for the example of Sec.~\ref{sec:T1} the single-log terms vanish identically, eq.~\eqref{eq:Yecancel}, and the leading order is $L^2$ to $L^3$. For T3 it is the single log of eq.~\eqref{eq:dim8sol} for light-flavoured operators and the double log of Eq.~\eqref{eq:T3toploop} for top-flavoured ones.

Furthermore, we include the interference between T1 and T3 as it carries no extra suppression in $\Lambda$: both amplitudes are proportional to $C_6^{\tau e}C_6^{\tau\mu}/\Lambda^4$. The cross-section of any $\mu \to e$ process is then 
\begin{align}
\mathcal{B} \;\propto\;
|T1|^2 + |T3|^2
+ 2\,{\rm Re}\!\left(T1\,T3^*\right).
\label{eq:interference}
\end{align}
The Wilson-coefficient phases cancel in $T1\,T3^*
\propto |C_6^{\tau e}|^2|C_6^{\tau\mu}|^2$, and the remaining factors are real, since all internal masses exceed the momentum transfer and the amplitudes have no absorptive parts (Sec.~\ref{sec:T2}). The relative phase is therefore $0$ or $\pi$ and the interference is maximal,
\begin{align}
2\,{\rm Re}\!\left(T1\,T3^*\right)
= \pm\, 2\,|T1|\,|T3|\,.
\label{eq:interference_sqrt}
\end{align}
The sign is fixed by the two calculations themselves: the RG evolution of Sec.~\ref{sec:T1}
determines the sign of the generated $\mu e$ coefficients (Eqs.\,\eqref{eq:T1eu} and \eqref{eq:T1phie}), and that of the effective coefficient used for T3 follows from Eqs.\,\eqref{eq:dim8RGE} and \eqref{eq:dim8top}. Where both mechanisms populate the same $\mu e$ operator, the sign in Eq.~\eqref{eq:interference_sqrt} is the relative sign of these
coefficients at the common scale $\mu$. We have verified it by evaluating both sets simultaneously, in which case the interference appears automatically in ${\cal B}$.

\subsection{Contribution from direct double insertion (T2)}\label{sec:T2}
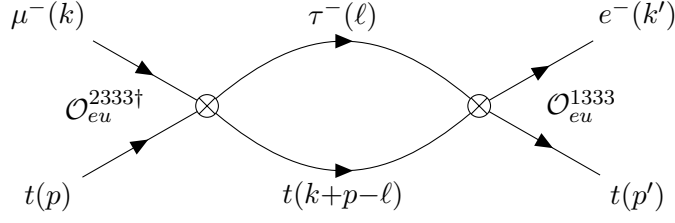
\begin{figure}[t]
\centering
\begin{tikzpicture}
\begin{feynman}
  \vertex[crossed dot] (vL) at (0,0) {};
  \vertex[crossed dot] (vR) at (3.6,0) {};

  \vertex (mu)   at (-2.1,  1.2) {$\mu^-(k)$};
  \vertex (tin)  at (-2.1, -1.2) {$t(p)$};
  \vertex (e)    at ( 5.7,  1.2) {$e^-(k')$};
  \vertex (tout) at ( 5.7, -1.2) {$t(p')$};

  \vertex (opL) at (-1.35, 0) {$\mathcal{O}_{eu}^{2333\dagger}$};
  \vertex (opR) at ( 4.95, 0) {$\mathcal{O}_{eu}^{1333}$};

  \diagram*{
    (mu)  -- [fermion] (vL),
    (tin) -- [fermion] (vL),
    (vL) -- [fermion, out=40, in=140, looseness=1.2,
             edge label=$\tau^-(\ell)$] (vR),
    (vL) -- [fermion, out=-40, in=-140, looseness=1.2,
             edge label'=$t(k{+}p{-}\ell)$] (vR),
    (vR) -- [fermion] (e),
    (vR) -- [fermion] (tout),
  };
\end{feynman}
\end{tikzpicture}
\caption{Representative one-loop ``fish'' diagram for the direct double
insertion (T2) of the two $\tau\to\ell$ operators, here
$\mathcal{O}_{eu}^{2333}$ and $\mathcal{O}_{eu}^{1333}$, generating a
$\mu\to e$ amplitude with an internal $(t,\tau)$ pair.}
\label{fig:doubleins_eu}
\end{figure}
Alongside the RG-induced contribution, a $\mu\to e$ transition receives
contributions from diagrams in which the two $\tau\to\ell$ operators appear
directly in the same amplitude, as shown in Fig.~\ref{fig:doubleins_eu}. As
discussed in Sec.~\ref{sec:framework}, T1 and T3 are the logarithms
accompanying the two divergences of these same graphs. We define T2 as the
remainder:
\begin{align}
T2 \;\equiv\;
\mathcal{A}_{\rm double}
\;-T1\;-T3\,,
\end{align}

that is, the finite part of the double insertion left over once 
the overall divergence has been absorbed into the dimension-8 counterterm.

Its evaluation proceeds in
four steps: (i) construct the one-loop amplitude with one insertion of each
operator; (ii) reduce the tensor integrals to scalar Passarino--Veltman
functions~\cite{Passarino:1978jh}; (iii) subtract the poles in the same scheme
used for the T1 running and the T3 matching; (iv) collect the remaining finite
terms.

We take the Warsaw-basis operators
\begin{align}
\mathcal{O}_{eu}^{\,2333}&=(\bar\mu\,\gamma_\alpha P_R \tau)\,
(\bar t\,\gamma^\alpha P_R t), &
\mathcal{O}_{eu}^{\,1333}&=(\bar e\,\gamma_\alpha P_R \tau)\,
(\bar t\,\gamma^\alpha P_R t),
\end{align}
with Wilson coefficients $C_{1333}, C_{2333}$ at the scale $\Lambda$. The
external legs are $t(p)\to t(p')$ and $\mu(k)\to e(k')$, and we define
$q^\alpha \equiv p'^\alpha - p^\alpha = k^\alpha - k'^\alpha$. Contracting the
$\tau$ line and one top line between the two insertions gives
\begin{align}
i\mathcal M_{T2}=
\frac{C_{1333}C_{2333}^*}{\Lambda^4}
\int\!\frac{\rd^{d}\ell}{(2\pi)^{d}}\,
\frac{\big[\bar u_e(k')\,\gamma^{\alpha}\slashed{\ell}\,\gamma^{\beta}P_R\,
u_\mu(k)\big]\,
\big[\bar u_t(p')\,\gamma_{\beta}(\slashed{\ell}-\slashed{q})\,\gamma_{\alpha}
P_R\,u_t(p)\big]}
{(\ell^{2}-m_\tau^{2})\,\big((\ell-q)^{2}-m_t^{2}\big)}\,,
\label{eq:Mloop_eu}
\end{align}
where $\ell$ is the momentum of the internal $\tau$, matching the routing of Fig.~\ref{fig:doubleins_eu}. Both operators carry right-handed currents, so the internal mass terms are projected out,
\begin{align}
\gamma^{\alpha}P_R(\slashed{\ell}+m_\tau)\gamma^{\beta}P_R
&=\gamma^{\alpha}\slashed{\ell}\,\gamma^{\beta}P_R\,, &
\gamma_{\beta}P_R(\slashed{\ell}-\slashed{q}+m_t)\gamma_{\alpha}P_R
&=\gamma_{\beta}(\slashed{\ell}-\slashed{q})\gamma_{\alpha}P_R\,,
\end{align}
and both fermion chains are purely vector. 
Reducing the numerator of Eq.~\eqref{eq:Mloop_eu} gives the two-point functions
$B_0$, $B_1$, $B_{00}$ and $B_{11}$ with arguments $(q^2; m_\tau^2, m_t^2)$.
After $\overline{\rm MS}$ subtraction the leading structure is a colour-singlet
right-handed vector interaction,
\begin{align}
\mathcal L_{T2}~\supset~
C_{T2}\,(\bar e\,\gamma^{\alpha}P_R\,\mu)\,(\bar t\,\gamma_{\alpha}P_R\,t),
\qquad
C_{T2}=\frac{C_{1333}C_{2333}^*}{\Lambda^4}\,\frac{1}{16\pi^2}\,
\overline{\mathcal{F}}(q^2;m_\tau^2,m_t^2)\,,
\label{eq:Leff7_eu}
\end{align}
where $\overline{\mathcal F}$ is the UV-finite part of the form factor and has
mass dimension two. On dimensional grounds it can be built only from $m_t^2$,
$m_\tau^2$ and $q^2$, multiplied by logarithms and $\mathcal{O}(1)$
coefficients; this is used in the power counting below. The pole removed in Eq.\,\eqref{eq:Leff7_eu} is absorbed by the dimension-8 counterterm
$\mathcal{O}^{(1)}_{e^2u^2 \phi^2}$, and the accompanying $\ln\mu$ gives the running of Eq.\,\eqref{eq:dim8RGE}.

All three contributions are proportional to
$C_6^{\tau e}C_6^{\tau\mu}/\Lambda^4$ and carry a factor $1/16\pi^2$ per loop.
They differ in the mass and logarithmic factors that accompany them. The
effective dimension-6 $\mu e$ coefficient generated by each mechanism is
\begin{align}
C^{\mu e}_{T1}&\sim\frac{v^{2}\,L^{n}}{(16\pi^2)^{n}}\,
(\text{couplings})\,\frac{C_6^{\tau e}C_6^{\tau\mu}}{\Lambda^{4}}\,,
\qquad n=1\text{--}3\,,\\
C^{\mu e}_{\rm T3} &= \frac{v_T^2}{2}\,\frac{C_8^{\mu e}}{\Lambda^4}
\;\sim\;\frac{m_t^2}{16\pi^2}\,L\;
\frac{C_6^{\tau e}C_6^{\tau\mu}}{\Lambda^4}\,,\\
C^{\mu e}_{\rm T2} &\sim \frac{\overline{\mathcal{F}}}{16\pi^2}\;
\frac{C_6^{\tau e}C_6^{\tau\mu}}{\Lambda^4}\,,
\end{align}
where $L=\ln(\Lambda/\mu)$, and we have used $v_T^2 y_t^2/2 = m_t^2$ in the T3
line. 

T1 and T3 are the coefficients of logarithms and are enhanced by powers of $L\simeq 3.8$ for $\Lambda = 4$~TeV. T2 is by definition the non-logarithmic remainder and receives no such
enhancement. It is a next-to-leading-logarithmic effect, and is therefore absent at the leading-log accuracy at which T1 and T3 are evaluated.

The terms in $\overline{\mathcal{F}}$ proportional to $m_t^2$ and $m_\tau^2$ are local. Using
$m_t^2 = y_t^2 v_T^2/2$ and $m_\tau^2 = y_\tau^2 v_T^2/2$, they are of the form
$v_T^2 \times (\text{Yukawa})^2$, so they are the finite matching contributions to the same dimension-8 operators that generate T3 and belong to the T3 sector. T2 is distinguished by its dependence on the momentum transfer, and is suppressed by

\begin{align}
\frac{\mathrm{T2}}{\mathrm{T3}}\;\sim\;\frac{q^2}{M^2\,L}\,,
\label{eq:T2T3ratio}
\end{align}

where $M$ is the lightest mass in the loop. The $\tau$ is always an internal line, since it is the contraction of the $\tau$ fields of the two operators that removes the $\tau$ flavour and produces a $\mu\to e$ amplitude. Hence $M \geq m_\tau$ for every operator pair considered. All $\mu\to e$ observables used in this work ($\mu\to e\gamma$, $\mu\to 3e$, coherent $\mu\to e$ conversion) have $|q|\lesssim m_\mu$, so that $|q^2| \leq m_\mu^2 \ll m_\tau^2$ and Eq.~\eqref{eq:T2T3ratio} gives

\begin{align}
\frac{\mathrm{T2}}{\mathrm{T3}}\;\lesssim\;
\frac{m_\mu^2}{m_\tau^2\,L}\;\approx\;10^{-3}
\qquad\text{(light internal quark)},
\qquad
\frac{\mathrm{T2}}{\mathrm{T3}}\;\sim\;
\frac{m_\mu^2}{m_t^2\,L}\;\approx\;10^{-7}
\qquad\text{(internal top)}.
\end{align}
The same counting applies to T2/T1 wherever T1 is nonzero, since T1 is
enhanced by additional logarithms.

T2 is not computed by our numerical pipeline. The package \texttt{wilson} implements the dimension-6
anomalous-dimension matrix, which is linear in the Wilson coefficients,
together with tree-level SMEFT$\to$WET matching. The product
$C_6^{\tau e}C_6^{\tau\mu}$ arises only through the iterated solution of the
coupled RG system and the accompanying rotation to the charged-lepton mass
basis, which is T1. In addition, T2 is a momentum-dependent form factor rather than a
constant, and cannot be represented by a local WET Wilson coefficient. T2 and
T3 are therefore obtained outside the running pipeline: T2 analytically, as
above, and T3 through the effective dimension-6 substitution as explained in Sec.~\ref{sec:T3}.

\section{Direct bound and constraints on LFV operators}
\label{directbound}

In this section, we present the current and projected future bounds on the SMEFT Wilson
coefficients for operators contributing to charged lepton flavor violation. The bounds are derived from the experimental constraints on the LFV observables listed in Table~\ref{tab:obsbound}, adopting a one-operator-at-a-time approach with the NP scale fixed at $\Lambda = 4$~TeV. Throughout, the quark-generation indices are chosen to display the most stringent available bound; the full set of generation-specific bounds is collected in Appendix~\ref{app:bounds}.

Fig.~\ref{fig:currentbound} shows the present upper limits on $|C_i|/\Lambda^2$ for the three LFV channels $\mu \to e$, $\tau \to e$, and $\tau \to \mu$ across all relevant operator classes in the Warsaw basis. The operators considered span dipole-type couplings ($\mathcal{O}_{eB}$, $\mathcal{O}_{eW}$), scalar and tensor four-fermion operators ($\mathcal{O}_{ed}$, $\mathcal{O}_{ee}$, $\mathcal{O}_{eu}$, $\mathcal{O}_{\ell d}$, $\mathcal{O}_{\ell e}$, $\mathcal{O}_{\ell e d q}$, $\mathcal{O}_{\ell e q u}^{(1)}$, $\mathcal{O}_{\ell e q u}^{(3)}$,
$\mathcal{O}_{\ell \ell}$), semi-leptonic vector operators ($\mathcal{O}_{\ell q}^{(1)}$, $\mathcal{O}_{\ell q}^{(3)}$, $\mathcal{O}_{\ell u}$, $\mathcal{O}_{qe}$), and Yukawa-like Higgs–lepton couplings ($\mathcal{O}_{\phi e}$, $\mathcal{O}_{\phi \ell}^{(1)}$,
$\mathcal{O}_{\phi \ell}^{(3)}$).

As is evident from Fig.~\ref{fig:currentbound}, there is a clear hierarchy among the three channels. The $\mu \to e$ channel yields by far the most stringent constraints, with bounds reaching as low as $|C| \sim 10^{-15}$--$10^{-14}$ for dipole and four-fermion operators, driven primarily by the extremely precise measurements of $\mu \to e \gamma$, $\mu \to 3e$, and $\mu$-to-$e$ conversion in nuclei. The $\tau \to e$ and $\tau \to \mu$ channels provide comparatively weaker bounds, typically in the range $|C| \sim 10^{-9}$--$10^{-8}$, reflecting the larger experimental uncertainties and lower luminosity accumulated in $\tau$-lepton searches at $B$-factories and the LHC. Across all channels, dipole operators $\mathcal{O}_{eB}$ and $\mathcal{O}_{eW}$ are particularly
strongly constrained through radiative charged-lepton flavour-violating decays, while semileptonic scalar and vector operators can also receive very stringent bounds from coherent $\mu\to e$ conversion and from lepton-flavour-violating hadronic $\tau$ decays
\cite{Crivellin:2013hpa,Husek:2020fru,Fernandez-Martinez:2024bxg}.

\begin{figure}
\centering
\includegraphics[width=0.98\linewidth]{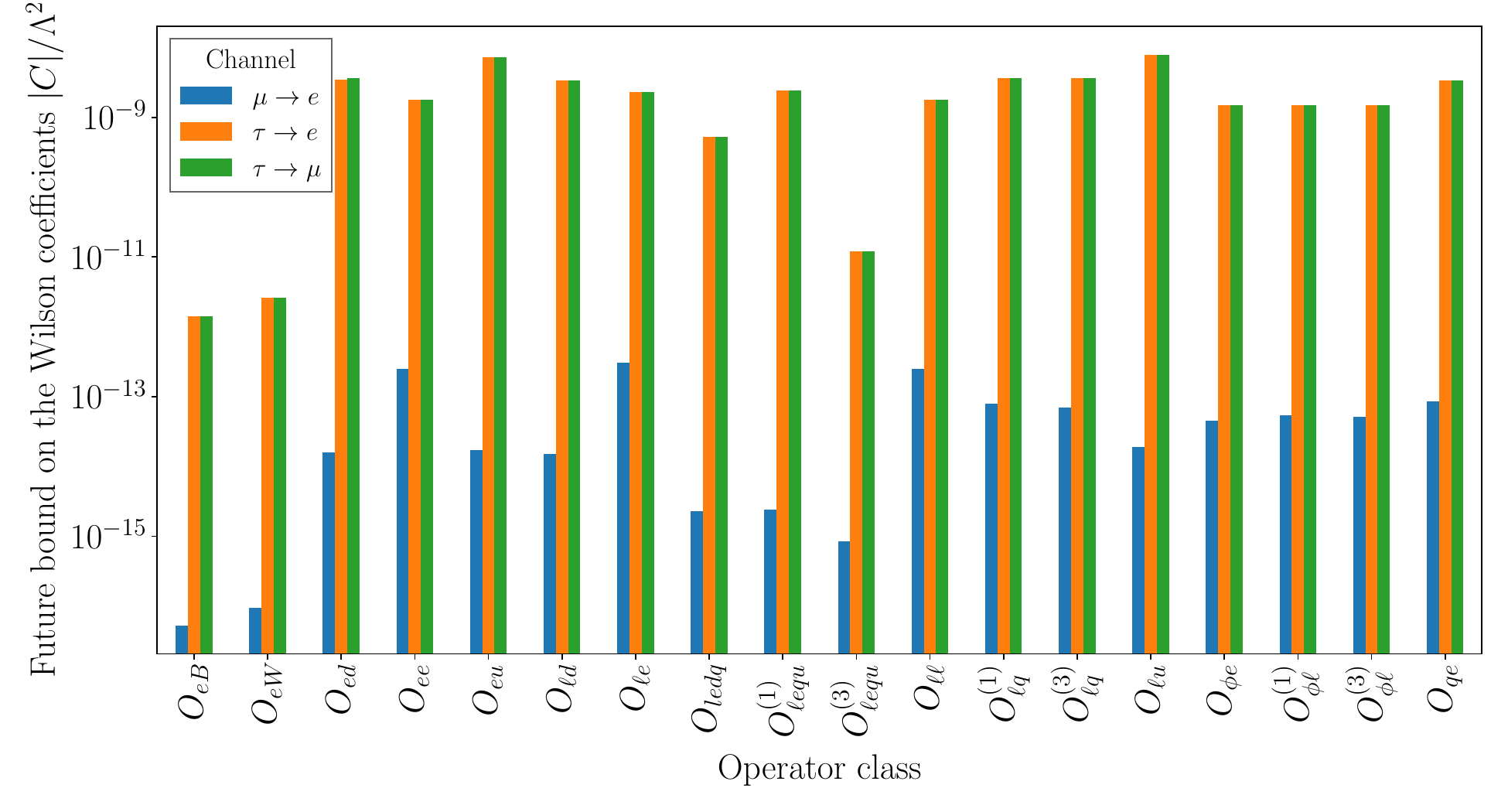}
\caption{Future bounds on LFV operator classes.}
\label{fig:futurebound}
\end{figure}

Fig.~\ref{fig:futurebound} displays the projected future bounds, obtained by rescaling the current experimental sensitivities according to the anticipated precision upgrades described in Table~\ref{tab:obsbound}. The qualitative pattern of the hierarchy between the three channels is preserved, but the absolute reach improves markedly.
 
The $\mu \to e$ channel benefits most dramatically from planned experiments. Next-generation
searches for $\mu \to e \gamma$ (MEG~II~\cite{MEGII:2018kmf,Baldini:2021elz}), $\mu \to 3e$ (Mu3e ~\cite{Arndt:2020mgu}), and $\mu$-to-$e$ conversion (Mu2e, COMET~\cite{Mu2e:2014fns,Adamov:2018vin}) are expected to improve current sensitivities by several orders of magnitude,
pushing the projected bounds on the corresponding WCs down to $|C| \sim 10^{-16}$--$10^{-15}$. This represents an improvement of roughly one to two orders of magnitude with respect to the current limits shown in Fig.~\ref{fig:currentbound}.
 
For the $\tau \to e$ and $\tau \to \mu$ channels, the anticipated improvement stems from the Belle~II programme and future lepton colliders, which are expected to accumulate $\mathcal{O}(50\text{--}100~\text{ab}^{-1})$ of data \cite{Belle-II:2018jsg}. This translates into projected bounds at the level of $|C| \sim 10^{-10}$--$10^{-9}$, an improvement of approximately one order of magnitude compared to present constraints.
 
The large gap between the $\mu \to e$ and $\tau \to \ell$ projected sensitivities underscores the complementary role of muon experiments as precision probes of charged-lepton flavour violation. Even in the most optimistic future scenario considered here, the direct constraints from the $\tau$ sector remain several orders of magnitude weaker than those from the $\mu-e$ sector for comparable operator structures. This difference has an important implication for the interpretation of lepton-flavour-violating new-physics scenarios: operators involving $\tau$ flavour can appear comparatively weakly constrained when tested only through direct $\tau$-decay observables, while the same operators can contribute indirectly to the much more precisely measured $\mu\to e$ processes.

This motivates us to investigate whether pairs of $\tau\to e$ and $\tau\to\mu$ operators can generate observable $\mu\to e$ transitions through higher-order effects. In the following section, we therefore study the indirect constraints on $\tau$-flavour-violating SMEFT operators arising from their combined contributions to the $\mu-e$ sector. We separate the contributions into the dimension-six operator-mixing mechanism T1 and the dimension-eight double-insertion mechanism T3, and identify representative scenarios in which either T1 or T3 dominates, as well as cases where the two contributions are comparable.

\begin{figure}
\centering
\includegraphics[width=0.98\linewidth]{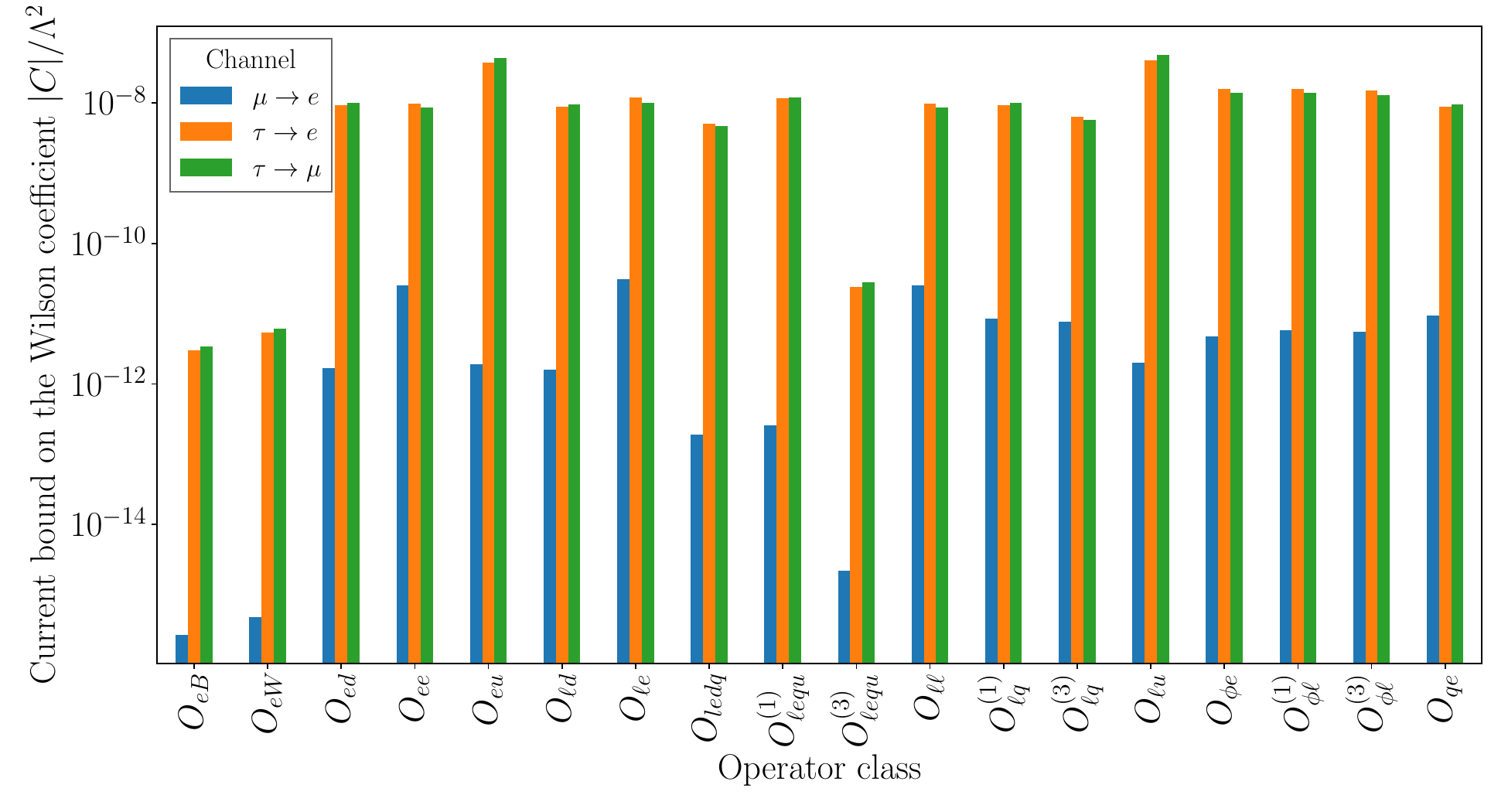}
\caption{Current bounds on LFV operator classes.}
\label{fig:currentbound}
\end{figure}

\section{Indirect bounds}\label{sec:indirect}

In the discussion so far we have seen that LFV operators inducing $\tau\to\mu$ and $\tau\to e$ transitions generate $\mu\to e$ transitions through RG evolution, double insertions, and higher-dimensional counterterms. Since $\mu\to e$ observables are probed with far greater precision than $\tau\to\ell$ processes, they provide a powerful indirect probe of LFV at high scales. As established in Sec.~\ref{sec:framework}, all three contributions to the $\mu\to e$ amplitude are proportional to $C_6^{\tau e}C_6^{\tau\mu}/\Lambda^4$. However, they differ in the logarithmic and coupling factors they carry. At leading-log accuracy the $\mu e$ coefficient generated by each mechanism takes the form
\begin{align}
  \frac{C^{\mu e}_{T1}(\mu)}{\Lambda^2}
  &= \frac{v^2}{\Lambda^4}\,\frac{a_n}{(16\pi^2)^n}\,L^{\,n}\;
     C^{\tau e}_6(\Lambda)\,C^{\tau\mu}_6(\Lambda),
  \nonumber\\
  \frac{C^{\mu e}_{T3}(\mu)}{\Lambda^2}
  &= \frac{v^2}{\Lambda^4}\,\frac{b}{16\pi^2}\,L\;
     C^{\tau e}_6(\Lambda)\,C^{\tau\mu}_6(\Lambda),
  \label{eq:T1T3param}
\end{align}
where $n$ is the number of one-loop RG steps in the T1 chain, and $a_n$, $b$ are $\Lambda$-independent combinations of Yukawa and gauge couplings and group factors fixed by the anomalous dimensions. The competition between T1 and T3 is therefore controlled by the length of the T1 chain and by the couplings entering $a_n$ and $b$. T1 dominates when its chain is short and top-Yukawa enhanced, T3 when the flavour structure of the initial operators makes the RG transfer inefficient, and the two are comparable in between. In the following we classify representative operator combinations into these three regimes.

\subsection{T1-dominated scenarios}
We first consider pairs for which the indirect bound is set by the RG-induced contribution
T1, with T3 subleading and T2 negligible. This occurs when the operator pair efficiently mixes into $\mu \to e$ operators via RG evolution, often enhanced by top-Yukawa effects. Representative examples include
\begin{align}
&{\rm 1.}~~C^{(1)1333}_{lequ},\; C^{(1)23}_{\phi l}, &
&{\rm 2.}~~C_{le}^{3313},\; C_{ledq}^{2321}, &
&{\rm 3.}~~C_{le}^{3313},\; C_{qe}^{1123}.
\end{align}
Below we summarize the dominant mixing chains for the these operator pairs.

\subsubsection{$C^{(1)1333}_{lequ},\; C^{(1)23}_{\phi l}$} \label{sec:411}
We begin the survey of T1-dominated pairs with the combination $C^{(1)1333}_{\ell equ}$ and $C^{(1)23}_{\phi\ell}$, which realises the T1 mechanism of Sec.~\ref{sec:T1} in its most efficient form: a single explicitly loop-suppressed step, enhanced by the top Yukawa coupling and by an uncompensated factor of $1/m_\tau$ from the rediagonalization.
The flavour flow is summarized by the chain
\begin{align}
C^{(1)1333}_{\ell equ}
\xrightarrow{\;N_c\,y_t\;}
\big\{\,[Y_e]_{13}\,,\;C^{13}_{e \phi}\,\big\}
\longrightarrow
\delta[M_e]_{13}
\xrightarrow[\;C_{\phi l}^{(1)23}\;]{\;\text{mass rotation}\;}
 C^{(1) 12}_{\phi \ell}\,,
\label{eq:chain_lequ_phil}
\end{align}
and the entire indirect $\mu\to e$ sensitivity of this pair is of type T1: as shown below, the T3 contribution vanishes at the order of our analysis.

In the first step, closing the scalar quark bilinear $\bar q_3 u_3$ of $O^{(1)1333}_{\ell equ}$ into a top loop with a single insertion of $y_t$ populates both sources of the charged-lepton mass matrix introduced in Sec.~\ref{sec:T1}. The first is the $e$--$\tau$ entry of the lepton Yukawa, generated through the $m_\phi^2$-proportional dimension-six correction to its running \cite{Jenkins:2013zja, Jenkins:2013wua},
\begin{align}
16\pi^2\,\mu\frac{d}{d\mu}[Y_e]_{13}
= 2\,m_\phi^2\, N_c\,y_t\,\frac{C^{(1)1333}_{\ell equ}}{\Lambda^2}\,,
\label{eq:Yeshift}
\end{align}
and the second is the coefficient $C^{13}_{e\phi}$, generated through ordinary dimension-six operator mixing~\cite{Jenkins:2013zja}. In the convention of Sec.~\ref{sec:T1} the two add up to the off-diagonal mass-matrix entry
\begin{align}
\delta[M_e]_{13}
&= \frac{v_T}{\sqrt{2}}\,\delta[Y_e]_{13}
- \frac{v_T^{3}}{2\sqrt{2}\,\Lambda^{2}}\,\delta C_{e \phi,13}
\notag\\
&= \frac{v_T}{\sqrt{2}}\,
\frac{2N_c\,y_t\,m_\phi^2}{16\pi^2}\,L\;\frac{C^{(1)1333}_{\ell equ}}{\Lambda^2}\,,
\label{eq:dM13}
\end{align}
Following the rediagonalization procedure of Sec.~\ref{sec:T1}, the misalignment is removed by a left-handed rotation in the $e$--$\tau$ block, which acts on the doublet index of the independent $\tau$--$\mu$ coefficient and generates the $\mu\to e$ Higgs-lepton current, with $\theta_{e\tau} = \frac{\delta[M_e]_{13}}{m_\tau}\,,$
\begin{align}
\frac{C^{(1)12}_{\phi\ell}}{\Lambda^2}\bigg|_{T_1}
= \theta_{e\tau}\,\frac{C^{(1)23}_{\phi\ell}}{\Lambda^2}
= \frac{\sqrt{2}\,N_c\,y_t\,v_T\,m_\phi^2}{16\pi^2\,m_\tau}\,L\;
\frac{C^{(1)1333}_{\ell equ}\,C^{(1)23}_{\phi\ell}}{\Lambda^4}\,,
\label{eq:T1_lequ_phil}
\end{align}
which matches onto flavour-changing $Z$ couplings and contributes at tree level to $\mu\to e$ conversion.

Eqs. \eqref{eq:Yeshift}--\eqref{eq:T1_lequ_phil} are leading-logarithmic bookkeeping expressions. The numerical results derived in \texttt{wilson}, include the complete anomalous-dimension matrix, the $C_{e \phi}$ back-reaction, the charged-lepton rediagonalisation, and the electroweak and low-energy matching, all of which are performed by the full SMEFT evolution.

We now turn to T3. For a four-fermion operator paired with a Higgs-lepton current this is the $P_6\times 4f_6\to 4f_8$ class of Ref.~\cite{Ardu:2022pzk}.
In $O^{(1)1333}_{\ell equ}$ the $\tau$ is the right-handed singlet $e_\tau$, whereas $O^{(1)23}_{\phi\ell}$ contains only the left-handed doublet $\ell_\tau$. Gauge interactions preserve chirality, so no one-loop diagram connects the two $\tau$ lines, and the corresponding anomalous dimension entry vanishes identically. With no dimension-eight coefficient seeded at the single-log level, the nested double-log tower of Sec.~\ref{sec:T3}, which produces the leading T3 for the top-flavoured pairs, has nothing to feed on, and T3 is absent for this pair altogether.

\begin{figure}[H]
\centering
\includegraphics[width=0.47\linewidth]{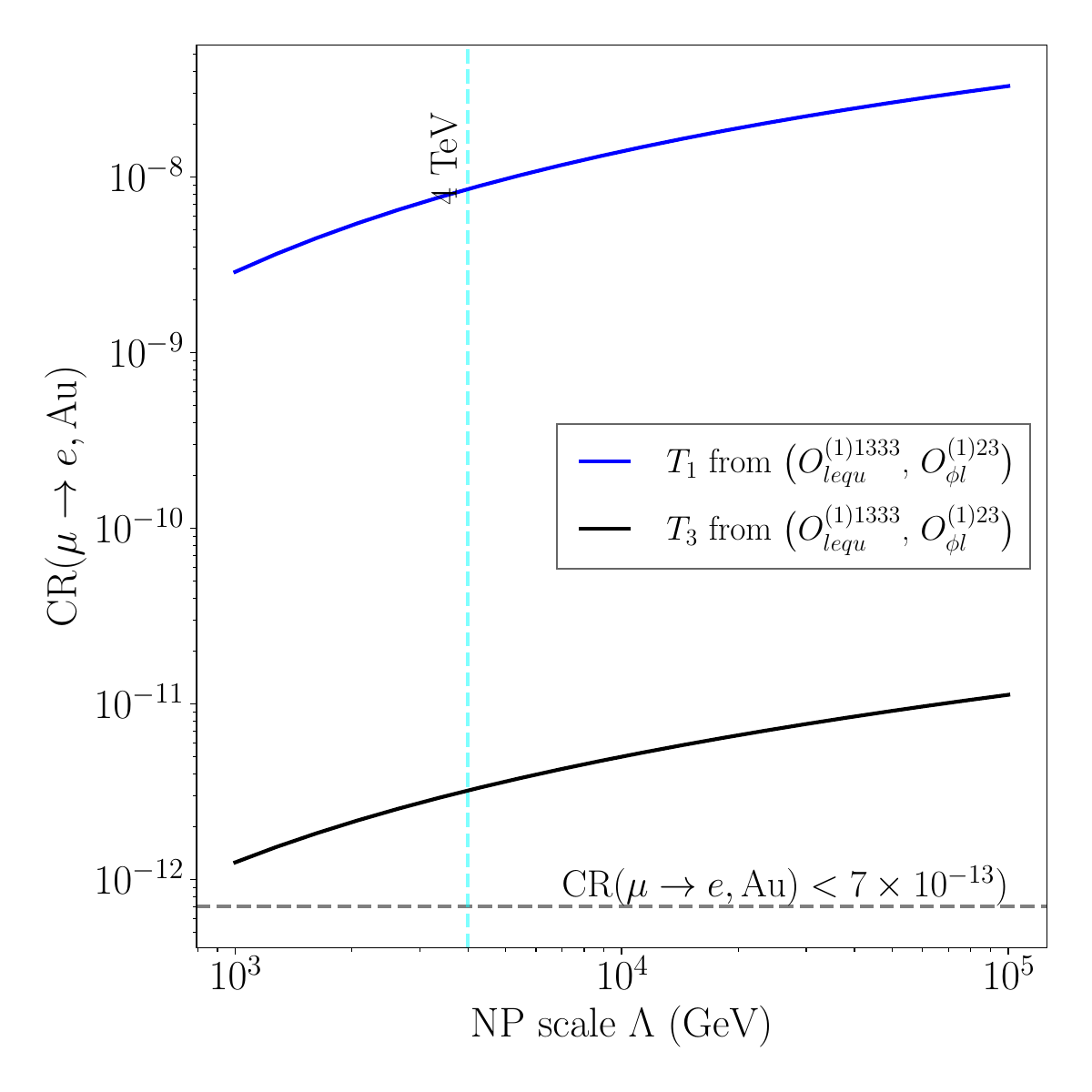}
\includegraphics[width=0.47\linewidth]{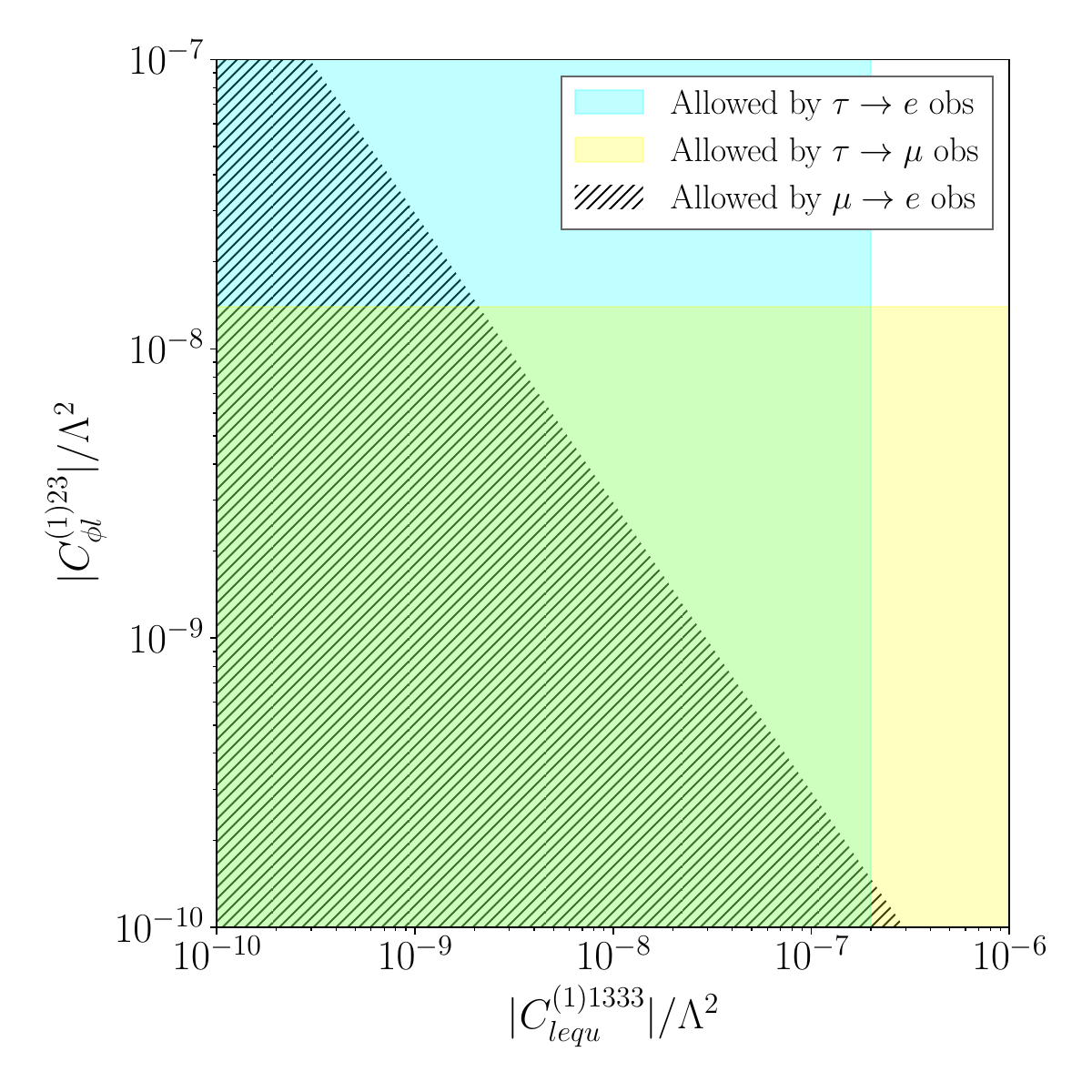}
\caption{Left: the $\mu\to e$ observable induced by the T1 (blue) and T3 (black) contributions and their computed combination, as functions of the new-physics scale $\Lambda$, with both Wilson coefficients fixed at their direct bounds. Right: the plane of the moduli $|C^{\tau e}|/\Lambda^2$ and $|C^{\tau\mu}|/\Lambda^2$ at $\Lambda=4$~TeV; the cyan and yellow bands are allowed by the direct $\tau\to e$ and $\tau\to\mu$ limits, and the hatched region by the $\mu\to e$ observable. Shown here for the T1-dominated pair $C^{(1)1333}_{\ell equ}$--$C^{(1)23}_{\phi\ell}$.}
\label{fig:411}
\end{figure}
The pair considered here is therefore a pure T1 scenario and and a book-keeping comparison of T1 and T3 contribution is shown in the left panel of Fig~\ref{fig:411}. We have presented the indirect bounds on this pair of coefficients achievable from $\mu - e$ sensitivity on the right panel in Fig.~\ref{fig:411}.
At the corner defined by the direct bounds, $|C^{(1)1333}_{\ell equ}|/\Lambda^2=2.0\times10^{-7}\,\mathrm{GeV}^{-2}$ and $|C^{(1)23}_{\phi\ell}|/\Lambda^2=1.4\times10^{-8}\,\mathrm{GeV}^{-2}$, the full evolution gives $\mathrm{CR}(\mu\to e,\mathrm{Au})=8.5\times10^{-9}$ at $\Lambda=4$~TeV, exceeding the SINDRUM~II limit by four orders of magnitude, so that the indirect bound on the coefficient product is stronger than the product of the direct $\tau\to\ell$ bounds by a factor of about $100$. The corresponding exclusion read from the plot as the un-dashed region.

\subsubsection{$C_{\ell e}^{3313}$, $C_{\ell edq}^{2321}$}\label{sec:412}

\begin{figure}[htbp]
\centering
\includegraphics[width=0.47\linewidth]{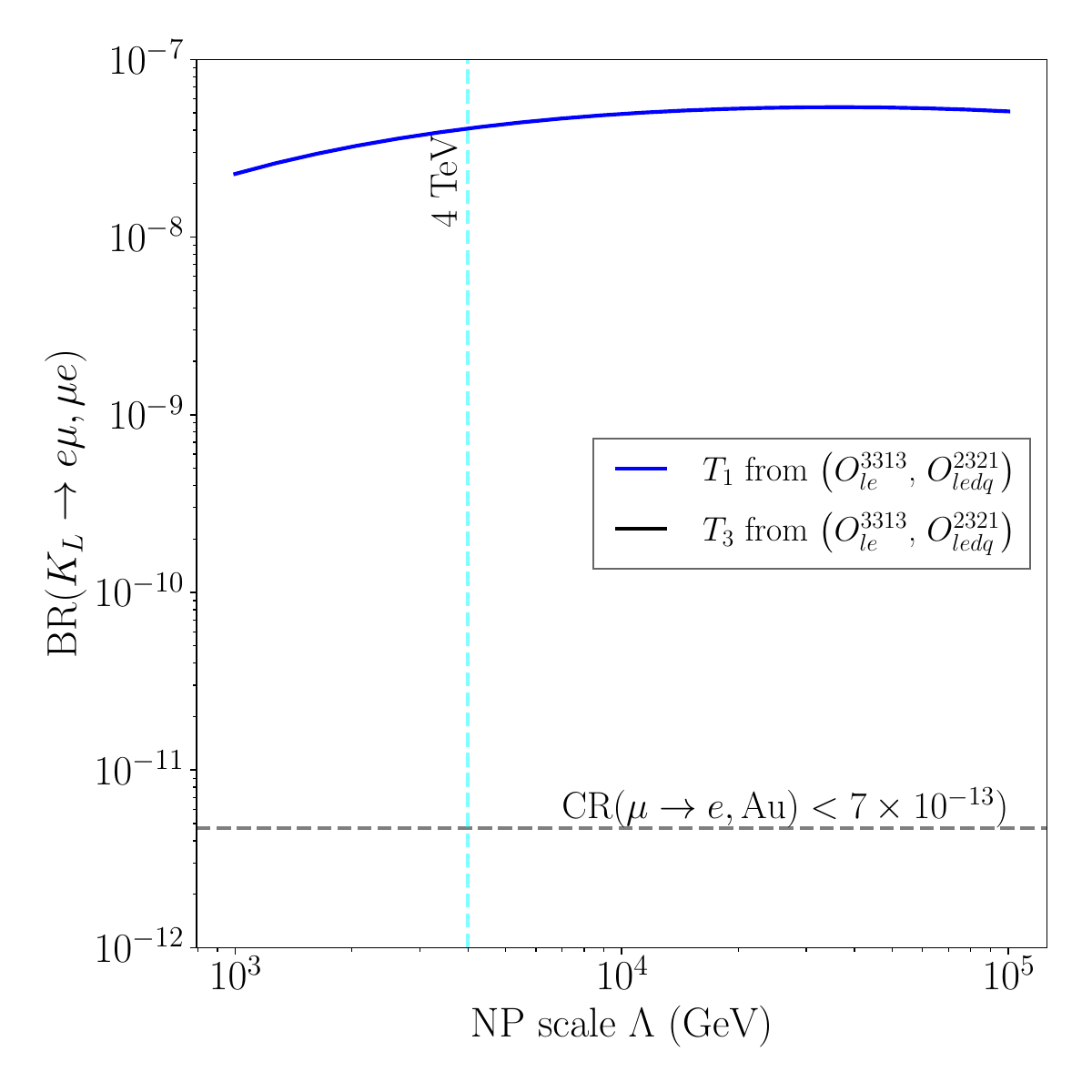}
\includegraphics[width=0.47\linewidth]{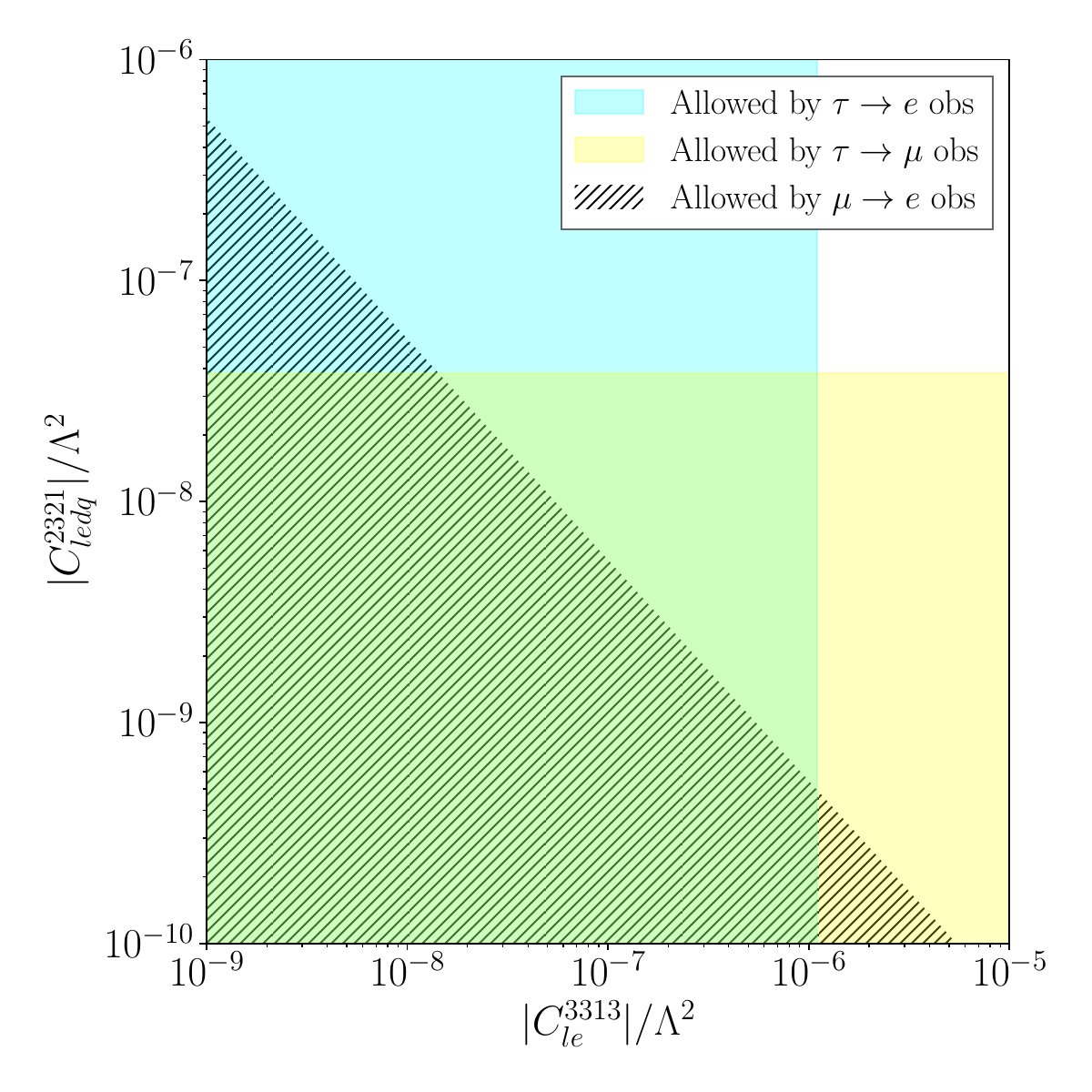}
\caption{Same as Fig.~\ref{fig:411} but for the T1-dominated pair $C_{\ell e}^{3313}$--$C_{\ell edq}^{2321}$.}
\label{fig:412}
\end{figure}

The second T1-dominated pair combines the four-lepton coefficient $C_{\ell e}^{3313}$, in which the $\tau\to e$ transition sits in the right-handed singlet current, with the scalar coefficient $C_{\ell edq}^{2321}$, which carries the $\tau\to\mu$ transition together with the quark bilinear $\bar d_2 q_1$. Similar to the previous example, for the T1 contribution, the flavour flow is summarised by the chain
\begin{align}
C^{3313}_{\ell e}
\xrightarrow{\;y_\tau\;}\
\big\{\,[Y_e]_{31}\,,\;C^{31}_{e \phi}\,\big\}
\longrightarrow
\delta[M_e]_{31}
\;\xrightarrow[\;C_{\ell edq}^{2321}\;]{\;\text{mass rotation}\;}C^{2121}_{\ell edq}\,,
\label{eq:chain_le_ledq}
\end{align}
and the generated $\mu\to e$ coefficient mediates $K_L\to e\mu$ at tree level through its $\bar s d$ quark content. 
The operator $O^{3313}_{\ell e}$ generates the $\tau_L$--$e_R$ entry of the mass matrix through the $m_\phi^2$-proportional correction to the Yukawa running~\cite{Jenkins:2013zja, Jenkins:2013wua}
\begin{align}
16\pi^2\,\mu\frac{d}{d\mu}[Y_e]_{31}
= 4\,m_\phi^2\,\,y_\tau\,\frac{\big(C^{3313}_{\ell e}\big)^{*}}{\Lambda^2}\,,
\label{eq:Yeshift_le}
\end{align}
together with the $y_\tau$-driven mixing into $C^{31}_{e \phi}$ \cite{Jenkins:2013zja}. In the convention of Sec.~\ref{sec:T1} the two add up to
\begin{align}
\delta[M_e]_{31}
&= \frac{v_T}{\sqrt{2}}\,\delta[Y_e]_{31}
- \frac{v_T^{3}}{2\sqrt{2}\,\Lambda^{2}}\,\delta C_{e \phi}^{31}
\notag\\
&= \frac{v_T}{\sqrt{2}}\,
\frac{4\,y_\tau\,m_\phi^2}{16\pi^2}\,L\;\frac{\big(C^{3313}_{\ell e}\big)^{*}}{\Lambda^2}\,.
\label{eq:dM31}
\end{align}
Following the same prescription as in Sec.~\ref{sec:411} for the rediagonalization, we get
\begin{align}
\frac{C^{2121}_{\ell edq}}{\Lambda^2}\bigg|_{T_1}
= \frac{4\,m_\phi^2}{16\pi^2}\,L\;
\frac{\big(C^{3313}_{\ell e}\big)^{*}\,C^{2321}_{\ell edq}}{\Lambda^4}\,.
\label{eq:T1_le_ledq}
\end{align}
The corresponding T1 contribution to $\mu -e$ transition observable is presented in Fig.~\ref{fig:412}, left panel.
In order to derive the T3 contribution, note that $O_{\ell e}$ is a four-lepton operator, so after the single contractible $\tau$ pair is closed the double insertion still carries six external fermion legs, and its local counter-terms are of dimension nine, not eight. No entry of the dimension-eight mixing classes of Ref.~\cite{Ardu:2022pzk} applies.

At the corner defined by the direct bounds, $|C^{3313}_{\ell e}|/\Lambda^2=1.1\times10^{-6}\,\mathrm{GeV}^{-2}$ and $|C^{2321}_{\ell edq}|/\Lambda^2=3.8\times10^{-8}\,\mathrm{GeV}^{-2}$, the full evolution generates $|C^{2121}_{\ell edq}|/\Lambda^2=2.3\times10^{-11}\,\mathrm{GeV}^{-2}$ and predicts $\mathrm{BR}(K_L\to e\mu,\mu e)=4.1\times10^{-8}$ at $\Lambda=4$~TeV, exceeding the experimental bound of $4.7\times10^{-12}$ by close to four orders of magnitude, so that the indirect bound on the coefficient product is stronger than the product of the direct $\tau\to\ell$ bounds by a factor of about $100$
The corresponding exclusion plot is shown in the right panel of Fig.~\ref{fig:412}.

\subsubsection{$C_{\ell e}^{3313}$, $C_{qe}^{1123}$}\label{sec:413}

The third T1-dominated pair combines the same four-lepton source $C_{\ell e}^{3313}$ as in Sec.~\ref{sec:412} with the coefficient $C_{qe}^{1123}$, which carries the $\tau\to\mu$ transition in the right-handed singlet current together with the flavour-diagonal first-generation quark doublet current. The mechanism is identical to that of Sec.~\ref{sec:412}, with the partner operator exchanged:
\begin{align}
C^{3313}_{\ell e}
\xrightarrow{\;y_\tau\;}
\big\{\,[Y_e]_{31}\,,\;C^{31}_{e \phi}\,\big\}
\longrightarrow
\delta[M_e]_{31}
\xrightarrow[C_{qe}^{1123}]{\;\text{mass rotation}\;}C^{1112}_{qe}\,.
\label{eq:chain_le_qe}
\end{align}
The misalignment $\delta[M_e]_{31}$ is exactly those of Eq.\,(\ref{eq:dM31}) and following the rediagonalization procedure explained in Sec.~\ref{sec:T1}, the generated $\mu \to e$ transition coefficient is
\begin{align}
\frac{C^{1112}_{qe}}{\Lambda^2}\bigg|_{T_1}
= \frac{4\,m_\phi^2}{16\pi^2}\,L\;
\frac{C^{3313}_{\ell e}\,\big(C^{1123}_{qe}\big)^{*}}{\Lambda^4}\,.
\label{eq:T1_le_qe}
\end{align}

\begin{figure}[htbp]
\centering
\includegraphics[width=0.47\linewidth]{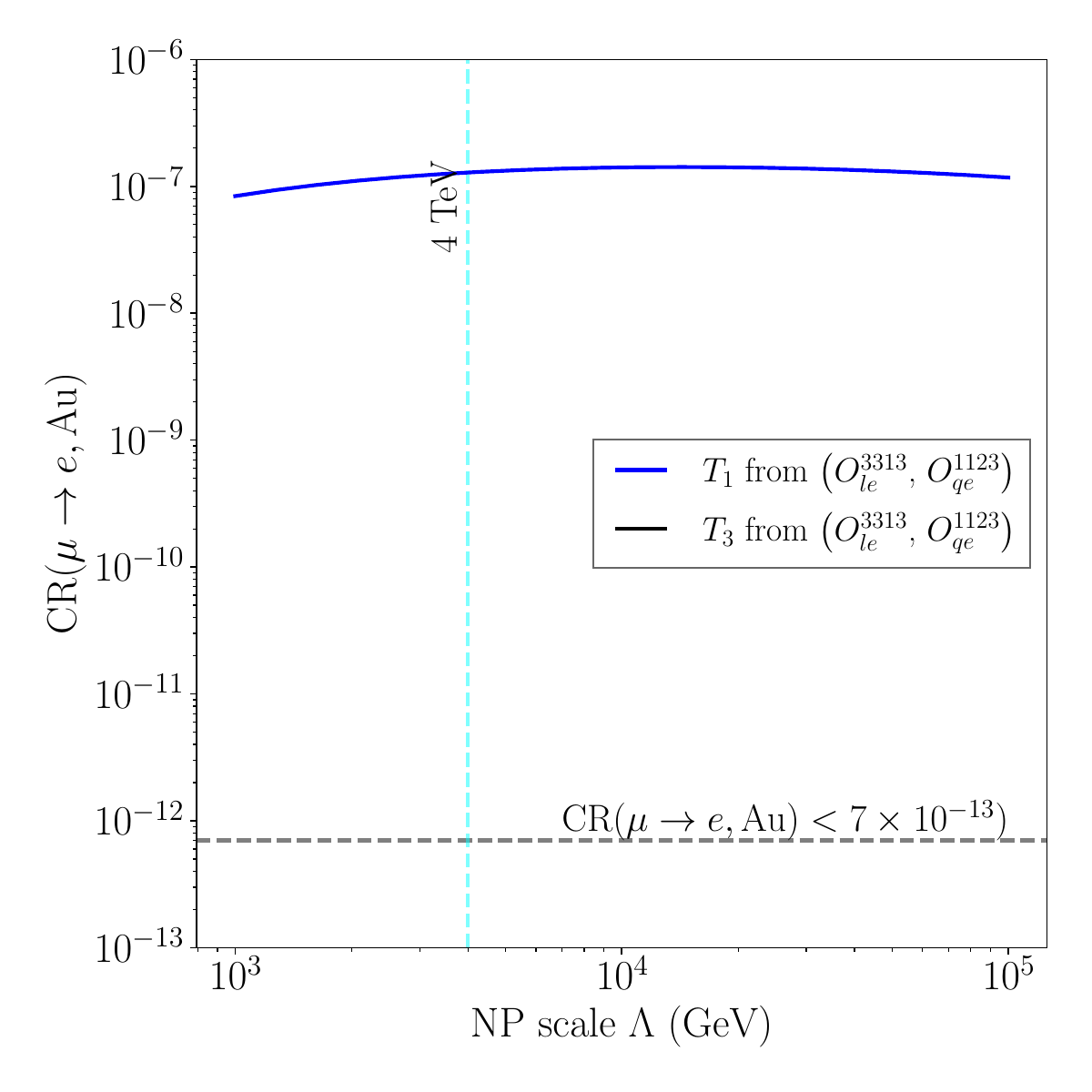}
\includegraphics[width=0.47\linewidth]{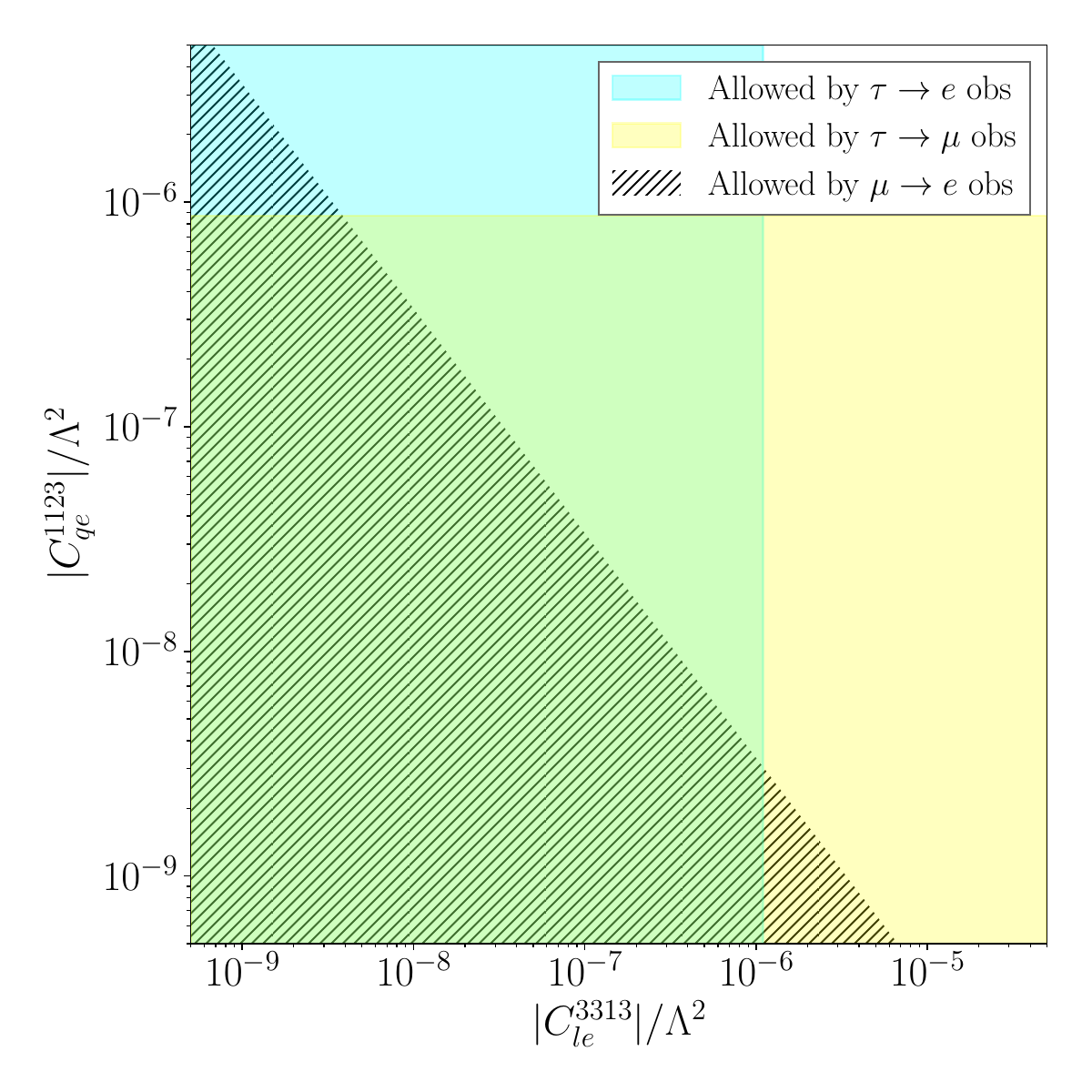}
\caption{Same as Fig.~\ref{fig:411} but for the T1-dominated pair $C_{\ell e}^{3313}$--$C_{qe}^{1123}$.}
\label{fig:413}
\end{figure}

The pair feeds $\mathrm{CR}(\mu\to e)$ at tree level  and inherits the exceptional reach of the conversion experiments. The left-panel of Fig.~\ref{fig:413} shows plot of the T1 contribution as funtion of $\Lambda$ scale. T3 vanishes by the same dimensional argument as in Sec.~\ref{sec:412}.

At the corner defined by the direct bounds, $|C^{3313}_{\ell e}|/\Lambda^2=1.1\times10^{-6}\,\mathrm{GeV}^{-2}$ and $|C^{1123}_{qe}|/\Lambda^2=8.7\times10^{-7}\,\mathrm{GeV}^{-2}$, the full evolution generates $|C^{1112}_{qe}|/\Lambda^2=4.2\times10^{-10}\,\mathrm{GeV}^{-2}$ and predicts $\mathrm{CR}(\mu\to e,\mathrm{Au})=1.3\times10^{-7}$ at $\Lambda=4$~TeV, exceeding the SINDRUM~II limit by more than five orders of magnitude, so that the indirect bound on the coefficient product is stronger than the product of the direct $\tau\to\ell$ bounds by a factor of about $400$, the largest gain among the T1-dominated pairs considered here. The corresponding exclusion plot is shown in the right panel of Fig.~\ref{fig:413}.

\subsection{T3-dominated scenarios}\label{sec:42}

We now consider operator pairs for which the indirect $\mu\to e$ bound is dominated by the dimension-eight double-insertion contribution T3. In this scenarios,  the T1 chain of Sec.~\ref{sec:T1} is either absent or inefficient as the flavour transfer requires several loop steps and the mass-basis rediagonalization. Whereas, the double insertion feeds a tree-level $\mu\to e$ interaction after fewer loops, so T3 contribution dominates despite its higher operator dimension. In all cases the flavour flow is
\begin{align}
C^{\tau e}_{6}\,,\;C^{\tau\mu}_{6}
\longrightarrow
C^{\mu e}_{8}
\longrightarrow
C^{\mu e}_{6,\,\mathrm{eff}}\,,
\label{eq:T3chain}
\end{align}
following the procedure of Sec.~\ref{sec:T3}. The first step is the one-loop double insertion, with the single-log anomalous dimensions taken from Ref.~\cite{Ardu:2022pzk}. The loop is closed by the top quark, with coefficient of order $y_t^2$, when the shared $\tau$ has matching chirality in the two operators, and by one $y_\tau$ and one $y_t$ insertion. The second step depends on the external quark flavour of $C^{\mu e}_{8}$. Light external quarks match directly onto an effective dimension-six coefficient by the $v_T^2/2\Lambda^2$ rule of Eq.\,(\ref{eq:effC6}), while top-flavoured operators have no low-energy image and instead reach an effective Higgs-lepton, Yukawa-like or dipole coefficient through the nested double log of Sec.~\ref{sec:T3}, injected at $\mu_{\rm inj}\simeq100$~GeV.

We consider the following three representative pairs for this scenario:
\begin{align}
&{\rm 1.}~~C_{eu}^{1333},C_{eu}^{2333}, &
&{\rm 2.}~~C_{qe}^{3313}, C_{qe}^{3323}, &
&{\rm 3.}~~C_{\ell u}^{1333},C_{qe}^{3323}.
\end{align}

The contributions to observables such as the coherent $\mu\to e$ conversion $\mathrm{CR}(\mu\to e,\mathrm{Au})$ and ${\rm BR}(\mu\to e\gamma)$, and the resulting indirect bounds on the $\tau-\ell$ operatros from each of the pair are discussed in the following subsections.

\subsubsection{$C_{eu}^{1333},~C_{eu}^{2333}$ \label{sec:421}}
The first T3-dominated example is the pair $C_{eu}^{1333}$ and $C_{eu}^{2333}$, which served as the worked example of Sec.~\ref{sec:framework}, so both mechanisms are already derived there and we only recall their structure. For T1, the single-log terms of the chain vanish identically, Eq.\,(\ref{eq:Yecancel}), and the leading contribution is the rediagonalization result of Eqs.\,(\ref{eq:T1eu}), (\ref{eq:T1phie}). The rotation angles $\theta_{1,2}$ generate $C^{1233}_{eu}$ at order $L^2$ and the Higgs-lepton current $C^{12}_{\phi e}$ at order $L^3$, carrying two and three loop factors respectively, both validated against the full evolution. For T3, the one-loop double insertion gives the single-log dimension-eight coefficient of Eq.\,(\ref{eq:dim8top}). Since both external up-type quarks are tops, its tree-level WET matching vanishes, and the leading contribution is the nested double log of Eq.\,(\ref{eq:T3toploop}), injected as the effective coefficient $C^{\mathrm{eff},12}_{\phi e}$ of Eq.\,(\ref{eq:effphie}) at $\mu_{\rm inj}\simeq 100$~GeV, where it reaches coherent $\mu\to e$ conversion at tree level through the flavour-changing $Z$ coupling.

\begin{figure}[ht]
\centering
\includegraphics[width=0.48\linewidth]{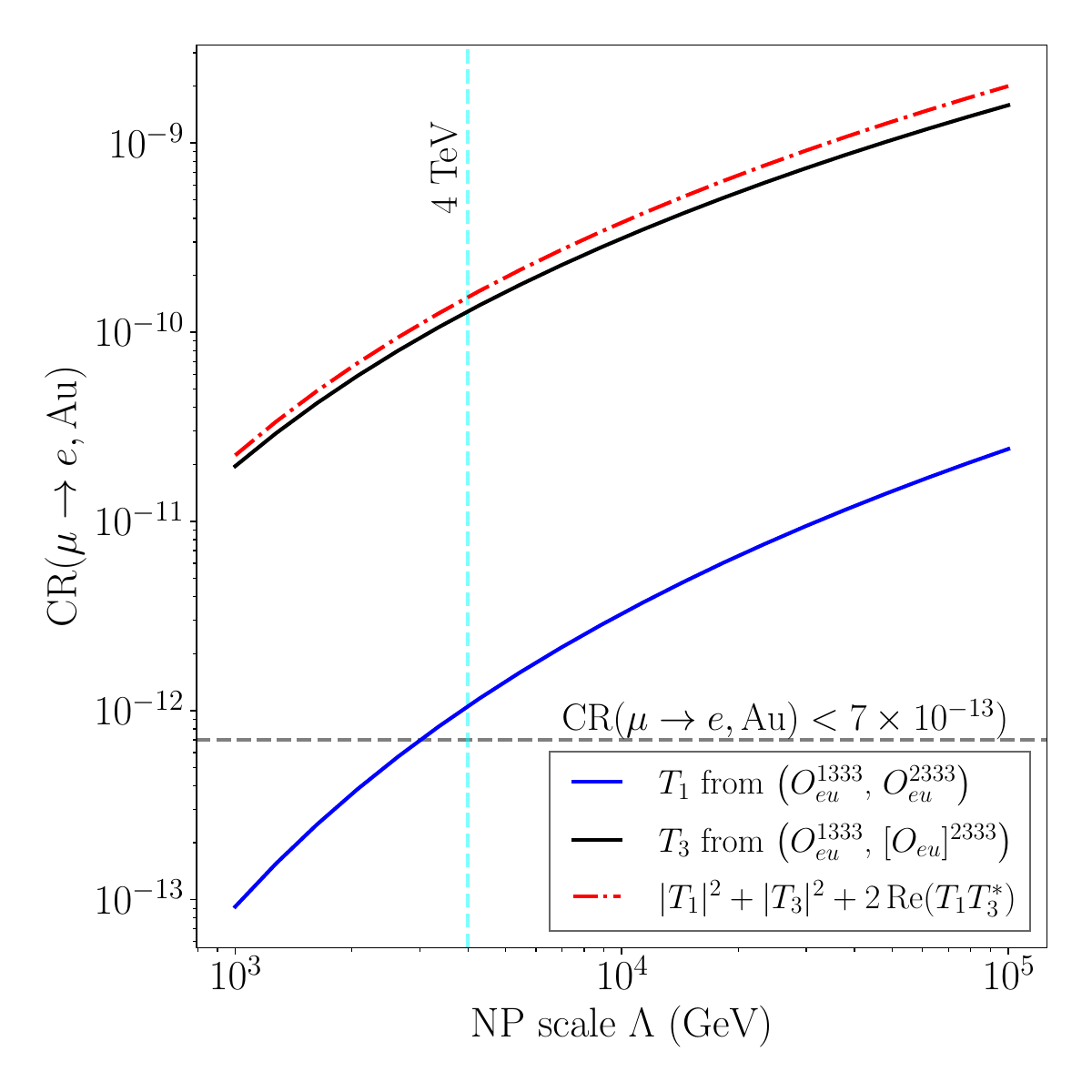}
\includegraphics[width=0.48\linewidth]{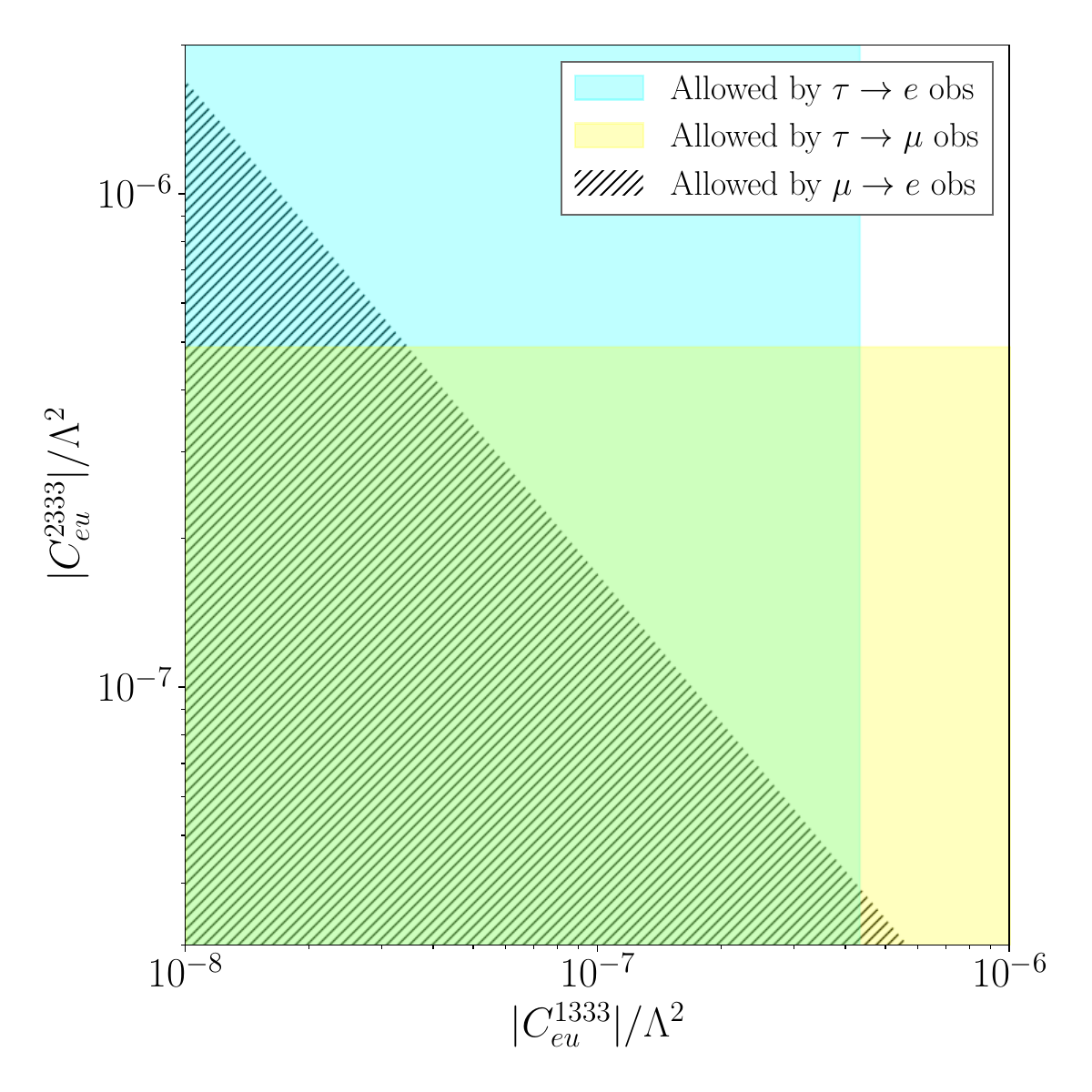}
\caption{Same as Fig.~\ref{fig:411} but for the T3-dominated pair $C_{eu}^{1333}$--$C_{eu}^{2333}$.}
\label{fig:421}
\end{figure}

At $\Lambda=4$~TeV with both coefficients at their direct bounds, $|C^{1333}_{eu}|/\Lambda^2=4.3\times10^{-7}\,\mathrm{GeV}^{-2}$ and $|C^{2333}_{eu}|/\Lambda^2=4.9\times10^{-7}\,\mathrm{GeV}^{-2}$, the full evolution gives $\mathrm{CR}(\mu\to e,\mathrm{Au})=1.1\times10^{-12}$ from T1 and $1.3\times10^{-10}$ from T3, and the same two-order hierarchy persists across the range $\Lambda=1$--$100$~TeV shown in the left panel of Fig.~\ref{fig:421}.
The dominance admits a compact analytical form. Because both mechanisms populate the same operator, the amplitude ratio is the ratio of the coefficients of Eq.\,(\ref{eq:T1phie}) and~(\ref{eq:effphie}), in which $N_c$, $y_t^4$ and $v_T^2$ cancel,
\begin{align}
\frac{|T_1|}{|T_3|}
= \frac{|C^{12}_{\phi e}\big|_{T_1}|}{|C^{\mathrm{eff},12}_{\phi e}\big|_{T_3}|}
= \,\frac{4\,\big(2N_c\,y_t^{2}-3\lambda\big)}{3\cdot 16\pi^{2}}\;L\,,
\label{eq:T1T3ratio}
\end{align}
so a single loop factor separates the two mechanisms and only one power of the logarithm works in favour of T1. The resulting exclusion is the hyperbola $|C^{1333}_{eu}C^{2333}_{eu}|/\Lambda^4=1.4\times10^{-14}\,\mathrm{GeV}^{-4}$, strengthening the bound on the coefficient product by a factor of about $15$ relative to the product of the direct $\tau\to\ell$ bounds. The corresponding plots are shown in the right panel of Fig.~\ref{fig:421}.

\subsubsection{$C_{qe}^{3313}$, $C_{qe}^{3323}$}
\label{sec:422}
The second example is the pair $C_{qe}^{3313}$ and $C_{qe}^{3323}$, with direct bounds $|C^{3313}_{qe}|/\Lambda^2=4.0\times10^{-7}\,\mathrm{GeV}^{-2}$ and $|C^{3323}_{qe}|/\Lambda^2=5.0\times10^{-7}\,\mathrm{GeV}^{-2}$, both from $\tau\to\ell\gamma$. This pair is similar to Sec.~\ref{sec:421}, with the lepton flavour carried by the right-handed singlet currents and the top singlet current replaced by the diagonal third-generation quark doublet current, and both mechanisms follow the corresponding chains of Sec.~\ref{sec:framework} step by step.
The T1 contribution proceeds through the sequence
\begin{align}
C^{33i3}_{qe}
\xrightarrow{\;2N_c(y_t^2-y_b^2)\;}
C^{i3}_{\phi e}
\xrightarrow{\;y_\tau\;}
\big\{\,[Y_e]\,,\;C_{e\phi}\,\big\}
\longrightarrow
\delta M_e
\xrightarrow{\;\text{mass rotation}\;}
C^{12}_{\phi e}\,,
\qquad i=1,2\,.
\label{eq:chain_qe_T1}
\end{align}
The first step is the single-log penguin mixing, given is Ref.~\cite{Jenkins:2013wua} as
\begin{align}
16\pi^2\,\mu\frac{d}{d\mu}\,C^{i3}_{\phi e}
\supset 2N_c\big(y_t^2-y_b^2\big)\,C^{33i3}_{qe}\,,
\label{eq:qeqe_penguin}
\end{align}
the up and down components of the doublet loop entering with opposite sign. The subsequent flavour transfer is the rediagonalization of Sec.~\ref{sec:T1}, generating the $\mu\to e$ Higgs-lepton current $C^{12}_{\phi e}$ at order $L^3$. The rotation also generates the four-fermion coefficient $C^{3312}_{qe}$ at order $L^2$, however, its contribution is negligible, as discussed below.

The T3 contribution follows the chain
\begin{align}
\big\{C^{3313}_{qe},\,C^{3323}_{qe}\big\}
\xrightarrow{\;\mp3y_t^2\;}
C^{(1),(2)e\mu 33}_{e^2q^2 \phi^2}
\xrightarrow{\;2N_c y_t^2\;}
C^{e\mu}_{e^2 \phi^4D}
\xrightarrow{\;v_T^2/2\;}
C^{\mathrm{eff},12}_{\phi e}
\label{eq:chain_qe_T3}
\end{align}
The first step is the one-loop double insertion, whose single-log solutions follow from \cite{Ardu:2022pzk}, evaluated at coincident third-generation quark indices,
\begin{align}
C^{(1)e\mu 33}_{e^2q^2 \phi^2}(\mu_{\rm inj})
= \frac{3\,y_t^{2}}{16\pi^2}\,L\;C^{3313}_{qe}\big(C^{3323}_{qe}\big)^{*}
= -\,C^{(2)e\mu 33}_{e^2q^2 \phi^2}(\mu_{\rm inj})\,,
\label{eq:qeqe_dim8}
\end{align}
The first approach of Sec.~\ref{sec:T3} therefore does not contribute as the direct matching of Eq.\,(\ref{eq:effC6}) would produce a $\mu e\,\bar t t$ contact interaction with vanishing low-energy image. The leading T3 arises through the second approach, the nested double log into the Higgs-lepton operator $O_{e^2 \phi^4D}$, with the top-loop anomalous dimension approximated by the up-component of the same dimension-six entry in Ref.~\cite{Jenkins:2013wua}. 
In analogy with Eq.\,(\ref{eq:effphie}) we get,
\begin{align}
\frac{C^{\mathrm{eff},12}_{\phi e}(\mu_{\rm inj})}{\Lambda^2}
= -\,\frac{3\,N_c\,v_T^{2}\,y_t^{4}}{(16\pi^2)^{2}}\,L^{2}\;
\frac{C^{3313}_{qe}\,\big(C^{3323}_{qe}\big)^{*}}{\Lambda^{4}}\,,
\label{eq:qeqe_T3}
\end{align}

\begin{figure}[ht]
\centering
\includegraphics[width=0.48\linewidth]{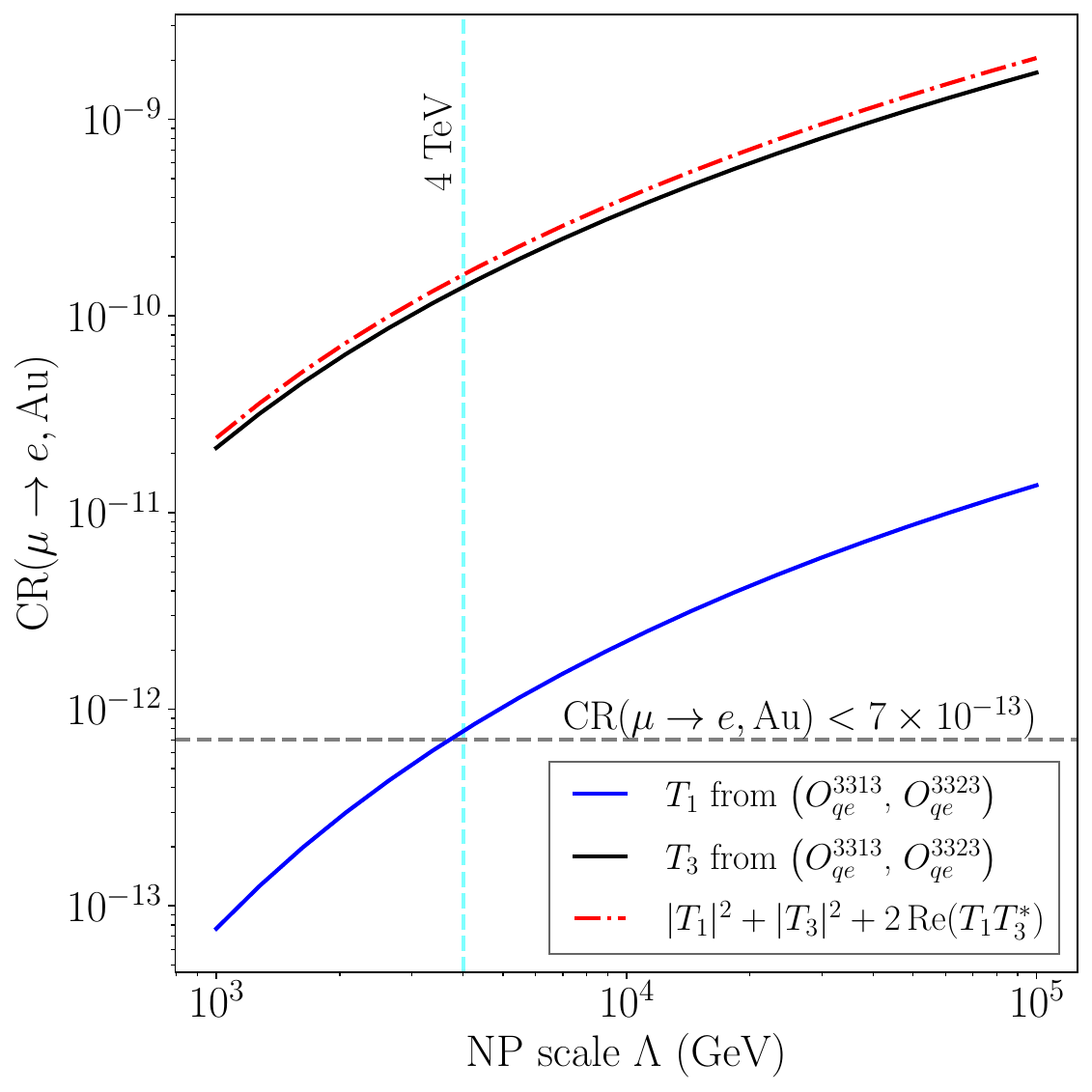}
\includegraphics[width=0.48\linewidth]{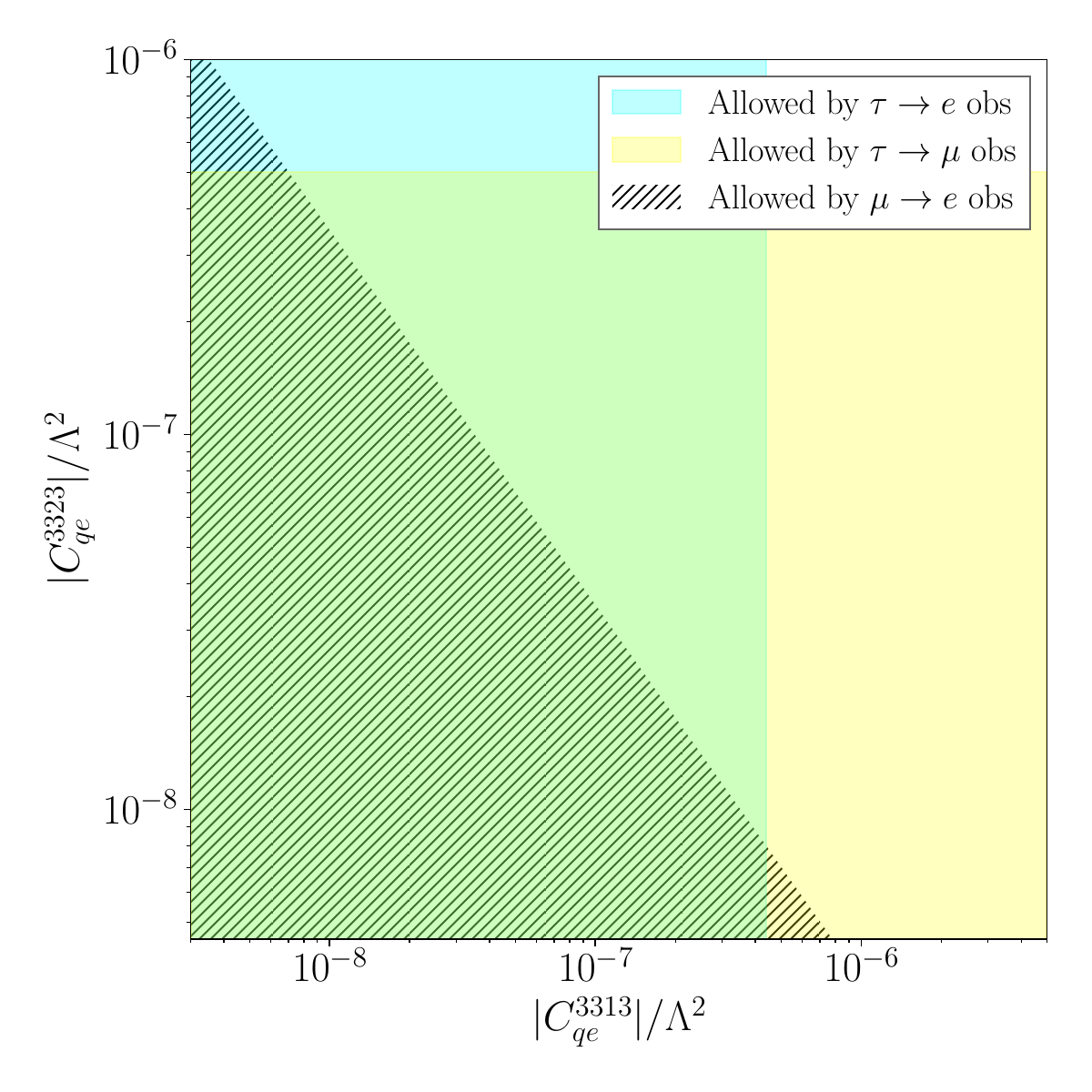}
\caption{Same as Fig.~\ref{fig:411} but for the T3-dominated pair $C_{qe}^{3313}$--$C_{qe}^{3323}$.}
\label{fig:422}
\end{figure}

Numerically, at $\Lambda=4$~TeV with both coefficients at their direct bounds, $C^{13}_{\phi e}=4.1\times10^{-8}\,\mathrm{GeV}^{-2}$ and $C^{23}_{\phi e}=5.1\times10^{-8}\,\mathrm{GeV}^{-2}$, the full evolution gives $C^{12}_{\phi e}=4.0\times10^{-12}\,\mathrm{GeV}^{-2}$ and $\mathrm{CR}(\mu\to e,\mathrm{Au})\big|_{T_1}=6.4\times10^{-13}$.
The injection of Eq.~\eqref{eq:qeqe_T3} evaluates to $-5.7\times10^{-11}\,\mathrm{GeV}^{-2}$ and gives $\mathrm{CR}(\mu\to e,\mathrm{Au})\big|_{T_3}=1.2\times10^{-10}$. Since both mechanisms generate $O^{12}_{\phi e}$, their ratio takes the closed form
\begin{align}
\frac{|T_1|}{|T_3|}
= \frac{|C^{12}_{\phi e}\big|_{T_1}|}{|C^{\mathrm{eff},12}_{\phi e}\big|_{T_3}|}
= \,\frac{4\,\big(y_t^{2}-y_b^{2}\big)\big(2N_c\,y_t^{2}-3\lambda\big)}{3\,y_t^{2}\cdot 16\pi^{2}}\;L\,,
\label{eq:qeqe_ratio}
\end{align}
resulting in $|T_1/T_3|=0.07$. T3 dominates the rate by a factor of $180$, uniformly in $\Lambda$ as shown in the left panel of Fig.~\ref{fig:422}.
The quoted indirect bound uses $\mathrm{CR}(\mu\to e,\mathrm{Au})=1.3\times10^{-10}$ at the direct-bound corner, exceeding the SINDRUM~II limit by a factor of about $200$; the resulting exclusion is the hyperbola $|C^{3313}_{qe}C^{3323}_{qe}|/\Lambda^4=1.4\times10^{-14}\,\mathrm{GeV}^{-4}$, strengthening the bound on the coefficient product by a factor of about $14$ relative to the product of the direct bounds. The corresponding plots are shown in the right panel of Fig.~\ref{fig:422}.

\subsubsection{$C_{\ell u}^{1333}$, $C_{qe}^{3323}$}
\label{sec:423}

The third example combines $C_{\ell u}^{1333}$, carrying the $\tau\to e$ transition on the lepton doublets together with the diagonal top singlet current, with $C_{qe}^{3323}$, carrying the $\tau\to\mu$ transition on the lepton singlets together with the diagonal third-generation quark doublet current. The pair instead populates the mixed class in which the $\tau$ line is closed by one $y_\tau$ insertion and the quark lines by one $y_t$ insertion, each supplying one external Higgs leg~\cite{Ardu:2022pzk}. Since both up-type indices of the sourced dimension-eight operator are in the third generation, the first approach of Sec.~\ref{sec:T3} does not contribute. The direct matching of Eq.\,(\ref{eq:effC6}) would produce a $\mu e\,\bar t t$ contact interaction with vanishing image in the low-energy theory. The T3 contribution follows the second approach, the nested double log, along the chain
\begin{align}
\big\{C^{1333}_{\ell u},\,C^{3323}_{qe}\big\}
\xrightarrow{\;y_\tau\,y_t\;}
C^{(3)e\mu 33}_{\ell equ \phi^2}
\xrightarrow{\;g_1 y_t,\;g_2 y_t\;}
C^{e\mu}_{\ell eB \phi^3}\,,\;C^{e\mu}_{\ell eW \phi^3}
\xrightarrow{\;v_T^2/2\;}
C^{12}_{eB}\,,\;C^{12}_{eW}
\longrightarrow
(\mu\to e\gamma)\,.
\label{eq:chain_lu_qe_T3}
\end{align}
The first step of Eq.~\eqref{eq:chain_lu_qe_T3} is the one-loop double insertion sourcing the top-flavoured dimension-eight tensor, given in ~\cite{Ardu:2022pzk} as 
\begin{align}
16\pi^2\,\mu\frac{d}{d\mu}\,C^{(3)e\mu 33}_{\ell equ \phi^2}
&= y_\tau\,y_t\;C^{1333}_{\ell u}\,\big(C^{3323}_{qe}\big)^{*}
\\
\frac{C^{(3)e\mu 33}_{\ell equ \phi^2}(\mu)}{\Lambda^4}
&= -\,\frac{y_\tau\,y_t}{16\pi^2}\,L\;
\frac{C^{1333}_{\ell u}\,\big(C^{3323}_{qe}\big)^{*}}{\Lambda^{4}}\,,.
\label{eq:luqe_dim8}
\end{align}
The second step is the mixing of the top-flavoured tensor into the dimension-eight dipole operators. A chirality-flipping tensor with top legs closes the top loop with a single Yukawa insertion and a gauge-boson emission, so the anomalous dimension is $m_t$-enhanced.
We approximate it by the corresponding dimension-six entries \cite{Jenkins:2013wua}
\begin{align}
\gamma_{eW} = -\,2\,N_c\,g_2\,y_t\,,
\qquad
\gamma_{eB} = +\,4\,N_c\,\big(\mathsf{y}_u+\mathsf{y}_q\big)\,g_1\,y_t
= +\,\tfrac{10}{3}\,N_c\,g_1\,y_t\,,
\label{eq:luqe_gammas}
\end{align}
with $\mathsf{y}_{u,q}$ the hypercharges.
The nested double-log solution, matched by the $v_T^2/2\Lambda^2$ rule, then gives the effective dipole coefficients injected at $\mu_{\rm inj}\simeq100$~GeV,
\begin{align}
\frac{C^{12}_{eB,eW}(\mu_{\rm inj})}{\Lambda^2}
= \frac{v_T^{2}}{2}\;
\frac{\gamma_{eB,eW}\;y_\tau\,y_t}{2\,(16\pi^2)^{2}}\,L^{2}\;
\frac{C^{1333}_{\ell u}\,\big(C^{3323}_{qe}\big)^{*}}{\Lambda^{4}}
= \big(+5.7,\,-6.1\big)\times10^{-14}\,\mathrm{GeV}^{-2}
\label{eq:luqe_dipole}
\end{align}
at $\Lambda=4$~TeV with both coefficients at their direct bounds.

\begin{figure}[ht]
\centering
\includegraphics[width=0.48\linewidth]{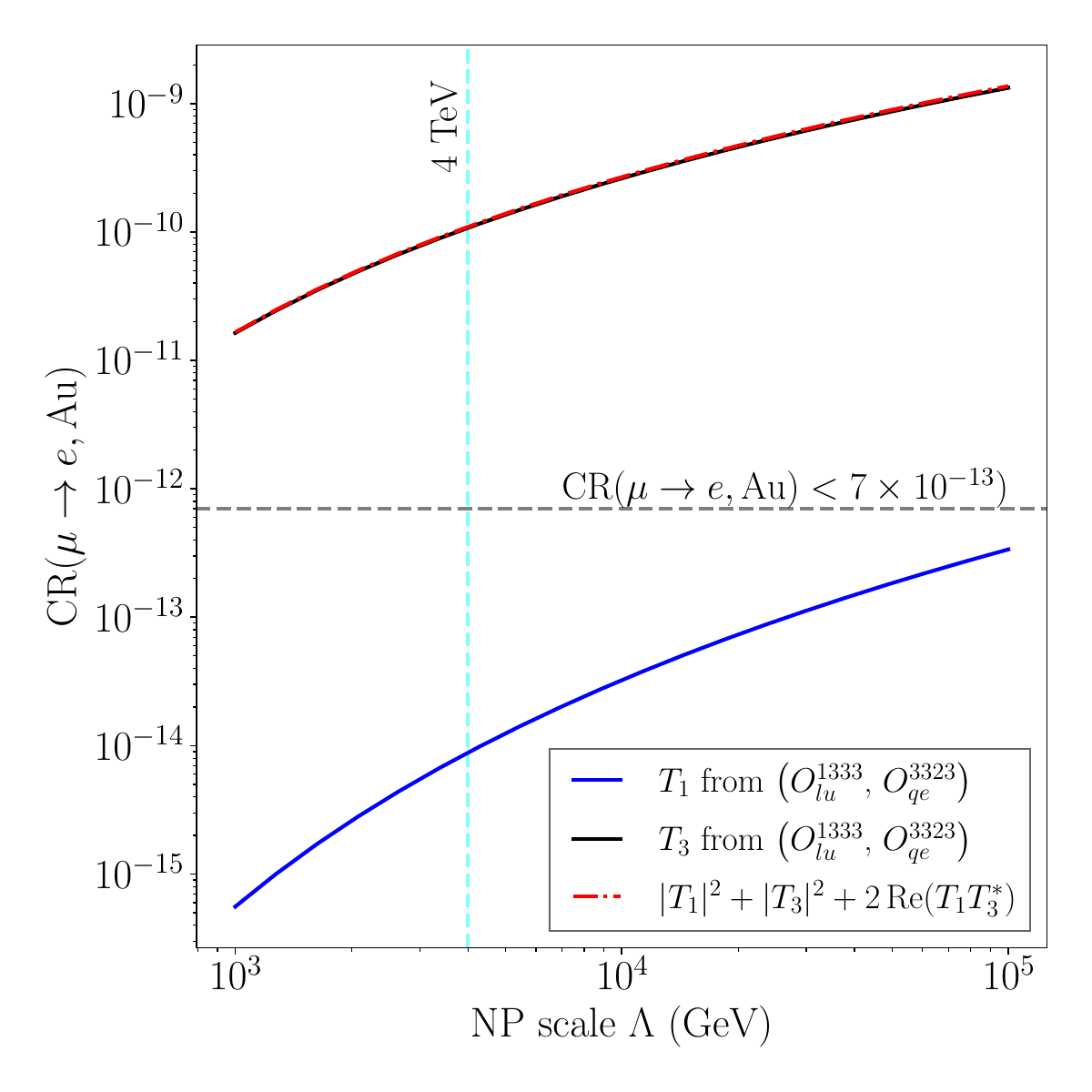}
\includegraphics[width=0.48\linewidth]{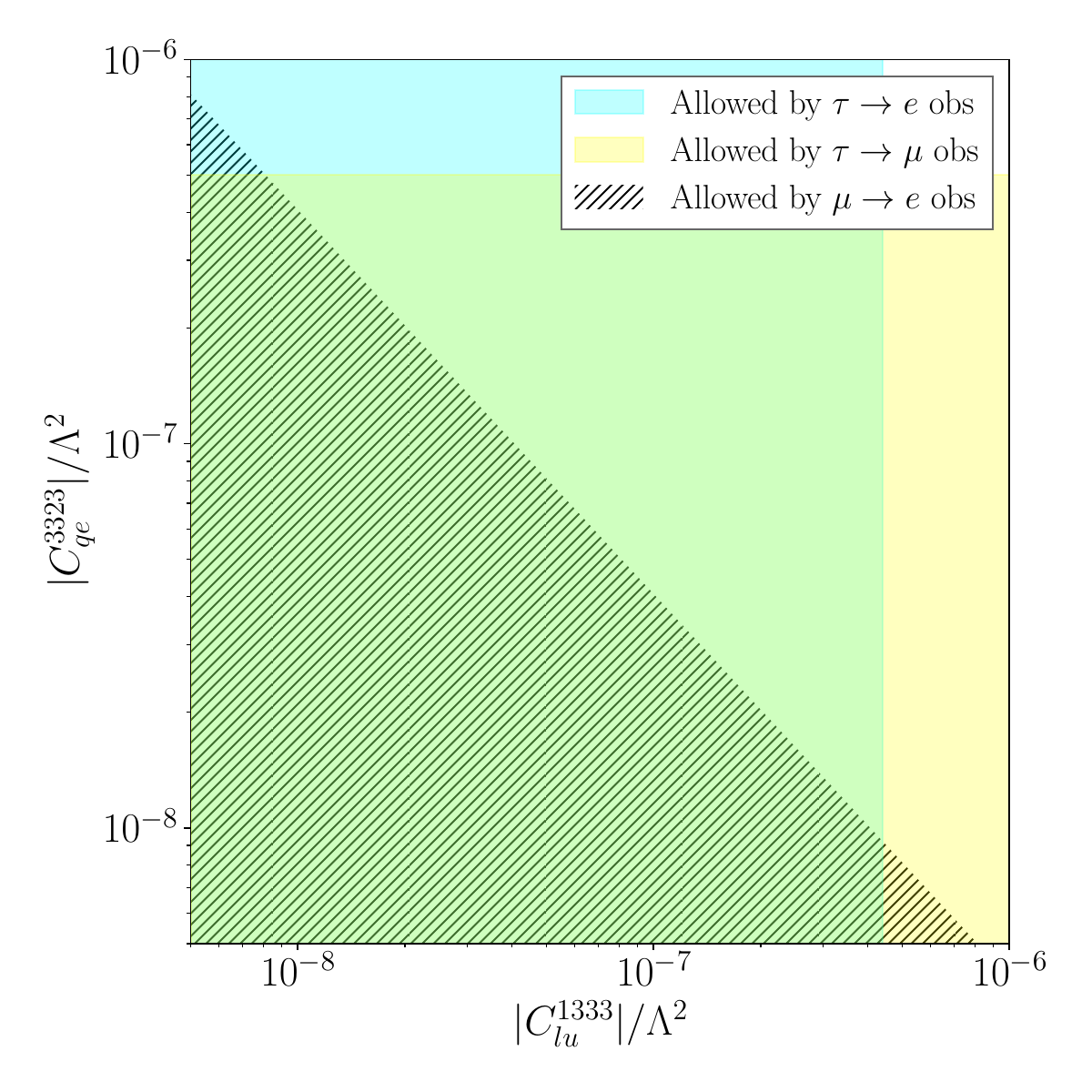}
\caption{Same as Fig.~\ref{fig:411} but for the T3-dominated pair $C_{\ell u}^{1333}$--$C_{qe}^{3323}$.}
\label{fig:423}
\end{figure}

The T1 contribution is multi-step: each operator closes its own diagonal top loop into a Higgs-lepton penguin at single-log strength, $C^{(1)13}_{\phi\ell}=4.3\times10^{-8}\,\mathrm{GeV}^{-2}$ and $C^{23}_{\phi e}=5.1\times10^{-8}\,\mathrm{GeV}^{-2}$, the $\mu\to\tau$ penguin feeds the mass matrix through its $y_\tau$-suppressed Yukawa mixing, and the rediagonalization of Sec.~\ref{sec:T1} generates $C^{(1)12}_{\phi\ell}=2.4\times10^{-13}\,\mathrm{GeV}^{-2}$, three explicitly loop-suppressed steps giving $\mathrm{BR}(\mu\to e\gamma)=5.4\times10^{-13}$.
The T3 dipole of Eq.~\eqref{eq:luqe_dipole} instead gives $\mathrm{BR}(\mu\to e\gamma)=5.4\times10^{-8}$, five orders of magnitude larger, so the pair is by far the most strongly T3-dominated combination in our analysis. The interference is irrelevant at this hierarchy , and the constructive envelope coincides with T3 to high accuracy. The radiative mode is the leading observable as the prediction exceeding the MEG limit of $4.2\times10^{-13}$ by a factor of $1.3\times10^{5}$. The indirect bound on the coefficient product is stronger than the product of the direct $\tau\to\ell$ bounds by a factor of about $360$. The corresponding plots are shown in Fig.~\ref{fig:423}.

\subsection{T1--T3 comparable scenarios}

Finally, we consider operator pairs for which the two mechanisms, T1 and T3, contribute at comparable strength, so that neither can be neglected in the indirect bound. Both contributions are at one-loop and single-log level. Here T1 comes from the rotation mechanism of Sec.~\ref{sec:T1}, in which one operator sources the charged-lepton mass misalignment and the resulting rotation acts on its partner and T3 comes from the light-flavour track of Sec.~\ref{sec:T3}, in which the double insertion matches directly onto an effective dimension-six coefficient. As a result, the electroweak factor $v_T^2$, the loop factor, the logarithm and the coefficient product all cancel in their ratio. With the treatment of the preceding sections, the criterion for this regime is structural rather than scale dependent.

When the amplitudes are this close, the interference term $2|T_1||T_3|$ modifies the combined rate, and hence the extracted bound, at the tens-of-percent level, and its treatment becomes quantitatively important. The representative example in the following, combines the four-fermion coefficient with the Higgs-lepton penguin 
:
\begin{align}
{\rm 1. }~C^{(1)1311}_{lq}, C^{(1)23}_{\phi l}.
\end{align}

\subsubsection{$C^{(1)1311}_{lq},~C^{(1)23}_{\phi l}$}\label{sec:431}

The pair $C^{(1)1311}_{\ell q}$ and $C^{(1)23}_{\phi\ell}$, with direct bounds $|C^{(1)1311}_{\ell q}|/\Lambda^2=5.7\times10^{-7}\,\mathrm{GeV}^{-2}$ from $\tau\to\rho e$ and $|C^{(1)23}_{\phi\ell}|/\Lambda^2=1.4\times10^{-8}\,\mathrm{GeV}^{-2}$ from $\tau\to3\mu$, illustrates the intermediate regime in which the combined bound gets contributions from both T1 and T3 mechanisms.

The T1 contribution follows the single-step rotation mechanism of Sec.~\ref{sec:412}, with the penguin as the source of the misalignment:
\begin{align}
C^{(1)23}_{\phi\ell}
\xrightarrow{\;y_\tau\;}
\big\{\,[Y_e]_{23}\,,\;C^{23}_{e\phi}\,\big\}
\longrightarrow
\delta[M_e]_{23}
\xrightarrow[C^{(1)1311}_{lq}]{\;\text{mass rotation}\;}
C^{(1)1211}_{\ell q}\,.
\label{eq:chain_lq_phil}
\end{align}
The induced misalignment of the charged-lepton mass matrix is
\begin{align}
\delta[M_e]_{23}
= \frac{v_T}{\sqrt{2}}\Big(\delta[Y_e]_{23} - \frac{v_T^{2}}{2}\,\delta C^{23}_{e \phi}\Big)
= \gamma_\theta\;y_\tau\,\frac{v_T}{\sqrt2}\,\frac{v_T^{2}}{2}\,\frac{L}{16\pi^2}\;C^{(1)23}_{\phi\ell}\,,
\label{eq:lqphil_theta}
\end{align}
where $\gamma_\theta=1.4$ collects the $\lambda$-, gauge- and Yukawa-weighted terms upon using the ADM elements for $[Y_e]_{23}$ and  $C^{23}_{e \phi}$ from \cite{Jenkins:2013zja, Jenkins:2013wua}.
The generated coefficient takes the single-log closed form
\begin{align}
\frac{C^{(1)e\mu 11}_{\ell q}(\mu_{\rm inj})}{\Lambda^2}
= -\,\gamma_\theta\;\frac{v_T^{2}}{2}\,\frac{L}{16\pi^2}\;
\frac{C^{(1)1311}_{\ell q}\,\big(C^{(1)23}_{\phi\ell}\big)^{*}}{\Lambda^{4}}\,,
\label{eq:lqphil_T1}
\end{align}

The T3 contribution is the one-loop double insertion, which for this pairing generates two dimension-eight structures~\cite{Ardu:2022pzk}
\begin{align}
16\pi^2\,\mu\frac{d}{d\mu}\,C^{(1)e\mu 11}_{\ell^2q^2 \phi^2}
= g'^2\;C^{(1)1311}_{\ell q}\big(C^{(1)23}_{\phi\ell}\big)^{*}\,,
\qquad
16\pi^2\,\mu\frac{d}{d\mu}\,C^{(4)e\mu 11}_{\ell^2q^2 \phi^2}
= 3g^2\;C^{(1)1311}_{\ell q}\big(C^{(1)23}_{\phi\ell}\big)^{*}\,,
\label{eq:lqphil_dim8}
\end{align}
Since the external quarks are first generation, the first approach of Sec.~\ref{sec:T3} applies, and the matching rule of Eq.\,(\ref{eq:effC6}) converts the leading-log solutions into effective dimension-six coefficients of singlet and triplet type at $\mu_{\rm inj}$. 
\begin{align}
\frac{C^{(1),\mathrm{eff}\,e\mu 11}_{\ell q}}{\Lambda^2}
&= -\,\frac{v_T^{2}\,g'^2}{2\cdot16\pi^2}\,L\;
\frac{C^{(1)1311}_{\ell q}\big(C^{(1)23}_{\phi\ell}\big)^{*}}{\Lambda^{4}}\,,
\\
\frac{C^{(3),\mathrm{eff}\,e\mu 11}_{\ell q}}{\Lambda^2}
&= -\,\frac{3\,v_T^{2}\,g^2}{2\cdot16\pi^2}\,L\;
\frac{C^{(1)1311}_{\ell q}\big(C^{(1)23}_{\phi\ell}\big)^{*}}{\Lambda^{4}}\,.
\label{eq:lqphil_eff}
\end{align}
Contribution from the triplet operator is suppressed due to the isospin structure and cancellation between protons and neutrons.

\begin{figure}[ht]
\centering
\includegraphics[width=0.48\linewidth]{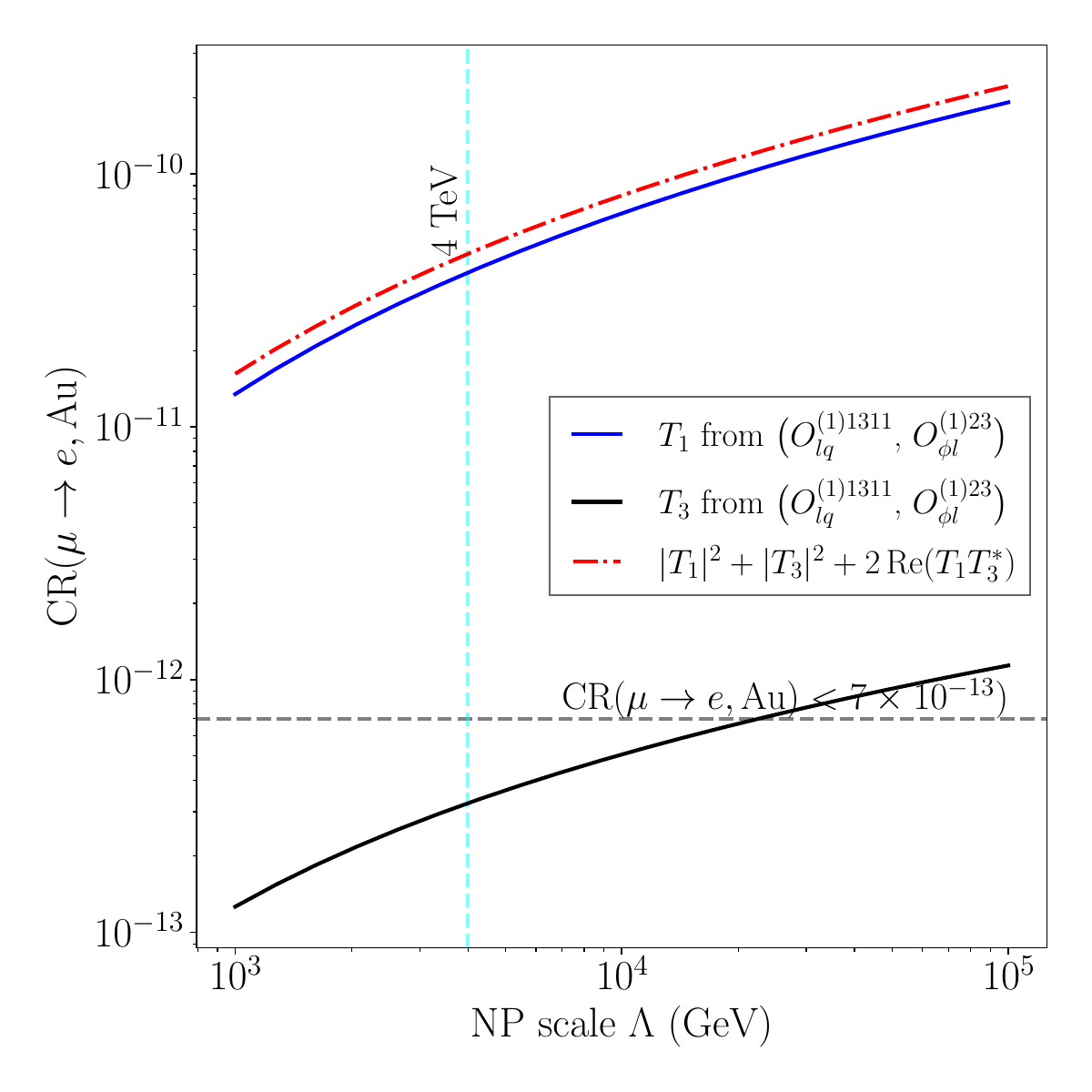}
\includegraphics[width=0.48\linewidth]{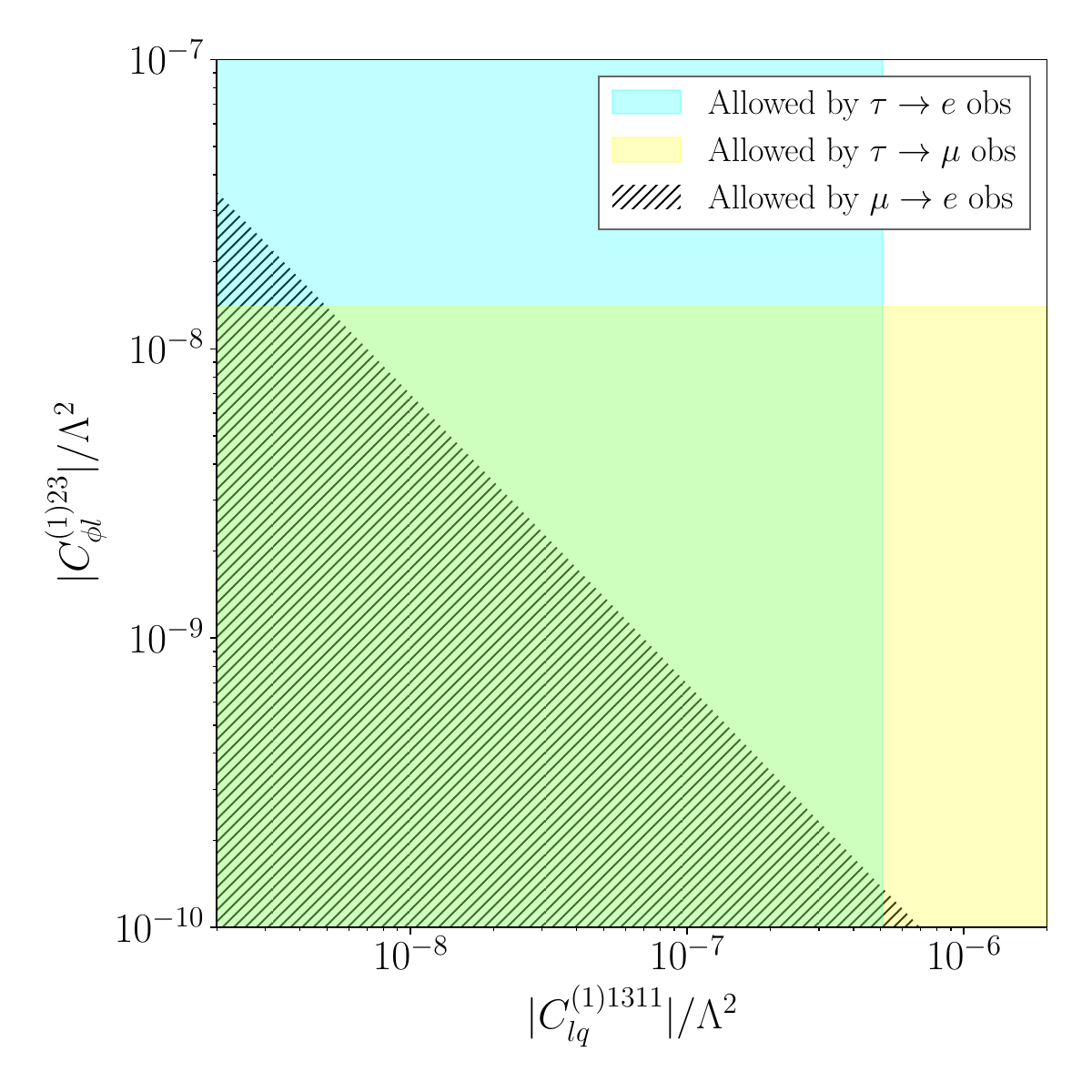}
\caption{Same as Fig.~\ref{fig:411} but for the T1 and T3 comparable pair $C_{lq}^{(1)1311}$--$C_{\phi l}^{(1)23}$.}
\label{fig:431}
\end{figure}

Both the mechanisms T1 and T3 generate the singlet operator $O^{(1)e\mu 11}_{\ell q}$, and dividing Eq.~\eqref{eq:lqphil_T1} by the first of Eqs.~\eqref{eq:lqphil_eff} gives the closed-form ratio
\begin{align}
\frac{T_1}{T_3}\bigg|_{O^{(1)}_{\ell q}}
= +\,\frac{\gamma_\theta}{g'^{2}}\,.
\label{eq:lqphil_ratio}
\end{align}
In contrast to the double-log pairs of Sec.~\ref{sec:42}, where the corresponding ratio grows with the logarithm, here the electroweak factor $v_T^2$, the loop factor, the logarithm and the coefficient product all cancel. The comparable character of this pair is therefore structural rather than accidental. Numerically, at $\Lambda=4$~TeV with both coefficients at their direct bounds, the full evolution gives the effective rotation angle $\theta_{\mu\tau}=1.4\times10^{-5}$ and $C^{(1)1211}_{\ell q}=8.0\times10^{-12}\,\mathrm{GeV}^{-2}$, carrying essentially the entire T1 rate, $\mathrm{CR}(\mu\to e,\mathrm{Au})\big|_{T_1}=5.1\times10^{-11}$; the penguin route yields $C^{(1)13}_{\phi\ell}=1.1\times10^{-9}\,\mathrm{GeV}^{-2}$ but only $C^{(1)12}_{\phi\ell}=1.9\times10^{-14}\,\mathrm{GeV}^{-2}$, confirming its irrelevance. The injections of Eq.~\eqref{eq:lqphil_eff} evaluate to $(-0.73,\,-7.1)\times10^{-12}\,\mathrm{GeV}^{-2}$ and give $\mathrm{CR}(\mu\to e,\mathrm{Au})\big|_{T_3}=1.3\times10^{-12}$, three times the value obtained from the singlet structure alone. With $\gamma_\theta=1.4$, Eq.~\eqref{eq:lqphil_ratio} evaluates to $10.9$ on the singlet coefficient, reduced to the amplitude ratio $|T_1/T_3|=6.4$ once the isospin-suppressed triplet piece is included. T1 dominates the rate by a factor of $40$, but the interference term $2|T_1||T_3|$ amounts to $31\%$ of the T1 rate. The comparison of T1, T3 and total including the interference over the scale are shown in the left panel of Fig.~\ref{fig:431}. The quoted indirect bound uses $\mathrm{CR}(\mu\to e,\mathrm{Au})=6.8\times10^{-11}$ at the corner, exceeding the SINDRUM~II limit by a factor of about $100$ and strengthening the bound on the coefficient product by a factor of about $10$ relative to the product of the direct bounds. The corresponding plots are shown in the right panel of Fig.~\ref{fig:431}.


\subsection{Projected sensitivity}
\label{sec:future}

It is instructive to ask how the indirect bounds evolve as the experimental
programme advances. A $\mu\to e$ observable mediated by a single $\mu e$ operator
scales as $|C_{\mu e}|^2/\Lambda^4$, whereas an observable induced by a double
insertion of two $\tau\to\ell$ operators scales as
$|C_{\tau e}|^2\,|C_{\tau\mu}|^2/\Lambda^8$. The $\tau\to e$ and $\tau\to\mu$ observable limits are expected to
improve by roughly one order of magnitude (Belle~II), while the $\mu\to e$ limits
are projected to tighten by three to four orders of magnitude
(MEG~II, Mu3e, Mu2e/COMET); see Table~\ref{tab:obsbound}.

\begin{figure}[H]
\centering
\includegraphics[width=0.55\linewidth]{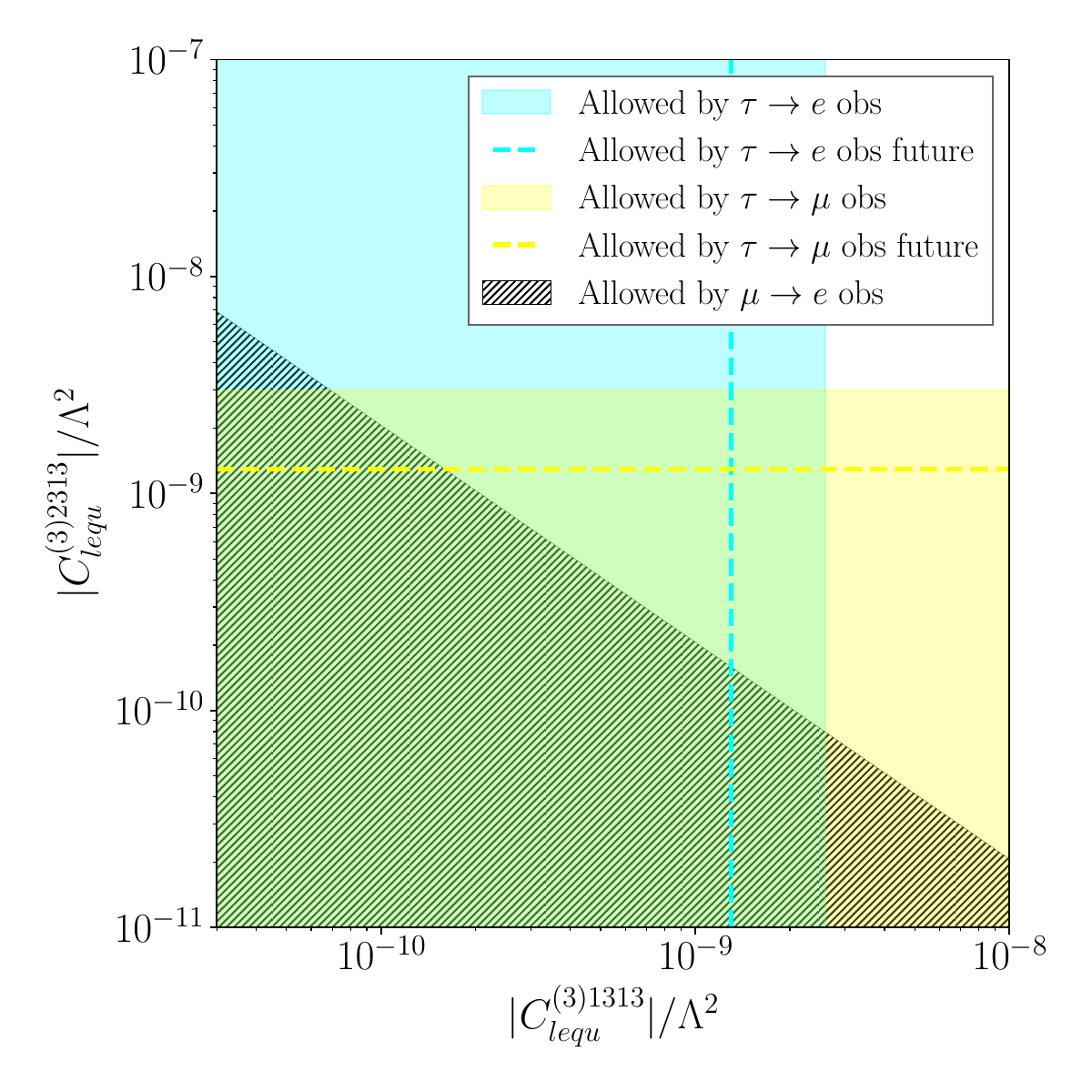}
\caption{Projected future indirect bound for the $C^{(3)1313}_{\ell equ}$--$C^{(3)2313}_{\ell equ}$ pair, compared with the current limits. The yellow and cyan color region corresponds to current limits. The colored parallel dashed lines region corresponds to future limits. Black contour with dashed lines represents the $\mu-e$ limits.}
\label{fig:futurebound_lequ_lequ}
\end{figure}

The direct bound on $\tau\to\ell$ coefficients decrease about less than one order from future sensitivity. For one order change in coefficient, branching ratios of $\mu\to e$ observable should decrease by an order of four.  The indirect bound on a product of $\tau\to\ell$  coefficients depends on the
 the $\mu\to e$ sensitivity for eg. $\rm CR(\mu \to e, Al)$, a four-order-of-magnitude decrement in the
$\mu\to e$ limit translates into only about two order of magnitude in the bound on
each $\tau\to\ell$ coefficient. For most operator pairs this is comparable to the
direct improvement expected from Belle~II, so the relative ranking of direct versus
indirect bounds is largely preserved. The exception is the set of operator pairs
involving $O_{\phi\ell}^{(1,3)}$ and $O_{\ell equ}^{(1,3)}$, for which the indirect bounds already
constrain a large portion of parameter space at current sensitivity; for these the
future $\mu\to e$ program pushes the constraints well beyond the direct reach.
Fig.~\ref{fig:futurebound_lequ_lequ} illustrates this for the
$C^{(3)1313}_{\ell equ}$--$C^{(3)2313}_{\ell equ}$ pair. 

At current sensitivity the strongest indirect constraints arise from
$\mathrm{CR}(\mu\to e,\mathrm{Au})$ acting on Higgs-current and two-quark--two-lepton
operators, in particular $O_{\ell edq}$ and $O_{\ell equ}^{(1,3)}$, with several other
two-quark--two-lepton operators receiving milder constraints. With future limits, the
``self-constraining'' $\tau\to\ell$ observables begin to dominate over
$\mathrm{CR}(\mu\to e)$ for some channels, so that the indirect reach grows fastest
precisely for the $O_{\phi\ell}^{(1,3)}$ and $O_{\ell equ}^{(1,3)}$ type pairs highlighted above.


\section{Concluding remarks} \label{conclusion}
We have studied how charged-lepton flavour violation generated purely in the $\tau$ sector propagates into $\mu\to e$ observables within the SMEFT. Assuming that new physics at a scale $\Lambda\gg v$ produces only a pair of dimension-6 operators mediating $\tau\to e$ and $\tau\to\mu$ transitions, the $\mu\to e$ amplitude first arises at order
$C_6^{\tau e}C_6^{\tau\mu}/\Lambda^4$, through the double insertion of the two operators, which we decompose into three contributions and compare on an equal footing: (T1) renormalisation-group mixing followed by the rotation to the charged-lepton mass basis, which populates dimension-6 $\mu\to e$ operators; (T2) the finite, non-logarithmic
remainder, numerically negligible for all observables considered; and (T3) the overall divergence, which renormalises dimension-8 operators contributing after electroweak symmetry breaking. Since T1 and T3 carry no relative power of $\Lambda$, their hierarchy is fixed by loop factors, logarithms and couplings alone: it is a property of the operator pair, not of the scale, and neither contribution can be dropped a priori.

The dominant mechanism follows from the length of the mixing chain. When one operator feeds the charged-lepton Yukawa sector at one loop, as for combinations of $O^{(1)}_{\ell equ}$, $O_{\ell e}$, $O_{\ell edq}$, $O_{qe}$ and $O^{(1)}_{\phi\ell}$, a single rotation transfers the flavour violation and T1 gives the strongest indirect bound. When the T1 chain needs several loop steps, or is absent, T3 dominates. For top-flavoured operators, whose direct matching onto the low-energy theory vanishes, the leading T3 effect is a double logarithm from mixing into Higgs--lepton or dipole operators. For $\big(C_{\ell u}^{1333},C_{qe}^{3323}\big)$ this proceeds through a top-flavoured tensor and the $m_t$-enhanced dipole, making $\mu\to e\gamma$ the leading observable and strengthening the bound on the coefficient product by more than two orders of magnitude. Finally, for pairs such as $\big(C^{(1)1311}_{\ell q},C^{(1)23}_{\phi\ell}\big)$ both
mechanisms are one-loop, single-logarithm effects, their ratio reduces to a scale-independent ratio of couplings.

The practical consequence is that the limit on a given pair depends on which contributions are retained. Omitting the dominant mechanism misses the bound by orders of magnitude, and omitting the subleading one shifts it by factors of two to three through the neglected ontribution and its interference. This is where our treatment differs from earlier work,
which retained the dimension-8 operators generated by double insertions~\cite{Ardu:2021koz,Ardu:2022pzk}. Since T1 is of the same order, we compare the two mechanisms for the same operator pairs, including the matching of the dimension-8
coefficients onto effective dimension-6 coefficients for light external quarks and its nested treatment for top-flavoured operators.

For $\Lambda\sim4$~TeV and coefficients at the current direct bounds, the induced $\mu\to e$ observables exceed the limits from MEG~II, SINDRUM and SINDRUM~II by up to five orders of magnitude, so current $\mu\to e$ data already constrain the products of $\tau\to\ell$ coefficients ten to several hundred times more strongly than the direct $\tau$-decay measurements. Mu2e and COMET, targeting $\mathrm{CR}(\mu\to e,\mathrm{Al})\sim10^{-17}$, Mu3e, aiming at $\mathrm{BR}(\mu\to eee)\sim10^{-16}$, and MEG~II at its final sensitivity of $6\times10^{-14}$ will tighten the indirect bounds on the products by up to two further orders of magnitude, while Belle~II improves the direct $\tau$-sector limits by about one order.

We consider representative pairs, assume that no genuine dimension-8 operators are generated at the matching scale, and work at one-loop, leading-logarithmic accuracy in $\overline{\rm MS}$, so that effects beyond the leading logarithms are not included; their study, and the inclusion of matching-scale dimension-8 contributions, is left for future work. The decomposition developed here provides the ingredients for a combined interpretation of the $\mu\to e$ and $\tau\to\ell$ sectors in a global SMEFT analysis, which must retain both the dimension-6 and the dimension-8 contributions, or it will misstate the sensitivity to third-generation lepton-flavour violation.

\acknowledgments

This research was supported in part by the International Centre for Theoretical Sciences (ICTS) for the program - ``School for Advanced Topics in Particle physics (SATPP): QCD, Effective Field Theories, and Nuclear Physics" (code: ICTS/SATPP2026/01). NR expresses his gratitude to CHEP, IISc, for the office space during his visit. S.K. acknowledges support from the Institute Postdoctoral Fellowship, IIT Kanpur.


\appendix

\section{Experimental bounds}\label{app:bounds}

\begin{longtable}{llll}
\hline
\textbf{Observable} & \textbf{Current bound} & \textbf{Future bound} & \textbf{Reference} \\
\hline

${BR}(B^+\to K e\mu)$                       & $7.0\times 10^{-9}$   & $7.0\times 10^{-10}$ & \cite{Aaij:2019okb,Belle-II:2018jsg,Calibbi:2017uvl} \\
${BR}(B^+\to K \mu e)$                      & $6.45\times 10^{-9}$  & $6.45\times 10^{-10}$ & \cite{Aaij:2019okb,Belle-II:2018jsg,Calibbi:2017uvl} \\
${BR}(B^0\to K^\ast e\mu)$                 & $5.7\times 10^{-9}$   & $5.7\times 10^{-10}$ & \cite{Aaij:2022mqx,Belle-II:2018jsg,Calibbi:2017uvl} \\
${BR}(B^0\to K^\ast \mu e)$                & $6.8\times 10^{-9}$   & $6.8\times 10^{-10}$ & \cite{Aaij:2022mqx,Belle-II:2018jsg,Calibbi:2017uvl} \\
${BR}(B_s\to e\mu,\,\mu e)$                & $5.4\times 10^{-9}$   & $3.0\times 10^{-10}$ & \cite{Aaij:2017cjb,Calibbi:2017uvl} \\
${BR}(B_s\to \phi \mu e)$                  & $1.6\times 10^{-8}$   & $1.6\times 10^{-9}$ & \cite{Aaij:2022mqx,Belle-II:2018jsg,Calibbi:2017uvl} \\
${BR}(Z\to e\mu)$                          & $7.5\times 10^{-7}$   & $1.0\times 10^{-9}$ & \cite{ATLAS:2014vur,Calibbi:2021pyh} \\
${BR}(\mu\to e\gamma)$                     & $4.2\times 10^{-13}$  & $6.0\times 10^{-14}$ & \cite{MEG:2016leq,MEGII:2018kmf,Baldini:2021elz} \\
${BR}(\mu\to eee)$                         & $1.0\times 10^{-12}$  & $1.0\times 10^{-16}$ & \cite{SINDRUM:1987nra,Blondel:2013ia,Arndt:2020mgu} \\
${CR}(\mu\to e,{Au})$                      & $7.0\times 10^{-13}$  & $6.0\times 10^{-17}$ & \cite{SINDRUMII:2006dvw,Mu2e:2014fns,Adamov:2018vin} \\
${BR}(J/\psi\to e\mu)$                     & $4.5\times 10^{-9}$   & $4.5\times 10^{-9}$ & \cite{Ablikim:2013wfx,Ablikim:2020hsk} \\
${BR}(\Upsilon(1S)\to e\mu)$               & $3.6\times 10^{-7}$   & $3.6\times 10^{-7}$ & \cite{Belle:2022fmx,BABAR:2021weg} \\
${BR}(\Upsilon(1S)\to e\mu\gamma)$         & $4.2\times 10^{-7}$   & $4.2\times 10^{-7}$ & \cite{Belle:2022fmx} \\
${BR}(K_L\to e\mu,\,\mu e)$                & $4.7\times 10^{-12}$  & $4.7\times 10^{-12}$ & \cite{BNL:1998apv,Calibbi:2017uvl} \\

\hline

${BR}(B^+\to K e\tau,\,\tau e)$            & $1.52\times 10^{-5}$  & $2.0\times 10^{-6}$ & \cite{Calibbi:2017uvl,Belle-II:2018jsg,Belle:2023ziz} \\
${BR}(B_s\to e\tau,\,\tau e)$              & $7.3\times 10^{-4}$   & $1.0\times 10^{-7}$ & \cite{PDG,Calibbi:2017uvl} \\
${BR}(B^0\to e\tau,\,\tau e)$              & $2.1\times 10^{-5}$   & $1.0\times 10^{-7}$ & \cite{BABAR:2004gsg,PDG,Calibbi:2017uvl} \\
${BR}(Z\to e\tau)$                         & $9.8\times 10^{-6}$   & $1.0\times 10^{-9}$ & \cite{ATLAS:2021bdj,Calibbi:2021pyh} \\
${BR}(\tau\to e\gamma)$                    & $3.3\times 10^{-8}$   & $8.0\times 10^{-9}$ & \cite{Aubert:2009ag,Belle-II:2018jsg,Banerjee:2022xuw} \\
${BR}(\tau\to e e e)$                      & $2.7\times 10^{-8}$   & $1.0\times 10^{-9}$ & \cite{Hayasaka:2010np,Belle-II:2018jsg,Banerjee:2022xuw} \\
${BR}(\tau\to e \mu\mu)$                   & $2.7\times 10^{-8}$   & $1.0\times 10^{-9}$ & \cite{Hayasaka:2010np,Belle-II:2018jsg,Banerjee:2022xuw} \\
${BR}(\tau\to \pi e)$                      & $8.0\times 10^{-8}$   & $3.0\times 10^{-9}$ & \cite{Belle:2007cio,HFLAV:2023pdp,Belle-II:2018jsg,Banerjee:2022xuw} \\
${BR}(\tau\to K e)$                        & $2.6\times 10^{-8}$   & $3.0\times 10^{-9}$ & \cite{Belle:2010rxj,HFLAV:2023pdp,Belle-II:2018jsg,Banerjee:2022xuw} \\
${BR}(\tau\to \phi e)$                     & $2.0\times 10^{-8}$   & $3.0\times 10^{-9}$ & \cite{Miyazaki:2011xe,Aubert:2009agV,Belle-II:2018jsg,Banerjee:2022xuw} \\
${BR}(\tau\to \rho e)$                     & $2.2\times 10^{-8}$   & $3.0\times 10^{-9}$ & \cite{Miyazaki:2011xe,Aubert:2009agV,Belle-II:2018jsg,Banerjee:2022xuw} \\
${BR}(J/\psi\to e\tau)$                    & $7.5\times 10^{-8}$   & $1.0\times 10^{-9}$ & \cite{BES:2004rku,Ablikim:2020hsk} \\
${BR}(\Upsilon(1S)\to e\tau)$              & $2.4\times 10^{-6}$   & $1.0\times 10^{-7}$ & \cite{Belle:2022fmx,Calibbi:2017uvl} \\
${BR}(\Upsilon(1S)\to e\tau\gamma)$        & $6.5\times 10^{-6}$   & $3.0\times 10^{-7}$ & \cite{Belle:2022fmx,Calibbi:2017uvl} \\
${BR}(\Upsilon(2S)\to e\tau)$              & $3.2\times 10^{-6}$   & $1.0\times 10^{-7}$ & \cite{PDG,Belle:2022fmx,Calibbi:2017uvl} \\
${BR}(\Upsilon(3S)\to e\tau)$              & $4.2\times 10^{-6}$   & $1.0\times 10^{-7}$ & \cite{PDG,Belle:2022fmx,Calibbi:2017uvl} \\

\hline

${BR}(B^+\to K \mu \tau)$                   & $5.9\times 10^{-6}$   & $2.0\times 10^{-6}$ & \cite{Aaij:2020mqb,Belle-II:2018jsg,Calibbi:2017uvl, Belle-II:2018jsg} \\
${BR}(B^+\to K \tau \mu)$                   & $2.45\times 10^{-5}$  & $2.0\times 10^{-6}$ & \cite{Aaij:2020mqb,Belle:2022pcr,Belle-II:2018jsg, Calibbi:2017uvl} \\
${BR}(B^0\to K^\ast \mu \tau)$             & $1.0\times 10^{-5}$   & $1.0\times 10^{-6}$ & \cite{Aaij:2022uow,Belle-II:2018jsg,Calibbi:2017uvl} \\
${BR}(B^0\to K^\ast \tau \mu)$             & $8.2\times 10^{-6}$   & $1.0\times 10^{-6}$ & \cite{Aaij:2022uow,Belle-II:2018jsg,Calibbi:2017uvl} \\
${BR}(B_s\to \mu\tau,\,\tau\mu)$           & $4.2\times 10^{-5}$   & $1.0\times 10^{-7}$ & \cite{Aaij:2019okbBsTauMu,Calibbi:2017uvl} \\
${BR}(Z\to \mu\tau)$                       & $1.2\times 10^{-5}$   & $1.0\times 10^{-9}$ & \cite{ATLAS:2021bdj,Calibbi:2021pyh} \\
${BR}(\tau\to \mu\gamma)$                  & $4.2\times 10^{-8}$   & $8.0\times 10^{-9}$ & \cite{Aubert:2009ag,Belle-II:2018jsg,Banerjee:2022xuw} \\
${BR}(\tau\to \mu\mu\mu)$                  & $2.1\times 10^{-8}$   & $1.0\times 10^{-9}$ & \cite{Hayasaka:2010np,Belle-II:2018jsg,Banerjee:2022xuw} \\
${BR}(\tau\to \mu e e)$                    & $1.8\times 10^{-8}$   & $1.0\times 10^{-9}$ & \cite{Hayasaka:2010np,Belle-II:2018jsg,Banerjee:2022xuw} \\
${BR}(\tau\to \pi \mu)$                    & $1.1\times 10^{-7}$   & $3.0\times 10^{-9}$ & \cite{Hayasaka:2007vc,HFLAV:2023pdp,Belle-II:2018jsg,Banerjee:2022xuw} \\
${BR}(\tau\to K \mu)$                      & $2.3\times 10^{-8}$   & $3.0\times 10^{-9}$ & \cite{Hayasaka:2007vc,HFLAV:2023pdp,Belle-II:2018jsg,Banerjee:2022xuw} \\
${BR}(\tau\to \phi \mu)$                   & $2.3\times 10^{-8}$   & $3.0\times 10^{-9}$ & \cite{Miyazaki:2011xe,Aubert:2009agV,Belle-II:2018jsg,Banerjee:2022xuw} \\
${BR}(\tau\to \rho \mu)$                   & $1.7\times 10^{-8}$   & $3.0\times 10^{-9}$ & \cite{Miyazaki:2011xe,Aubert:2009agV,Belle-II:2018jsg,Banerjee:2022xuw} \\
${BR}(J/\psi\to \mu\tau)$                  & $2.0\times 10^{-6}$   & $1.0\times 10^{-9}$ & \cite{BES:2004rku,Ablikim:2020hsk} \\
${BR}(\Upsilon(1S)\to \mu\tau)$            & $2.6\times 10^{-6}$   & $1.0\times 10^{-7}$ & \cite{BaBar:2010vxb,Belle:2022fmx,Love:2008aa,Calibbi:2017uvl} \\
${BR}(\Upsilon(1S)\to \mu\tau\gamma)$      & $6.1\times 10^{-6}$   & $3.0\times 10^{-7}$ & \cite{Belle:2022fmx,Calibbi:2017uvl} \\
${BR}(\Upsilon(2S)\to \mu\tau)$            & $3.3\times 10^{-6}$   & $1.0\times 10^{-7}$ & \cite{PDG,Belle:2022fmx,Calibbi:2017uvl} \\
${BR}(\Upsilon(3S)\to \mu\tau)$            & $3.1\times 10^{-6}$   & $1.0\times 10^{-7}$ & \cite{PDG,Belle:2022fmx,Calibbi:2017uvl} \\

\hline
\caption{Current bounds and future sensitivities for the $\mu\to e$, $\tau\to e$ and $\tau\to\mu$ lepton-flavour violating observables used in this work.}
\label{tab:obsbound}
\end{longtable}

\section{Interesting combinations based on T1}

\renewcommand{\arraystretch}{1.15}
\setlength{\tabcolsep}{3pt}
\setlength{\LTleft}{0pt}
\setlength{\LTright}{0pt}

\begin{longtable}{@{}p{0.23\textwidth}p{0.25\textwidth}p{0.20\textwidth}p{0.15\textwidth}p{0.15\textwidth}@{}}
	\caption{Interesting pairs for T1, excluding pairs involving \(O_{e\phi}\).}
	\label{tab:T1T2woephi_consistent}\\
	\toprule
	\textbf{WC 1} & \textbf{WC 2} & \textbf{Observable} & \textbf{Prediction} & \textbf{Bound}\\
	\midrule
	\endfirsthead
	
	\toprule
	\textbf{WC 1} & \textbf{WC 2} & \textbf{Observable} & \textbf{Prediction} & \textbf{Bound}\\
	\midrule
	\endhead
	
	\midrule
	\multicolumn{5}{r}{\footnotesize Continued on next page}\\
	\endfoot
	
	\bottomrule
	\endlastfoot
	${C_{\phi\ell}^{(1)}}^{13}=1\times10^{-8}$ & $C_{\ell q}^{(1){2311}}=1\times10^{-7}$ & $\mathrm{CR}(\mu\to e,\mathrm{Au})$ & $7.9\times10^{-13}$ & $7\times10^{-13}$ \\
	  $C_{\ell e}^{1333}=5.6\times10^{-7}$ & $C_{\ell q}^{(1){2311}}=2.9\times10^{-7}$ & $\mathrm{CR}(\mu\to e,\mathrm{Au})$ & $1.5\times10^{-9}$ & $7\times10^{-13}$ \\
	$C_{\ell e}^{1333}=1\times10^{-6}$ & $C_{\ell d}^{2311}=1\times10^{-8}$ & $\mathrm{CR}(\mu\to e,\mathrm{Au})$ & $4.5\times10^{-12}$ & $7\times10^{-13}$ \\
	$C_{\ell e}^{1333}=1\times10^{-6}$ & $C_{\ell d}^{2321}=3.3\times10^{-8}$ & $\mathrm{BR}(K_L\to e\mu,\mu e)$ & $9.9\times10^{-12}$ & $4.7\times10^{-12}$ \\
	$C_{\ell e}^{1333}=1\times10^{-6}$ & $C_{\ell u}^{2311}=1\times10^{-8}$ & $\mathrm{CR}(\mu\to e,\mathrm{Au})$ & $3\times10^{-12}$ & $7\times10^{-13}$ \\
	$C_{\ell e}^{3313}=1\times10^{-6}$ & $C_{eB}^{23}=1\times10^{-12}$ & $\mathrm{BR}(\mu\to e\gamma)$ & $1.1\times10^{-12}$ & $4.2\times10^{-13}$ \\
	  $C_{\ell e}^{3313}=3.2\times10^{-7}$ & $C_{qe}^{1123}=4\times10^{-7}$ & $\mathrm{CR}(\mu\to e,\mathrm{Au})$ & $1.4\times10^{-9}$ & $7\times10^{-13}$ \\
	$C_{\ell e}^{3313}=1\times10^{-6}$ & $C_{qe}^{3323}=1\times10^{-7}$ & $\mathrm{CR}(\mu\to e,\mathrm{Au})$ & $7.4\times10^{-13}$ & $7\times10^{-13}$ \\
	$C_{\ell e}^{3313}=1\times10^{-6}$ & $C_{ed}^{2311}=1\times10^{-8}$ & $\mathrm{CR}(\mu\to e,\mathrm{Au})$ & $4.1\times10^{-12}$ & $7\times10^{-13}$ \\
	$C_{\ell e}^{3313}=1\times10^{-6}$ & $C_{ed}^{2321}=3.4\times10^{-8}$ & $\mathrm{BR}(K_L\to e\mu,\mu e)$ & $9.8\times10^{-12}$ & $4.7\times10^{-12}$ \\
	  $C_{\ell e}^{3313}=3.2\times10^{-6}$ & $C_{ledq}^{2311}=6.2\times10^{-7}$ & $\mathrm{CR}(\mu\to e,\mathrm{Au})$ & $2\times10^{-10}$ & $7\times10^{-13}$ \\
	$C_{\ell e}^{3313}=1\times10^{-6}$ & $C_{ledq}^{2312}=1\times10^{-9}$ & $\mathrm{BR}(K_L\to e\mu,\mu e)$ & $2.3\times10^{-11}$ & $4.7\times10^{-12}$ \\
	  $C_{\ell e}^{3313}=1\times10^{-6}$ & $C_{ledq}^{2321}=3.8\times10^{-8}$ & $\mathrm{BR}(K_L\to e\mu,\mu e)$ & $3.4\times10^{-8}$ & $4.7\times10^{-12}$ \\
	$C_{\ell e}^{3313}=5.5\times10^{-7}$ & $C_{ledq}^{2322}=9.2\times10^{-8}$ & $\mathrm{CR}(\mu\to e,\mathrm{Au})$ & $4.3\times10^{-11}$ & $7\times10^{-13}$ \\
	$C_{\ell e}^{3313}=1\times10^{-6}$ & $C_{lequ}^{(3){2312}}=1\times10^{-8}$ & $\mathrm{BR}(\mu\to e\gamma)$ & $7.5\times10^{-12}$ & $4.2\times10^{-13}$ \\
	$C_{\ell e}^{3313}=1\times10^{-6}$ & $C_{lequ}^{(3){2313}}=1\times10^{-9}$ & $\mathrm{BR}(\mu\to e\gamma)$ & $1.4\times10^{-12}$ & $4.2\times10^{-13}$ \\
	$C_{\ell e}^{3313}=1\times10^{-6}$ & $C_{lequ}^{(3){2321}}=1\times10^{-8}$ & $\mathrm{CR}(\mu\to e,\mathrm{Au})$ & $1.1\times10^{-12}$ & $7\times10^{-13}$ \\
	$C_{\ell e}^{3313}=1\times10^{-6}$ & $C_{lequ}^{(3){2322}}=1\times10^{-9}$ & $\mathrm{BR}(\mu\to e\gamma)$ & $1.4\times10^{-12}$ & $4.2\times10^{-13}$ \\
	$C_{\ell e}^{3313}=1\times10^{-6}$ & $C_{lequ}^{(3){2333}}=1\times10^{-11}$ & $\mathrm{BR}(\mu\to e\gamma)$ & $1.6\times10^{-12}$ & $4.2\times10^{-13}$ \\
	$C_{\ell e}^{3313}=1\times10^{-6}$ & $C_{eu}^{2311}=1\times10^{-8}$ & $\mathrm{CR}(\mu\to e,\mathrm{Au})$ & $3.6\times10^{-12}$ & $7\times10^{-13}$ \\
	$C_{\ell e}^{3313}=1\times10^{-6}$ & $C_{eu}^{2333}=1\times10^{-7}$ & $\mathrm{CR}(\mu\to e,\mathrm{Au})$ & $9\times10^{-13}$ & $7\times10^{-13}$ \\
	${C_{\ell q}^{(1){1311}}}=1.5\times10^{-8}$ & $C_{ledq}^{2333}=2.1\times10^{-7}$ & $\mathrm{CR}(\mu\to e,\mathrm{Au})$ & $3.4\times10^{-12}$ & $7\times10^{-13}$ \\
	$C_{\ell q}^{(1){1333}}=1\times10^{-7}$ & $C_{\ell q}^{(1){2311}}=1\times10^{-7}$ & $\mathrm{CR}(\mu\to e,\mathrm{Au})$ & $1.1\times10^{-12}$ & $7\times10^{-13}$ \\
	$C_{qe}^{3313}=1\times10^{-7}$ & $C_{qe}^{1123}=1\times10^{-7}$ & $\mathrm{CR}(\mu\to e,\mathrm{Au})$ & $7.9\times10^{-13}$ & $7\times10^{-13}$ \\
	$C_{qe}^{3313}=1\times10^{-7}$ & $C_{ledq}^{2321}=3.8\times10^{-8}$ & $\mathrm{BR}(K_L\to e\mu,\mu e)$ & $1.9\times10^{-11}$ & $4.7\times10^{-12}$ \\
	$C_{ed}^{1321}=2.9\times10^{-7}$ & $C_{\ell e}^{3323}=1\times10^{-7}$ & $\mathrm{BR}(K_L\to e\mu,\mu e)$ & $7.2\times10^{-12}$ & $4.7\times10^{-12}$ \\
	$C_{ledq}^{1311}=1\times10^{-8}$ & $C_{\ell e}^{3323}=1\times10^{-7}$ & $\mathrm{CR}(\mu\to e,\mathrm{Au})$ & $2\times10^{-12}$ & $7\times10^{-13}$ \\
	  $C_{ledq}^{1321}=3.5\times10^{-8}$ & $C_{\ell e}^{3323}=6.1\times10^{-7}$ & $\mathrm{BR}(K_L\to e\mu,\mu e)$ & $2.9\times10^{-10}$ & $4.7\times10^{-12}$ \\
	$C_{ledq}^{1321}=3.5\times10^{-8}$ & $C_{qe}^{3323}=1\times10^{-7}$ & $\mathrm{BR}(K_L\to e\mu,\mu e)$ & $1.6\times10^{-11}$ & $4.7\times10^{-12}$ \\
	$C_{ledq}^{1321}=3.5\times10^{-8}$ & $C_{ledq}^{2333}=2.1\times10^{-7}$ & $\mathrm{BR}(K_L\to e\mu,\mu e)$ & $1.1\times10^{-11}$ & $4.7\times10^{-12}$ \\
	$C_{ledq}^{1321}=3.5\times10^{-8}$ & $C_{eu}^{2333}=1\times10^{-7}$ & $\mathrm{BR}(K_L\to e\mu,\mu e)$ & $2\times10^{-11}$ & $4.7\times10^{-12}$ \\
	$C_{ledq}^{1322}=1.2\times10^{-5}$ & $C_{\ell e}^{3323}=1\times10^{-8}$ & $\mathrm{CR}(\mu\to e,\mathrm{Au})$ & $7.4\times10^{-13}$ & $7\times10^{-13}$ \\
	  $C_{ledq}^{1333}=2.9\times10^{-7}$ & $C_{\ell q}^{(1){2311}}=6\times10^{-7}$ & $\mathrm{CR}(\mu\to e,\mathrm{Au})$ & $2.8\times10^{-10}$ & $7\times10^{-13}$ \\
	$C_{ledq}^{1333}=2.9\times10^{-7}$ & $C_{\ell d}^{2311}=1\times10^{-8}$ & $\mathrm{CR}(\mu\to e,\mathrm{Au})$ & $8.8\times10^{-13}$ & $7\times10^{-13}$ \\
	$C_{eu}^{1333}=1\times10^{-7}$ & $C_{qe}^{1123}=1\times10^{-7}$ & $\mathrm{CR}(\mu\to e,\mathrm{Au})$ & $9.6\times10^{-13}$ & $7\times10^{-13}$ \\
	$C_{eu}^{1333}=1\times10^{-7}$ & $C_{ledq}^{2321}=3.8\times10^{-8}$ & $\mathrm{BR}(K_L\to e\mu,\mu e)$ & $2.3\times10^{-11}$ & $4.7\times10^{-12}$ \\
	$C_{\ell u}^{1333}=1\times10^{-7}$ & $C_{\ell q}^{(1){2311}}=1\times10^{-7}$ & $\mathrm{CR}(\mu\to e,\mathrm{Au})$ & $8.9\times10^{-13}$ & $7\times10^{-13}$ \\
    $C_{qe}^{3313}=4\times 10^{-7}$ & $C_{qe}^{3323}=5\times 10^{-7}$ & ${\rm CR}(\mu \to e, {\rm Au})$ & $1.9 \times 10^{-12}$ & $7\times 10^{-13}$ \\
    $C^{(1)1311}_{lq}=5.7\times 10^{-7}$ & $C^{(1)23}_{\phi l}=1.4\times 10^{-8}$ & ${\rm CR}(\mu \to e, {\rm Au})$ & $5.1 \times 10^{-11}$ & $7\times 10^{-13}$ \\
    $C_{eu}^{1333} = 4.3 \times 10^{-7}$ & $C_{eu}^{2333} = 4.9 \times 10^{-7}$ & ${\rm CR}(\mu \to e, {\rm Au})$ & $1.05 \times 10^{-12}$ & $7\times 10^{-13}$ \\
    $C_{lequ}^{(1){1333}} = 2 \times 10^{-7}$ & $C_{\phi l}^{(1){23}} = 1.4 \times 10^{-8}$ & ${\rm CR}(\mu \to e, {\rm Au})$ & $8.5 \times 10^{-9}$ & $7\times 10^{-13}$ \\
    $C_{le}^{3313} = 1\times 10^{-6}$ & $C_{ledq}^{2321} =1\times 10^{-8}$ & ${\rm BR}(K_L \to e\mu,\mu e)$ & $4.09 \times 10^{-8}$ & $4.7\times 10^{-12}$ \\
    $C_{le}^{3313} = 1.1 \times 10^{-6}$ & $C_{qe}^{1123} = 8.7 \times 10^{-7}$ & ${\rm CR}(\mu \to e, {\rm Au})$ & $1.3 \times 10^{-7}$ & $7\times 10^{-13}$ \\
    $C_{eu}^{1323} = 4.3 \times 10^{-7}$  & $C_{eu}^{2332} = 4.9\times 10^{-7}$ & ${\rm CR}(\mu \to e, {\rm Au})$ & $4.8 \times 10^{-12}$ & $7\times 10^{-13}$ \\
    $C^{(3)1313}_{lq} = 6.6 \times 10^{-8}$ & $C^{(3)2331}_{lequ} = 3 \times 10^{-9}$ & ${\rm CR}(\mu \to e, {\rm Au})$ & $1.9 \times 10^{-12}$ & $7\times 10^{-13}$ \\
    $C^{(3)1313}_{lq}= 6.6 \times 10^{-8}$ & $C_{ledq}^{2313} = 1.3 \times 10^{-8}$ & ${\rm CR}(\mu \to e, {\rm Au})$ & $3.8 \times 10^{-11}$ & $7\times 10^{-13}$ \\ 
    $C_{lu}^{1333} = 4.4 \times 10^{-7}$  & $C_{qe}^{3323} = 5 \times 10^{-7}$ & ${\rm CR}(\mu \to e, {\rm Au})$  & $2.5 \times 10^{-12}$ & $7\times 10^{-13}$ \\
    $C_{lequ}^{(1)1323}= 2 \times 10^{-7}$ & $C_{eu}^{2332} = 5 \times 10^{-7}$ & ${\rm CR}(\mu \to e, {\rm Au})$  & $1.4 \times 10^{-8}$ & $7\times 10^{-13}$ \\
\end{longtable}

\begin{longtable}{@{}p{0.25\textwidth}p{0.25\textwidth}p{0.18\textwidth}p{0.15\textwidth}p{0.15\textwidth}@{}}
	\caption{Interesting pairs for T1 for fixed \(C_{e\phi}^{13}=1\times10^{-8}\).}
	\label{tab:T1T2ephi13_consistent}\\
	\toprule
	\textbf{WC 1} & \textbf{WC 2} & \textbf{Observable} & \textbf{Prediction} & \textbf{Bound}\\
	\midrule
	\endfirsthead
	
	\toprule
	\textbf{WC 1} & \textbf{WC 2} & \textbf{Observable} & \textbf{Prediction} & \textbf{Bound}\\
	\midrule
	\endhead
	
	\midrule
	\multicolumn{5}{r}{\footnotesize Continued on next page}\\
	\endfoot
	
	\bottomrule
	\endlastfoot
	$C_{e\phi}^{13}=1\times10^{-8}$ & $C_{\phi\ell}^{(1)\,23}=1\times10^{-9}$ & $\mathrm{BR}(\mu\to eee)$ & $1.5\times10^{-12}$ & $1\times10^{-12}$ \\
	$C_{e\phi}^{13}=1\times10^{-8}$ & $C_{\phi\ell}^{(1)\,23}=1\times10^{-9}$ & $\mathrm{CR}(\mu\to e,\mathrm{Au})$ & $5.7\times10^{-11}$ & $7\times10^{-13}$ \\
	$C_{e\phi}^{13}=1\times10^{-8}$ & $C_{\ell \ell}^{1123}=1\times10^{-9}$ & $\mathrm{BR}(\mu\to eee)$ & $4.3\times10^{-12}$ & $1\times10^{-12}$ \\
	$C_{e\phi}^{13}=1\times10^{-8}$ & $C_{\ell \ell}^{2333}=1\times10^{-8}$ & $\mathrm{CR}(\mu\to e,\mathrm{Au})$ & $2.9\times10^{-12}$ & $7\times10^{-13}$ \\
	$C_{e\phi}^{13}=1\times10^{-8}$ & $C_{\ell e}^{2311}=1\times10^{-9}$ & $\mathrm{BR}(\mu\to eee)$ & $2.8\times10^{-12}$ & $1\times10^{-12}$ \\
	$C_{e\phi}^{13}=1\times10^{-8}$ & $C_{\ell e}^{2333}=1\times10^{-8}$ & $\mathrm{CR}(\mu\to e,\mathrm{Au})$ & $7.6\times10^{-13}$ & $7\times10^{-13}$ \\
	  $C_{e\phi}^{13}=1\times10^{-8}$ & $C_{\ell q}^{(1)\,2311}=1\times10^{-8}$ & $\mathrm{CR}(\mu\to e,\mathrm{Au})$ & $2.2\times10^{-7}$ & $7\times10^{-13}$ \\
	$C_{e\phi}^{13}=1\times10^{-8}$ & $C_{\ell q}^{(1)\,2312}=1\times10^{-9}$ & $\mathrm{CR}(\mu\to e,\mathrm{Au})$ & $2.7\times10^{-11}$ & $7\times10^{-13}$ \\
	$C_{e\phi}^{13}=1\times10^{-8}$ & $C_{\ell q}^{(1)\,2313}=1\times10^{-8}$ & $\mathrm{CR}(\mu\to e,\mathrm{Au})$ & $7.4\times10^{-13}$ & $7\times10^{-13}$ \\
	$C_{e\phi}^{13}=1\times10^{-8}$ & $C_{\ell q}^{(1)\,2321}=1\times10^{-9}$ & $\mathrm{CR}(\mu\to e,\mathrm{Au})$ & $2.7\times10^{-11}$ & $7\times10^{-13}$ \\
	$C_{e\phi}^{13}=1\times10^{-8}$ & $C_{\ell q}^{(1)\,2322}=1\times10^{-8}$ & $\mathrm{CR}(\mu\to e,\mathrm{Au})$ & $1.2\times10^{-10}$ & $7\times10^{-13}$ \\
	$C_{e\phi}^{13}=1\times10^{-8}$ & $C_{\ell q}^{(1)\,2331}=1\times10^{-8}$ & $\mathrm{CR}(\mu\to e,\mathrm{Au})$ & $7.4\times10^{-13}$ & $7\times10^{-13}$ \\
	$C_{e\phi}^{13}=1\times10^{-8}$ & $C_{\ell q}^{(1)\,2333}=1\times10^{-9}$ & $\mathrm{CR}(\mu\to e,\mathrm{Au})$ & $7.7\times10^{-13}$ & $7\times10^{-13}$ \\
	$C_{e\phi}^{13}=1\times10^{-8}$ & $C_{\ell q}^{(3)\,2312}=1\times10^{-9}$ & $\mathrm{CR}(\mu\to e,\mathrm{Au})$ & $3.3\times10^{-11}$ & $7\times10^{-13}$ \\
	$C_{e\phi}^{13}=1\times10^{-8}$ & $C_{\ell q}^{(3)\,2313}=1\times10^{-8}$ & $\mathrm{CR}(\mu\to e,\mathrm{Au})$ & $9.7\times10^{-13}$ & $7\times10^{-13}$ \\
	$C_{e\phi}^{13}=1\times10^{-8}$ & $C_{\ell q}^{(3)\,2321}=1\times10^{-9}$ & $\mathrm{CR}(\mu\to e,\mathrm{Au})$ & $3.3\times10^{-11}$ & $7\times10^{-13}$ \\
	$C_{e\phi}^{13}=1\times10^{-8}$ & $C_{\ell q}^{(3)\,2322}=1\times10^{-8}$ & $\mathrm{CR}(\mu\to e,\mathrm{Au})$ & $1\times10^{-10}$ & $7\times10^{-13}$ \\
	$C_{e\phi}^{13}=1\times10^{-8}$ & $C_{\ell u}^{2311}=1\times10^{-9}$ & $\mathrm{CR}(\mu\to e,\mathrm{Au})$ & $4.5\times10^{-10}$ & $7\times10^{-13}$ \\
	$C_{e\phi}^{13}=1\times10^{-8}$ & $C_{\ell u}^{2322}=1\times10^{-8}$ & $\mathrm{CR}(\mu\to e,\mathrm{Au})$ & $2.8\times10^{-12}$ & $7\times10^{-13}$ \\
	$C_{e\phi}^{13}=1\times10^{-8}$ & $C_{\ell u}^{2333}=1\times10^{-9}$ & $\mathrm{CR}(\mu\to e,\mathrm{Au})$ & $8.9\times10^{-13}$ & $7\times10^{-13}$ \\
	$C_{e\phi}^{13}=1\times10^{-8}$ & $C_{\ell d}^{2311}=1\times10^{-10}$ & $\mathrm{CR}(\mu\to e,\mathrm{Au})$ & $6.7\times10^{-12}$ & $7\times10^{-13}$ \\
	$C_{e\phi}^{13}=1\times10^{-8}$ & $C_{\ell d}^{2322}=1\times10^{-8}$ & $\mathrm{CR}(\mu\to e,\mathrm{Au})$ & $8\times10^{-13}$ & $7\times10^{-13}$ \\
\end{longtable}

\begin{longtable}{@{}p{0.25\textwidth}p{0.25\textwidth}p{0.18\textwidth}p{0.15\textwidth}p{0.15\textwidth}@{}}
	\caption{Interesting pairs for T1 for fixed $C_{e\phi}^{31}=1\times10^{-8}$.}
	\label{tab:T1T2ephi31_consistent}\\
	\toprule
	\textbf{WC 1} & \textbf{WC 2} & \textbf{Observable} & \textbf{Prediction} & \textbf{Bound}\\
	\midrule
	\endfirsthead
	
	\toprule
	\textbf{WC 1} & \textbf{WC 2} & \textbf{Observable} & \textbf{Prediction} & \textbf{Bound}\\
	\midrule
	\endhead
	
	\midrule
	\multicolumn{5}{r}{\footnotesize Continued on next page}\\
	\endfoot
	
	\bottomrule
	\endlastfoot
	$C_{e\phi}^{31}=1\times10^{-8}$ & $C_{eW}^{23}=1\times10^{-13}$ & $\mathrm{BR}(\mu\to e\gamma)$ & $4.9\times10^{-11}$ & $4.2\times10^{-13}$ \\
	$C_{e\phi}^{31}=1\times10^{-8}$ & $C_{eB}^{23}=1\times10^{-13}$ & $\mathrm{BR}(\mu\to e\gamma)$ & $1.6\times10^{-10}$ & $4.2\times10^{-13}$ \\
	$C_{e\phi}^{31}=1\times10^{-8}$ & $C_{\phi e}^{23}=1\times10^{-9}$ & $\mathrm{BR}(\mu\to eee)$ & $1.5\times10^{-12}$ & $1\times10^{-12}$ \\
	$C_{e\phi}^{31}=1\times10^{-8}$ & $C_{\phi e}^{23}=1\times10^{-9}$ & $\mathrm{CR}(\mu\to e,\mathrm{Au})$ & $8.1\times10^{-11}$ & $7\times10^{-13}$ \\
	$C_{e\phi}^{31}=1\times10^{-8}$ & $C_{\ell e}^{1123}=1\times10^{-9}$ & $\mathrm{BR}(\mu\to eee)$ & $2.8\times10^{-12}$ & $1\times10^{-12}$ \\
	$C_{e\phi}^{31}=1\times10^{-8}$ & $C_{ee}^{1123}=1\times10^{-9}$ & $\mathrm{BR}(\mu\to eee)$ & $4.3\times10^{-12}$ & $1\times10^{-12}$ \\
	$C_{e\phi}^{31}=1\times10^{-8}$ & $C_{ee}^{2333}=1\times10^{-8}$ & $\mathrm{CR}(\mu\to e,\mathrm{Au})$ & $2.8\times10^{-12}$ & $7\times10^{-13}$ \\
    $C_{e\phi}^{31}=1\times10^{-8}$ & $C_{ledq}^{2311}=1\times10^{-9}$ & $\mathrm{CR}(\mu\to e,\mathrm{Au})$ & $3\times10^{-8}$ & $7\times10^{-13}$ \\
	$C_{e\phi}^{31}=1\times10^{-8}$ & $C_{ledq}^{2322}=1\times10^{-8}$ & $\mathrm{CR}(\mu\to e,\mathrm{Au})$ & $7.6\times10^{-9}$ & $7\times10^{-13}$ \\
	$C_{e\phi}^{31}=1\times10^{-8}$ & $C_{ledq}^{2333}=1\times10^{-7}$ & $\mathrm{BR}(\mu\to e\gamma)$ & $3.6\times10^{-12}$ & $4.2\times10^{-13}$ \\
    $C_{e\phi}^{31}=1\times10^{-8}$ & $C_{lequ}^{(1)\,2311}=1\times10^{-9}$ & $\mathrm{CR}(\mu\to e,\mathrm{Au})$ & $2.7\times10^{-8}$ & $7\times10^{-13}$ \\
	$C_{e\phi}^{31}=1\times10^{-8}$ & $C_{lequ}^{(1)\,2312}=1\times10^{-8}$ & $\mathrm{BR}(\mu\to e\gamma)$ & $6.3\times10^{-12}$ & $4.2\times10^{-13}$ \\
	$C_{e\phi}^{31}=1\times10^{-8}$ & $C_{lequ}^{(1)\,2313}=1\times10^{-8}$ & $\mathrm{BR}(\mu\to e\gamma)$ & $1.9\times10^{-11}$ & $4.2\times10^{-13}$ \\
	$C_{e\phi}^{31}=1\times10^{-8}$ & $C_{lequ}^{(1)\,2321}=1\times10^{-9}$ & $\mathrm{CR}(\mu\to e,\mathrm{Au})$ & $1.4\times10^{-9}$ & $7\times10^{-13}$ \\
	$C_{e\phi}^{31}=1\times10^{-8}$ & $C_{lequ}^{(1)\,2322}=1\times10^{-8}$ & $\mathrm{BR}(\mu\to e\gamma)$ & $1.2\times10^{-10}$ & $4.2\times10^{-13}$ \\
	$C_{e\phi}^{31}=1\times10^{-8}$ & $C_{lequ}^{(1)\,2323}=1\times10^{-8}$ & $\mathrm{BR}(\mu\to e\gamma)$ & $3.9\times10^{-10}$ & $4.2\times10^{-13}$ \\
	$C_{e\phi}^{31}=1\times10^{-8}$ & $C_{lequ}^{(1)\,2323}=1\times10^{-8}$ & $\mathrm{BR}(\mu\to eee)$ & $2.5\times10^{-12}$ & $1\times10^{-12}$ \\
	$C_{e\phi}^{31}=1\times10^{-8}$ & $C_{lequ}^{(1)\,2323}=1\times10^{-8}$ & $\mathrm{CR}(\mu\to e,\mathrm{Au})$ & $1.5\times10^{-12}$ & $7\times10^{-13}$ \\
	$C_{e\phi}^{31}=1\times10^{-8}$ & $C_{lequ}^{(1)\,2331}=1\times10^{-8}$ & $\mathrm{CR}(\mu\to e,\mathrm{Au})$ & $4\times10^{-11}$ & $7\times10^{-13}$ \\
	$C_{e\phi}^{31}=1\times10^{-8}$ & $C_{lequ}^{(1)\,2333}=1\times10^{-10}$ & $\mathrm{BR}(\mu\to e\gamma)$ & $2.2\times10^{-11}$ & $4.2\times10^{-13}$ \\
	$C_{e\phi}^{31}=1\times10^{-8}$ & $C_{lequ}^{(3)\,2311}=1\times10^{-10}$ & $\mathrm{CR}(\mu\to e,\mathrm{Au})$ & $3.2\times10^{-11}$ & $7\times10^{-13}$ \\
	$C_{e\phi}^{31}=1\times10^{-8}$ & $C_{lequ}^{(3)\,2312}=1\times10^{-9}$ & $\mathrm{BR}(\mu\to e\gamma)$ & $1.1\times10^{-9}$ & $4.2\times10^{-13}$ \\
	$C_{e\phi}^{31}=1\times10^{-8}$ & $C_{lequ}^{(3)\,2312}=1\times10^{-9}$ & $\mathrm{BR}(\mu\to eee)$ & $7\times10^{-12}$ & $1\times10^{-12}$ \\
	$C_{e\phi}^{31}=1\times10^{-8}$ & $C_{lequ}^{(3)\,2312}=1\times10^{-9}$ & $\mathrm{CR}(\mu\to e,\mathrm{Au})$ & $4\times10^{-12}$ & $7\times10^{-13}$ \\
	$C_{e\phi}^{31}=1\times10^{-8}$ & $C_{lequ}^{(3)\,2313}=1\times10^{-10}$ & $\mathrm{BR}(\mu\to e\gamma)$ & $2\times10^{-10}$ & $4.2\times10^{-13}$ \\
	$C_{e\phi}^{31}=1\times10^{-8}$ & $C_{lequ}^{(3)\,2313}=1\times10^{-10}$ & $\mathrm{BR}(\mu\to eee)$ & $1.3\times10^{-12}$ & $1\times10^{-12}$ \\
	$C_{e\phi}^{31}=1\times10^{-8}$ & $C_{lequ}^{(3)\,2313}=1\times10^{-10}$ & $\mathrm{CR}(\mu\to e,\mathrm{Au})$ & $8\times10^{-13}$ & $7\times10^{-13}$ \\
	$C_{e\phi}^{31}=1\times10^{-8}$ & $C_{lequ}^{(3)\,2321}=1\times10^{-9}$ & $\mathrm{CR}(\mu\to e,\mathrm{Au})$ & $1.7\times10^{-10}$ & $7\times10^{-13}$ \\
	$C_{e\phi}^{31}=1\times10^{-8}$ & $C_{lequ}^{(3)\,2322}=1\times10^{-10}$ & $\mathrm{BR}(\mu\to e\gamma)$ & $2.1\times10^{-10}$ & $4.2\times10^{-13}$ \\
	$C_{e\phi}^{31}=1\times10^{-8}$ & $C_{lequ}^{(3)\,2322}=1\times10^{-10}$ & $\mathrm{BR}(\mu\to eee)$ & $1.3\times10^{-12}$ & $1\times10^{-12}$ \\
	$C_{e\phi}^{31}=1\times10^{-8}$ & $C_{lequ}^{(3)\,2322}=1\times10^{-10}$ & $\mathrm{CR}(\mu\to e,\mathrm{Au})$ & $7.5\times10^{-13}$ & $7\times10^{-13}$ \\
	$C_{e\phi}^{31}=1\times10^{-8}$ & $C_{lequ}^{(3)\,2323}=1\times10^{-11}$ & $\mathrm{BR}(\mu\to e\gamma)$ & $4.2\times10^{-11}$ & $4.2\times10^{-13}$ \\
	$C_{e\phi}^{31}=1\times10^{-8}$ & $C_{lequ}^{(3)\,2331}=1\times10^{-8}$ & $\mathrm{CR}(\mu\to e,\mathrm{Au})$ & $4.7\times10^{-12}$ & $7\times10^{-13}$ \\
	$C_{e\phi}^{31}=1\times10^{-8}$ & $C_{lequ}^{(3)\,2332}=1\times10^{-9}$ & $\mathrm{BR}(\mu\to e\gamma)$ & $4.2\times10^{-11}$ & $4.2\times10^{-13}$ \\
	$C_{e\phi}^{31}=1\times10^{-8}$ & $C_{lequ}^{(3)\,2333}=1\times10^{-12}$ & $\mathrm{BR}(\mu\to e\gamma)$ & $2.3\times10^{-10}$ & $4.2\times10^{-13}$ \\
	$C_{e\phi}^{31}=1\times10^{-8}$ & $C_{lequ}^{(3)\,2333}=1\times10^{-12}$ & $\mathrm{BR}(\mu\to eee)$ & $1.5\times10^{-12}$ & $1\times10^{-12}$ \\
	$C_{e\phi}^{31}=1\times10^{-8}$ & $C_{lequ}^{(3)\,2333}=1\times10^{-12}$ & $\mathrm{CR}(\mu\to e,\mathrm{Au})$ & $9.1\times10^{-13}$ & $7\times10^{-13}$ \\
	$C_{e\phi}^{31}=1\times10^{-8}$ & $C_{qe}^{1123}=1\times10^{-8}$ & $\mathrm{CR}(\mu\to e,\mathrm{Au})$ & $2.1\times10^{-7}$ & $7\times10^{-13}$ \\
	$C_{e\phi}^{31}=1\times10^{-8}$ & $C_{qe}^{1223}=1\times10^{-9}$ & $\mathrm{CR}(\mu\to e,\mathrm{Au})$ & $2.2\times10^{-11}$ & $7\times10^{-13}$ \\
	$C_{e\phi}^{31}=1\times10^{-8}$ & $C_{qe}^{2223}=1\times10^{-8}$ & $\mathrm{CR}(\mu\to e,\mathrm{Au})$ & $1\times10^{-10}$ & $7\times10^{-13}$ \\
	$C_{e\phi}^{31}=1\times10^{-8}$ & $C_{qe}^{3323}=1\times10^{-9}$ & $\mathrm{CR}(\mu\to e,\mathrm{Au})$ & $1.1\times10^{-12}$ & $7\times10^{-13}$ \\
	$C_{e\phi}^{31}=1\times10^{-8}$ & $C_{eu}^{2311}=1\times10^{-10}$ & $\mathrm{CR}(\mu\to e,\mathrm{Au})$ & $5.4\times10^{-12}$ & $7\times10^{-13}$ \\
	$C_{e\phi}^{31}=1\times10^{-8}$ & $C_{eu}^{2322}=1\times10^{-8}$ & $\mathrm{CR}(\mu\to e,\mathrm{Au})$ & $3.1\times10^{-12}$ & $7\times10^{-13}$ \\
	$C_{e\phi}^{31}=1\times10^{-8}$ & $C_{eu}^{2333}=1\times10^{-9}$ & $\mathrm{CR}(\mu\to e,\mathrm{Au})$ & $1.3\times10^{-12}$ & $7\times10^{-13}$ \\
	$C_{e\phi}^{31}=1\times10^{-8}$ & $C_{ed}^{2311}=1\times10^{-9}$ & $\mathrm{CR}(\mu\to e,\mathrm{Au})$ & $6.2\times10^{-10}$ & $7\times10^{-13}$ \\
	$C_{e\phi}^{31}=1\times10^{-8}$ & $C_{ed}^{2322}=1\times10^{-8}$ & $\mathrm{CR}(\mu\to e,\mathrm{Au})$ & $7.9\times10^{-13}$ & $7\times10^{-13}$ \\
\end{longtable}

\begin{longtable}{@{}p{0.25\textwidth}p{0.25\textwidth}p{0.18\textwidth}p{0.15\textwidth}p{0.15\textwidth}@{}}
	\caption{Interesting pairs for T1 for fixed \(C_{e\phi}^{23}=1\times10^{-8}\).}
	\label{tab:T1T2ephi23_consistent}\\
	\toprule
	\textbf{WC 1} & \textbf{WC 2} & \textbf{Observable} & \textbf{Prediction} & \textbf{Bound}\\
	\midrule
	\endfirsthead
	
	\toprule
	\textbf{WC 1} & \textbf{WC 2} & \textbf{Observable} & \textbf{Prediction} & \textbf{Bound}\\
	\midrule
	\endhead
	
	\midrule
	\multicolumn{5}{r}{\footnotesize Continued on next page}\\
	\endfoot
	
	\bottomrule
	\endlastfoot
	$C_{e\phi}^{23}=1\times10^{-8}$ & $C_{eW}^{31}=1\times10^{-13}$ & $\mathrm{BR}(\mu\to e\gamma)$ & $4.9\times10^{-11}$ & $4.2\times10^{-13}$ \\
	$C_{e\phi}^{23}=1\times10^{-8}$ & $C_{eB}^{13}=1\times10^{-13}$ & $\mathrm{BR}(\mu\to e\gamma)$ & $5.6\times10^{-13}$ & $4.2\times10^{-13}$ \\
	$C_{e\phi}^{23}=1\times10^{-8}$ & $C_{eB}^{31}=1\times10^{-13}$ & $\mathrm{BR}(\mu\to e\gamma)$ & $1.6\times10^{-10}$ & $4.2\times10^{-13}$ \\
	$C_{e\phi}^{23}=1\times10^{-8}$ & $C_{\phi\ell}^{(1)\,13}=1\times10^{-9}$ & $\mathrm{BR}(\mu\to eee)$ & $1.5\times10^{-12}$ & $1\times10^{-12}$ \\
	$C_{e\phi}^{23}=1\times10^{-8}$ & $C_{\phi\ell}^{(1)\,13}=1\times10^{-9}$ & $\mathrm{CR}(\mu\to e,\mathrm{Au})$ & $5.7\times10^{-11}$ & $7\times10^{-13}$ \\
	$C_{e\phi}^{23}=1\times10^{-8}$ & $C_{\phi\ell}^{(3)\,13}=1\times10^{-9}$ & $\mathrm{BR}(\mu\to eee)$ & $1.7\times10^{-12}$ & $1\times10^{-12}$ \\
	$C_{e\phi}^{23}=1\times10^{-8}$ & $C_{\phi\ell}^{(3)\,13}=1\times10^{-9}$ & $\mathrm{CR}(\mu\to e,\mathrm{Au})$ & $6.3\times10^{-11}$ & $7\times10^{-13}$ \\
	$C_{e\phi}^{23}=1\times10^{-8}$ & $C_{\ell\ell}^{1333}=1\times10^{-8}$ & $\mathrm{CR}(\mu\to e,\mathrm{Au})$ & $2.9\times10^{-12}$ & $7\times10^{-13}$ \\
	$C_{e\phi}^{23}=1\times10^{-8}$ & $C_{\ell e}^{1311}=1\times10^{-9}$ & $\mathrm{BR}(\mu\to eee)$ & $2.8\times10^{-12}$ & $1\times10^{-12}$ \\
	$C_{e\phi}^{23}=1\times10^{-8}$ & $C_{\ell e}^{1333}=1\times10^{-8}$ & $\mathrm{CR}(\mu\to e,\mathrm{Au})$ & $7.6\times10^{-13}$ & $7\times10^{-13}$ \\
	$C_{e\phi}^{23}=1\times10^{-8}$ & $C_{\ell q}^{(1)\,1311}=1\times10^{-8}$ & $\mathrm{CR}(\mu\to e,\mathrm{Au})$ & $2.2\times10^{-7}$ & $7\times10^{-13}$ \\
	$C_{\ell q}^{(1)1311}=1\times10^{-8}$ & $C_{lequ}^{(1)2323}=1\times10^{-8}$ & $\mathrm{CR}(\mu\to e,\mathrm{Au})$ & $1.5\times10^{-11}$ & $7\times10^{-13}$ \\
	$C_{e\phi}^{23}=1\times10^{-8}$ & $C_{\ell q}^{(1)\,1312}=1\times10^{-9}$ & $\mathrm{CR}(\mu\to e,\mathrm{Au})$ & $2.7\times10^{-11}$ & $7\times10^{-13}$ \\
	$C_{e\phi}^{23}=1\times10^{-8}$ & $C_{\ell q}^{(1)\,1321}=1\times10^{-9}$ & $\mathrm{CR}(\mu\to e,\mathrm{Au})$ & $2.7\times10^{-11}$ & $7\times10^{-13}$ \\
	$C_{e\phi}^{23}=1\times10^{-8}$ & $C_{\ell q}^{(1)\,1333}=1\times10^{-8}$ & $\mathrm{BR}(\mu\to eee)$ & $2.1\times10^{-12}$ & $1\times10^{-12}$ \\
	$C_{e\phi}^{23}=1\times10^{-8}$ & $C_{\ell q}^{(1)\,1333}=1\times10^{-8}$ & $\mathrm{CR}(\mu\to e,\mathrm{Au})$ & $7.8\times10^{-11}$ & $7\times10^{-13}$ \\
	$C_{e\phi}^{23}=1\times10^{-8}$ & $C_{\ell q}^{(3)\,1312}=1\times10^{-9}$ & $\mathrm{CR}(\mu\to e,\mathrm{Au})$ & $3.4\times10^{-11}$ & $7\times10^{-13}$ \\
	$C_{e\phi}^{23}=1\times10^{-8}$ & $C_{\ell q}^{(3)\,1321}=1\times10^{-9}$ & $\mathrm{CR}(\mu\to e,\mathrm{Au})$ & $3.4\times10^{-11}$ & $7\times10^{-13}$ \\
	$C_{e\phi}^{23}=1\times10^{-8}$ & $C_{\ell u}^{1311}=1\times10^{-9}$ & $\mathrm{CR}(\mu\to e,\mathrm{Au})$ & $4.5\times10^{-10}$ & $7\times10^{-13}$ \\
	$C_{e\phi}^{23}=1\times10^{-8}$ & $C_{\ell u}^{1322}=1\times10^{-8}$ & $\mathrm{CR}(\mu\to e,\mathrm{Au})$ & $2.9\times10^{-12}$ & $7\times10^{-13}$ \\
	$C_{e\phi}^{23}=1\times10^{-8}$ & $C_{\ell u}^{1333}=1\times10^{-8}$ & $\mathrm{BR}(\mu\to eee)$ & $2\times10^{-12}$ & $1\times10^{-12}$ \\
	$C_{e\phi}^{23}=1\times10^{-8}$ & $C_{\ell u}^{1333}=1\times10^{-8}$ & $\mathrm{CR}(\mu\to e,\mathrm{Au})$ & $9\times10^{-11}$ & $7\times10^{-13}$ \\
	$C_{e\phi}^{23}=1\times10^{-8}$ & $C_{\ell d}^{1311}=1\times10^{-9}$ & $\mathrm{CR}(\mu\to e,\mathrm{Au})$ & $6.8\times10^{-10}$ & $7\times10^{-13}$ \\
	$C_{e\phi}^{23}=1\times10^{-8}$ & $C_{qe}^{1113}=1\times10^{-8}$ & $\mathrm{CR}(\mu\to e,\mathrm{Au})$ & $7.5\times10^{-10}$ & $7\times10^{-13}$ \\
	$C_{e\phi}^{23}=1\times10^{-8}$ & $C_{qe}^{3313}=1\times10^{-8}$ & $\mathrm{BR}(\mu\to e\gamma)$ & $1.9\times10^{-11}$ & $4.2\times10^{-13}$ \\
	$C_{e\phi}^{23}=1\times10^{-8}$ & $C_{ledq}^{1311}=1\times10^{-9}$ & $\mathrm{CR}(\mu\to e,\mathrm{Au})$ & $1.1\times10^{-10}$ & $7\times10^{-13}$ \\
	$C_{e\phi}^{23}=1\times10^{-8}$ & $C_{ledq}^{1322}=1\times10^{-8}$ & $\mathrm{CR}(\mu\to e,\mathrm{Au})$ & $2.7\times10^{-11}$ & $7\times10^{-13}$ \\
	$C_{e\phi}^{23}=1\times10^{-8}$ & $C_{ledq}^{3111}=1\times10^{-9}$ & $\mathrm{CR}(\mu\to e,\mathrm{Au})$ & $3\times10^{-8}$ & $7\times10^{-13}$ \\
	$C_{ledq}^{3111}=1\times10^{-9}$ & $C_{lequ}^{(1)\,2323}=1\times10^{-8}$ & $\mathrm{CR}(\mu\to e,\mathrm{Au})$ & $2.1\times10^{-12}$ & $7\times10^{-13}$ \\
	$C_{e\phi}^{23}=1\times10^{-8}$ & $C_{ledq}^{3122}=1\times10^{-8}$ & $\mathrm{CR}(\mu\to e,\mathrm{Au})$ & $7.6\times10^{-9}$ & $7\times10^{-13}$ \\
	$C_{e\phi}^{23}=1\times10^{-8}$ & $C_{lequ}^{(1)\,1311}=1\times10^{-9}$ & $\mathrm{CR}(\mu\to e,\mathrm{Au})$ & $9.6\times10^{-11}$ & $7\times10^{-13}$ \\
	$C_{e\phi}^{23}=1\times10^{-8}$ & $C_{lequ}^{(1)\,1321}=1\times10^{-9}$ & $\mathrm{CR}(\mu\to e,\mathrm{Au})$ & $5.1\times10^{-12}$ & $7\times10^{-13}$ \\
	$C_{e\phi}^{23}=1\times10^{-8}$ & $C_{lequ}^{(1)\,1323}=1\times10^{-8}$ & $\mathrm{BR}(\mu\to e\gamma)$ & $1.4\times10^{-12}$ & $4.2\times10^{-13}$ \\
	$C_{lequ}^{(1)\,1323}=1\times10^{-8}$ & $C_{\ell q}^{(1)\,2311}=1\times10^{-8}$ & $\mathrm{CR}(\mu\to e,\mathrm{Au})$ & $1.5\times10^{-11}$ & $7\times10^{-13}$ \\
	$C_{e\phi}^{23}=1\times10^{-8}$ & $C_{lequ}^{(1)\,3111}=1\times10^{-9}$ & $\mathrm{CR}(\mu\to e,\mathrm{Au})$ & $2.7\times10^{-8}$ & $7\times10^{-13}$ \\
	$C_{lequ}^{(1)\,3111}=1\times10^{-9}$ & $C_{lequ}^{(1)\,2323}=1\times10^{-8}$ & $\mathrm{CR}(\mu\to e,\mathrm{Au})$ & $1.9\times10^{-12}$ & $7\times10^{-13}$ \\
	$C_{e\phi}^{23}=1\times10^{-8}$ & $C_{lequ}^{(1)\,3112}=1\times10^{-8}$ & $\mathrm{BR}(\mu\to e\gamma)$ & $6.3\times10^{-12}$ & $4.2\times10^{-13}$ \\
	$C_{e\phi}^{23}=1\times10^{-8}$ & $C_{lequ}^{(1)\,3113}=1\times10^{-8}$ & $\mathrm{BR}(\mu\to e\gamma)$ & $1.9\times10^{-11}$ & $4.2\times10^{-13}$ \\
	$C_{e\phi}^{23}=1\times10^{-8}$ & $C_{lequ}^{(1)\,3121}=1\times10^{-9}$ & $\mathrm{CR}(\mu\to e,\mathrm{Au})$ & $1.5\times10^{-9}$ & $7\times10^{-13}$ \\
	$C_{e\phi}^{23}=1\times10^{-8}$ & $C_{lequ}^{(1)\,3123}=1\times10^{-8}$ & $\mathrm{BR}(\mu\to e\gamma)$ & $4\times10^{-10}$ & $4.2\times10^{-13}$ \\
	$C_{e\phi}^{23}=1\times10^{-8}$ & $C_{lequ}^{(1)\,3123}=1\times10^{-8}$ & $\mathrm{BR}(\mu\to eee)$ & $2.5\times10^{-12}$ & $1\times10^{-12}$ \\
	$C_{e\phi}^{23}=1\times10^{-8}$ & $C_{lequ}^{(1)\,3123}=1\times10^{-8}$ & $\mathrm{CR}(\mu\to e,\mathrm{Au})$ & $1.6\times10^{-12}$ & $7\times10^{-13}$ \\
	$C_{lequ}^{(1)\,3123}=1\times10^{-8}$ & $C_{qe}^{1123}=1\times10^{-8}$ & $\mathrm{CR}(\mu\to e,\mathrm{Au})$ & $7.4\times10^{-13}$ & $7\times10^{-13}$ \\
	$C_{e\phi}^{23}=1\times10^{-8}$ & $C_{lequ}^{(1)\,3131}=1\times10^{-8}$ & $\mathrm{CR}(\mu\to e,\mathrm{Au})$ & $4\times10^{-11}$ & $7\times10^{-13}$ \\
	$C_{e\phi}^{23}=1\times10^{-8}$ & $C_{lequ}^{(1)\,3133}=1\times10^{-10}$ & $\mathrm{BR}(\mu\to e\gamma)$ & $2.2\times10^{-11}$ & $4.2\times10^{-13}$ \\
	$C_{e\phi}^{23}=1\times10^{-8}$ & $C_{lequ}^{(3)\,1312}=1\times10^{-9}$ & $\mathrm{BR}(\mu\to e\gamma)$ & $4\times10^{-12}$ & $4.2\times10^{-13}$ \\
	$C_{e\phi}^{23}=1\times10^{-8}$ & $C_{lequ}^{(3)\,1313}=1\times10^{-10}$ & $\mathrm{BR}(\mu\to e\gamma)$ & $7.2\times10^{-13}$ & $4.2\times10^{-13}$ \\
	$C_{e\phi}^{23}=1\times10^{-8}$ & $C_{lequ}^{(3)\,1322}=1\times10^{-10}$ & $\mathrm{BR}(\mu\to e\gamma)$ & $7.5\times10^{-13}$ & $4.2\times10^{-13}$ \\
	$C_{e\phi}^{23}=1\times10^{-8}$ & $C_{lequ}^{(3)\,1333}=1\times10^{-12}$ & $\mathrm{BR}(\mu\to e\gamma)$ & $8.3\times10^{-13}$ & $4.2\times10^{-13}$ \\
	$C_{e\phi}^{23}=1\times10^{-8}$ & $C_{lequ}^{(3)\,3111}=1\times10^{-10}$ & $\mathrm{CR}(\mu\to e,\mathrm{Au})$ & $3.2\times10^{-11}$ & $7\times10^{-13}$ \\
	$C_{e\phi}^{23}=1\times10^{-8}$ & $C_{lequ}^{(3)\,3112}=1\times10^{-9}$ & $\mathrm{BR}(\mu\to e\gamma)$ & $1.1\times10^{-9}$ & $4.2\times10^{-13}$ \\
	$C_{e\phi}^{23}=1\times10^{-8}$ & $C_{lequ}^{(3)\,3112}=1\times10^{-9}$ & $\mathrm{BR}(\mu\to eee)$ & $7\times10^{-12}$ & $1\times10^{-12}$ \\
	$C_{e\phi}^{23}=1\times10^{-8}$ & $C_{lequ}^{(3)\,3112}=1\times10^{-9}$ & $\mathrm{CR}(\mu\to e,\mathrm{Au})$ & $4\times10^{-12}$ & $7\times10^{-13}$ \\
	$C_{e\phi}^{23}=1\times10^{-8}$ & $C_{lequ}^{(3)\,3113}=1\times10^{-10}$ & $\mathrm{BR}(\mu\to e\gamma)$ & $2.1\times10^{-10}$ & $4.2\times10^{-13}$ \\
	$C_{e\phi}^{23}=1\times10^{-8}$ & $C_{lequ}^{(3)\,3113}=1\times10^{-10}$ & $\mathrm{BR}(\mu\to eee)$ & $1.3\times10^{-12}$ & $1\times10^{-12}$ \\
	$C_{e\phi}^{23}=1\times10^{-8}$ & $C_{lequ}^{(3)\,3113}=1\times10^{-10}$ & $\mathrm{CR}(\mu\to e,\mathrm{Au})$ & $8\times10^{-13}$ & $7\times10^{-13}$ \\
	$C_{e\phi}^{23}=1\times10^{-8}$ & $C_{lequ}^{(3)\,3121}=1\times10^{-9}$ & $\mathrm{CR}(\mu\to e,\mathrm{Au})$ & $1.7\times10^{-10}$ & $7\times10^{-13}$ \\
	$C_{e\phi}^{23}=1\times10^{-8}$ & $C_{lequ}^{(3)\,3122}=1\times10^{-10}$ & $\mathrm{BR}(\mu\to e\gamma)$ & $2.1\times10^{-10}$ & $4.2\times10^{-13}$ \\
	$C_{e\phi}^{23}=1\times10^{-8}$ & $C_{lequ}^{(3)\,3122}=1\times10^{-10}$ & $\mathrm{BR}(\mu\to eee)$ & $1.3\times10^{-12}$ & $1\times10^{-12}$ \\
	$C_{e\phi}^{23}=1\times10^{-8}$ & $C_{lequ}^{(3)\,3122}=1\times10^{-10}$ & $\mathrm{CR}(\mu\to e,\mathrm{Au})$ & $7.5\times10^{-13}$ & $7\times10^{-13}$ \\
	$C_{e\phi}^{23}=1\times10^{-8}$ & $C_{lequ}^{(3)\,3123}=1\times10^{-11}$ & $\mathrm{BR}(\mu\to e\gamma)$ & $4.3\times10^{-11}$ & $4.2\times10^{-13}$ \\
	$C_{e\phi}^{23}=1\times10^{-8}$ & $C_{lequ}^{(3)\,3131}=1\times10^{-8}$ & $\mathrm{CR}(\mu\to e,\mathrm{Au})$ & $4.7\times10^{-12}$ & $7\times10^{-13}$ \\
	$C_{e\phi}^{23}=1\times10^{-8}$ & $C_{lequ}^{(3)\,3132}=1\times10^{-9}$ & $\mathrm{BR}(\mu\to e\gamma)$ & $4.2\times10^{-11}$ & $4.2\times10^{-13}$ \\
	$C_{e\phi}^{23}=1\times10^{-8}$ & $C_{lequ}^{(3)\,3133}=1\times10^{-12}$ & $\mathrm{BR}(\mu\to e\gamma)$ & $2.4\times10^{-10}$ & $4.2\times10^{-13}$ \\
	$C_{e\phi}^{23}=1\times10^{-8}$ & $C_{lequ}^{(3)\,3133}=1\times10^{-12}$ & $\mathrm{BR}(\mu\to eee)$ & $1.5\times10^{-12}$ & $1\times10^{-12}$ \\
	$C_{e\phi}^{23}=1\times10^{-8}$ & $C_{lequ}^{(3)\,3133}=1\times10^{-12}$ & $\mathrm{CR}(\mu\to e,\mathrm{Au})$ & $9.2\times10^{-13}$ & $7\times10^{-13}$ \\
\end{longtable}

\bibliographystyle{JHEP}
\bibliography{LFV}
\end{document}